\documentclass[a4paper,11pt]{article}
\pdfoutput=1
\usepackage{jcappub}

\usepackage[english]{babel}

\usepackage{graphicx}
\usepackage{amsmath,amssymb,bm}
\usepackage{hyperref}
\usepackage{braket}
\usepackage{subfigure}
\usepackage{float}
\usepackage[dvipsnames]{xcolor}
\usepackage{mathrsfs}
\usepackage{comment}
\usepackage{multirow}
\usepackage{soul}
\usepackage{mathtools}
\usepackage{mleftright}
\usepackage{sidecap}

\def\H{\mathcal{H}}

\def\S{\mathcal{S}}

\def\T{\mathcal{T}}

\graphicspath{{Plots/}}

\title{Wave-optics limit of the stochastic gravitational wave background}
\author[a]{Alice Garoffolo}
\emailAdd{garoffolo@lorentz.leidenuniv.nl }
\affiliation[a]{Institute Lorentz, Leiden University, PO Box 9506, Leiden 2300 RA, The Netherlands}

\abstract{The stochastic gravitational wave background (SGWB)
is a rich resource of cosmological information, encoded both in its source statistics and anisotropies induced by propagation effects. We provide a theoretical description of it, without employing theoretical tools which rely on the geometric optics limit. 
Our formalism is based on the so-called {\it classical matter approximation} and it is able to capture wave-optics effects, such as interference and diffraction. 
We show that the interaction between the gravitational waves and the cosmic structures along the line-of-sight produce observable scalar and vector polarization modes, on top of modulating the tensorial ones. We build the two point correlation function describing the statistics of the SGWB, and introduce the intensity and gravitational Stokes parameters for all its components. In the case of an unpolarized, Gaussian and statistically homogeneous SGWB, we show that the interaction with matter modulates its intensity and does not generate a net difference between left- and right- helicity tensor and vector modes, as expected. We demonstrate that, in order to produce $Q_T$- and $U_T$- tensor polarization modes the background must posses an hexadecapole anisotropy, while a quandrupole anisotropy can source the vector $Q_V$- and $U_V$- Stokes parameters. 
}

\begin{document}

\maketitle
\flushbottom

%\tableofcontents

\section{Introduction}

The incoherent superposition of many gravitational waves (GW) builds up the so-called stochastic gravitational wave background (SGWB). 
The stochastic nature of the background can be traced back to its source (e.g. inflationary tensor modes~\cite{Bartolo:2016ami,Caprini:2018mtu,Guzzetti:2016mkm,Buonanno:2003th}, cosmic strings and phase transitions~\cite{Auclair:2019wcv,Jenkins:2018nty,Kuroyanagi:2016ugi,Geller:2018mwu,Caprini:2015zlo}, scalar induced GW~\cite{Baumann:2007zm,Ananda:2006af,Kohri:2018awv,Domenech:2021ztg,Fumagalli:2020nvq,Saito:2009jt,Bartolo:2018evs,Pi:2020otn}) or to limitations of specific instruments employed to detect GWs (e.g. cumulative signal from unresolved astrophysical sources as: compact objects binaries~\cite{Regimbau:2011rp,Renzini:2022alw,Farmer:2003pa,Amaro-Seoane:2022rxf}, stellar core collapse events~\cite{Crocker:2015taa}).  
The search for the stochastic background~\cite{Allen:1996gp,Taylor:2013esa} is already ongoing with the online GW observatories~\cite{LIGOScientific:2016fpe,KAGRA:2021mth,NANOGrav:2020bcs}.
%
%The first category is referred to as the cosmological background, while the second one as the astrophysical one. 
%
Exactly as the Cosmic Microwave Background (CMB), the SGWB contains a formidable amount of cosmological information, both in its frequency profile and in the angular power spectrum describing its anisotropies~\cite{LISACosmologyWorkingGroup:2022jok}, justifying the importance of having different technologies to probe it~\cite{Romano:2016dpx,Regimbau:2011rp,Renzini:2021iim,Hotinli:2019tpc}. 
Interferometry is divided in four main observation channels: ground-based detectors sensitive to the $1- 100 \sim$ Hz band (LIGO/Virgo and KAGRA already operating~\cite{LIGOScientific:2016aoc,KAGRA:2020cvd} and the future Einstein Telescope (ET) and Cosmic Explorer~\cite{Maggiore:2019uih,Reitze:2019iox}), space-based observers in the $\sim 10^{-3}$ Hz frequency band (Laser Interferometer Space Antenna (LISA), the Big Bang Observer~\cite{LISA:2017,Crowder:2005nr}, TAiji~\cite{Ruan:2018tsw} and TianQin~\cite{TianQin:2015yph}), pulsar timing arrays (PTA) monitored by radio telescopes~\cite{manchester2013international,NANOGrav:2015aud,Weltman:2018zrl} in the $10^{-6}- 10^{-9}$ Hz band or the CMB itself, which is sensitive to even lower frequencies~\cite{Renzini:2022alw}. 
As the GW spectrum spans more than 20 orders of magnitude in frequency, multi-band GW observation will start a new observational phase for cosmology: the different behaviors in each energy regime can be used to break degeneracies between intrinsic and induced properties of the SGWB ~\cite{DallArmi:2022wnq}, or track different source populations~\cite{Raccanelli:2016cud,Jenkins:2018kxc,Cusin:2019jpv,Cusin:2019jhg,Scelfo:2018sny,Bavera:2021wmw,Wang:2021djr}. 
The SGWB carries cosmological information on two levels: through its sources and through propagation effects, deforming the observed signal as the GWs travel through the inhomogeneous matter distribution.
In case of the astrophysical background, the sources are additionally tracers of the underlying dark matter distribution while it is also known that large primordial non-Gaussianities in the squeezed limit can imprint anisotropies in the primordial SGWB~\cite{Dimastrogiovanni:2021mfs,Alba:2015cms,Dimastrogiovanni:2019bfl}.

\smallskip
The main goal of the present paper is to describe the interactions between the propagating GWs and the matter structures present in the Universe, going beyond the standard treatments and, hence, providing a novel and interesting cosmological probe. 
The current state of the art in literature, describes the SGWB through the characterization of the {\it GW energy density spectrum}, $\Omega_{\rm GW} (f)$, which can either be directly related to properties of the GW's sources~\cite{Pitrou:2019rjz,Cusin:2017fwz,Cusin:2017mjm}, or computed through a Boltzmann equation approach ~\cite{Contaldi:2016koz,Bartolo:2019yeu,LISACosmologyWorkingGroup:2022kbp}.
Line-of-sight effects are then accounted for by analyzing the propagation of GWs through the cosmic web. These include, e.g. Doppler, volume and weak gravitational lensing~\cite{Bartolo:2019yeu,LISACosmologyWorkingGroup:2022kbp,Cusin:2017fwz,ValbusaDallArmi:2020ifo,Cusin:2018rsq,Pitrou:2019rjz,Cusin:2017mjm,Cusin:2019jpv,Jenkins:2018nty,Bertacca:2019fnt} and numerical tools to compute them are already being developed~\cite{Bellomo:2021mer}. 
Because the SGWB anisotropies trace the dark matter distribution, they can be used in combination with galaxy and weak lensing surveys (e.g. KiDS~\cite{KiDS:2013}, DES~\cite{DES:2016jjg}, {\it Euclid}~\cite{Laureijs2011}, the Vera C. Rubin Observatory~\cite{LSST:2009} and Nancy Grace Roman Space Telescope \cite{2015arXiv150303757S}) or the CMB itself, to constrain the various parameters of the cosmological model~\cite{Ricciardone:2021kel,Canas-Herrera:2019npr,Cusin:2018rsq,Adshead:2020bji}.
%Despite the exciting prospects~\cite{Baker:2019ync}, the poor angular resolution typical of GW detectors still poses a serious challenge in accessing the high multipoles of the SGWB anisotropies (ET is expected to probe multipoles up to $\ell \lesssim 50$, while LISA $\ell \lesssim 20$).
%%%%%%%%%%%%%%%%%%%%%%%%%%%%%%%%%%%%%%%%%%%%%%%%%%%%%%%%%%%%%%%%%%%%%%%%
%Here transition to wave-optics
Both the Boltzmann equation and the line-of-sight treatment previously mentioned rely on the 
%As powerful the BE is, it comes with its assumptions, questioning its applicability in more general situation. The principal hypothesis it relies on is the definition of a phase-space for gravitons, namely 
the possibility of describing effectively the GWs as particles, with a well-defined momentum and position at each time, a condition which is met only in the high-frequency limit~\cite{Isaacson:1968hbi,Isaacson:1968zza,Harte:2018wni}. As a drawback, the wave nature of the GWs is neglected and no polarization effects can be included.
Additionally, in this regime, also called the {\it geometric optics limit}, the frequency dependence of the interaction of the SGWB and the cosmological structures is washed away.
%: each high frequency wave is affected equally by the matter overdensities.
%as GWs can be treated effectively as a stream of particles propagating along rays which are geodesics of the background spacetime. Methods that relies on the Boltzmann equation are, therefore, making the BE intrinsically unable to describe the SGWB across its entire frequency spectrum.  
When addressing the opposite situation, namely the {\it wave-optics limit}, the oscillatory nature of the GWs is fully taken into account and frequency dependent interference or diffraction effects arise.  The possibility of capturing these effects is interesting for two main reasons: a resonantly positive interference can boost the GW signal, making the detection possibly more feasible, and the interaction between the waves and the matter structures, in this limit, can polarize the stochastic background.
Even if the observed GW signal doesn't result amplified, having a precise prediction for the line-of-sight effect is important to extract the correct information out of the SGWB.
%
%%%%%%%%%%%%%%%%%%%%%%%%%%%%%%%%%%%%%%%%%%%%%%%%%%%%%%%%%%%%%%%%%%%%%%%%
%Here talk about wave-optics
The literature of wave-optics effects in gravitational lensing of gravitational waves started to flourish after their first detection, and it is starting to draw much attention.
%A great effort in literature has been dedicated to the description of wave-optics effects in gravitational lensing. 
%Depending on the ratio between the GW wavelength and typical curvature scale of the lens gravitational potential, lensing can induce small distortions in the waveform or produce multiple images, which might interfere giving rise to the already mentioned wave-optics effects~\cite{GravitationalLenses}. 
The interference and diffraction induced by a spherically symmetric and static lens has been addressed starting from the pioneering work~\cite{Nakamura1999WaveOI}, and proceeding with~\cite{Nambu:2012wa,Takahashi:2003ix,Takahashi:2005sxa,Oguri:2022zpn,GravitationalLenses}, showing that these become important when the mass of the lens is such that
\begin{equation}
M_L \lesssim 10^8 M_{\odot} \left( \, \frac{f}{mHz} \right)^{-1}\,,
\end{equation}
where $f$ is the frequency of the GW.
Such papers were then generalized to the case of multi-lens systems~\cite{Yamamoto:2003cd}, or lenses composed of binary objects ~\cite{Feldbrugge:2020ycp} and expanding backgrounds~\cite{Takahashi:2005ug}.   Wave-optics phenomena are expected to occur, for instance, in micro-lensing events due to substructures~\cite{Diego:2019lcd,Meena:2019ate} for GWs in the LIGO-Virgo frequency band~\cite{LIGOScientific:2021djp}. Interestingly, it was proposed to use such events to discover unknown objects, such as intermediate mass black holes~\cite{Lai:2018rto}, more exotic forms of compact dark matter~\cite{Jung:2017flg}, or low-mass dark matter halos~\cite{Guo:2022dre} and primordial black holes~\cite{Oguri:2020ldf}. In the case of resolved GW events observed by LISA, it has been assessed that over $(0.1 - 1.6) \%$ of massive black hole binaries in the range of $10^5 - 10^{6.5}$ solar masses will display wave-optics effects~\cite{Gao:2021sxw,LISACosmologyWorkingGroup:2022jok}, while, in the frequency band of the ground-based detectors, it is expected that such events will be visible for sources up to redshift $z_s = 2 - 4$ with third generation observatories~\cite{Dai:2018enj}. 
Since the SGWB contains all the possible frequencies, the probability of having a wave-optics event in this case is 1.
Contrary to lensing in geometric optics, wave-optics effects are frequency dependent and induce non-trivial deformation in the GWs amplitude, making these events very promising candidates to infer properties of lenses~\cite{Caliskan:2022hbu}.
Despite the exciting prospects, all of these works rely on the major assumption of treating the GWs as scalar waves, hence neglecting their tensorial nature. 
%
%
%
%This number refers to resolved events, in the case of the SGWB the probability of having lensing effects in the suitable range for wave-optics phenomena to arise is 1.
%Additionally, subtleties in the transitions between the geometric- and wave- optics regime were addressed in~\cite{Bulashenko:2021fes}, where the authors claim that the standard prescription of comparing the wavelength of the GW and the Schwarzschild radius of the lens might break down near the caustics of the specific lens model taken into account. 
%In general, wave-optics effects in gravitational lensing do not need to be small, as there is no requirement for the lensing convergence to be $\ll 1$ as in weak lensing studies.
%These works come with their assumptions as well, the main one being the treatment of the GWs as scalar waves, hence neglecting their tensorial nature. 
%Additionally, all of these references choose an empty background spacetime except for one localized lens, which is not a proper setting for the purpose of this work. In~\cite{Takahashi:2005ug} the authors consider an expanding Universe but, assuming that the GW wavelength is smaller than the Hubble radius, they effectively neglect the expansion term.
%%%%%%%%%%%%%%%%%%%%%%%%%%%%%%%%%%%%%%%%%%%%%%%%
%Here talk about the polarization and MG
%An optimal description of the SGWB must be able to account for its polarization content as well. Indeed, 
%
%
Leveraging on a parallelism with the CMB, we expect secondary effects to introduce a polarization pattern on the SGWB, tracking the profile of the potential wells along the line-of-sight~\cite{Lewis:2006fu,Dai:2013nda,Hu:2000ee,Durrer:2008eom}. 
Similar effects have been investigated in the case of high-frequency GWs~\cite{Dalang2022PolarizationDO}, where the authors claim that, already at the first order away from the geometric optics limit, new components in the GWs content arise.  
Following this logic, we expect that in the wave-optics regime, the polarization content of the SGWB should become a powerful resource of cosmological information, already in General Relativity. Still, it is generally believed that the detection of a non-trivial GW's polarization content is a sign of breaking of the standard gravitational theory; either because of the presence of parity violating GW sources~\cite{Callister:2017ocg,Omiya:2020fvw,Belgacem:2020nda,Sorbo:2011rz} or of extra propagating degrees of freedom~\cite{Eardley1973GravitationalwaveOA,Nunes:2020rmr,Horndeski:1974}.  
However, the direct detectability of the additional mediator has already been questioned in light of the screening mechanisms characterizing these theories~\cite{Garoffolo:2021fxj}.
It then becomes imperative to disentangle the polarization content arising from new physics from the one due to secondary anisotropies, even more so since efforts in extracting physics from the SGWB polarization content~\cite{Orlando:2022rih} and in characterizing its detection prospects~\cite{Seto:2008sr,Hu:2022byd,Callister:2017ocg} are already taking place.  
%%%%%%%%%%%%%%%%%%%%%%%%%%%%%%%%%%%%%%%%%%%%%%%%
%Here talk about this  paper

\smallskip
Given all of these considerations, the aim of this paper is to provide a formalism describing the SGWB across the entire frequency spectrum, thus without relying on the geometric optics approximation. Attempts to go in this direction can be found in~\cite{Pizzuti:2022nnj,Cusin:2018avf}, though the authors still employ a Boltzmann equation formalism and account for the wave effects by introducing suitable collision terms~\cite{Matzner:1977,Westervelt:1971,Peters:1976,Sanchez1976ScatteringOS,Dolan:2008}. 
%Additionally, these authors do not account for the impact of cosmic structures on the polarization content of the SGWB, which we keep in high consideration.
%Here we opt for a different route since a phase-space formulation for GWs is unclear unless in the high frequency limit. 
On the contrary, we follow closely the logic of~\cite{Takahashi:2005sxa}, and work directly at the level of the linearized Einstein's equations, finding an explicit solution for the GWs. 
In order to do so, we introduce the so-called {\it classical matter approximation}, which consist in considering matter exclusively as a source for the background curvature, thus neglecting its response to the presence of the SGWB. This is a good approximation in the late time Universe, where the relativistic species are subdominant.
%Also, since we are interested in describing only the SGWB, considering a non-dynamical matter content effectively means neglecting its back-reaction.
We solve the linearized Einstein's equations and find a solution for the GW containing scalar, vector and tensor modes. 
From this result, we build the two point functions of the polarization components of the first order Riemann tensor, and relate them to the Stokes parameters. Finally, we provide some numerical results in the simple scenario of a Gaussian,  unpolarized, homogeneous and isotropic zeroth order SGWB.

\smallskip
This paper is organized as follows: in Section~\ref{sec:EOM} we derive the linearized Einstein's equations and introduce the classical matter approximation. In Section~\ref{Section:PerturbedUniverse} we choose as background a perturbed Friedmann-Lema\^{i}tre-Robertson-Walker (FLRW) Universe, and we introduce the iterative scheme we employ to solve the GW's equations of motion, which we explicitly solve in Sections~\ref{sec:0order} and~\ref{sec:1order}. We link the found solution to the electric part of the Riemann tensor, regulating the geodesic motions of a ring of test particles, in Section~\ref{sec:geodesicRiemann} to show that all the modes (scalar, vector and tensor) are observables. 
Finally, in Section~\ref{sec:densityMatrix}, we build the two point correlation function and define the Stokes parameters of the SGWB, and we compute their theoretical predictions in the case of an unpolarized, Gaussian and statistically homogeneous unperturbed SGWB.  Finally, we investigate numerically our findings adding also the assumption of statistical isotropy.

%%%%%%%%%%%%%%%%%%%%%%%%%%%%%%%%%%%%%%%%%%%%%%%%%%%%%%%%%%%%%%%%%%%%%%%%%%%%%%%%%%%%%%%%%%%%%%%%%%%%%%%%%%%%
\section{Equations of motion of metric perturbation}\label{sec:EOM}
We consider a Universe whose geometry is described  by the metric $\bar g_{\mu\nu}$, and consider the propagation of GWs on top of it. 
For our purposes, we consider the background entirely fixed by the matter content of the Universe.  Eventually, we will choose $\bar g_{\mu\nu}$ to represent a perturbed cosmological background, therefore including also matter structures. 
%Even though in this first part of the paper we work in full generality, without ever specifying $\bar g_{\mu\nu}$, we keep in mind the situation where the background Universe is described by an expanding spacetime with cosmological structures as our goal is to study the dynamical properties of a propagating GW through the matter inhomogeneities. 
For this, we introduce two expansion parameters: $\alpha$, tracking the perturbation order of the gravitational wave around $\bar g_{\mu\nu}$, and $\epsilon$, accounting for the cosmic structures around the homogeneous and isotropic spacetime. We will make use of $\epsilon$ only from Section \ref{Section:PerturbedUniverse} onward.

%where $\bar g_{\mu\nu} $ represents the background spacetime. 
%As previously stated, at this point $\bar g_{\mu\nu}$ could be any desired spacetime and from Section \ref{Section:PerturbedUniverse} onward we will choose $\bar g_{\mu\nu} $ to be an FLRW spacetime plus inhomogeneities and anisotropies. 
%We stress again, that in this work the large scale structures of the  Universe are accommodated in  the background metric $\bar g_{\mu\nu}$, while $h_{\mu\nu}$ in Eq.~\eqref{eq:MetricExpansion} is the gravitational wave.

\subsection{Perturbed Einstein's equations: generic stress-energy tensor}
Expanding the total metric as
\begin{equation}\label{eq:MetricExpansion}
    g_{\mu\nu} = \bar g_{\mu\nu} + \alpha\, h_{\mu\nu}\,, \qquad g^{\mu\nu} = \bar g^{\mu\nu} - \alpha\, h^{\mu\nu}
\end{equation} 
we find the equations of motion of the GW can by perturbing Einstein's equations,
\begin{equation}\label{eq:PerturbedEFE}
    \,\delta_{ \alpha}  \bigg[\, G_{\mu\nu} - 8 \pi G \, T_{\mu\nu}\, \bigg]= 0\,.
\end{equation}
In \eqref{eq:PerturbedEFE}, $T_{\mu\nu}$ is the matter stress-energy tensor and $\delta_{ \alpha}$ means linearization to first order in $\alpha$.
Choosing the de-Donder gauge, 
\begin{equation}\label{eq:GaugeDeDonder}
    \bar \nabla^\mu \tilde h_{\mu\nu} = 0\,,
\end{equation}
where $ \tilde h_{\mu\nu} = h_{\mu\nu} - h \bar g_{\mu\nu} / 2$, the perturbation of the Einstein tensor reads
\begin{equation}
     \delta_{\alpha} G_{\mu\nu} = - \frac{\alpha}{ 2 } \left[\bar \Box{\tilde h_{\mu\nu}} +2 \bar R_{\lambda\mu\alpha\nu} h^{\lambda\alpha} - h^\lambda_\nu \bar R_{\lambda \mu} - h^\lambda_\mu \bar R_{\lambda \nu} + h_{\mu\nu} \bar R - \bar g_{\mu\nu} h^{\alpha\beta} \bar R_{\alpha \beta } \right]\,.
\end{equation}
Assume that the background field configuration is on-shell, $\bar G_{\mu\nu} = 8 \pi G \, \bar T_{\mu\nu}$, we rewrite equation \eqref{eq:PerturbedEFE} as
\begin{equation}\label{eq:PerturbedEFEtracereversetransverse}
    \bar \Box{\tilde h_{\mu\nu}} + 2 \bar R_{\lambda\mu\alpha\nu} h^{\lambda\alpha}  - \bar g_{\mu\nu} h^{\alpha\beta} \bar R_{\alpha \beta }    + 8 \pi G  \,  \delta_\alpha \Theta_{\mu\nu} = 0\,,
\end{equation}
where
\begin{equation}\label{eq:Theta}
    \delta_\alpha  \Theta_{\mu\nu} \equiv \frac{1}{2 \alpha } \left[ \bar g_{\mu \sigma} 
 \left( \delta_\alpha  T^\sigma_\nu \right)+  \bar g_{\nu \sigma} \left( \delta_\alpha  T^\sigma_\mu  \right)\right]\,.
\end{equation}

\subsection{Classical matter approximation} 
We regard matter as an external field whose role is {\it exclusively} to curve the background spacetime and unresponsive to the passage of the GWs. Indeed, we are interested in describing the propagation effects on the GW properties (such as strain, polarization and frequency) due to the structures in the Universe, and not how cosmic structures are affected by them. 
%
%
%Accounting for perturbations in the distributions of matter and energy would be equivalent to study the  formation of structures induced by $h_{\mu\nu}$, on top of those already included in $\bar g_{\mu\nu}$. 
%In other words, we do not wish to describe the coupled dynamical system of matter/energy and metric fluctuations, as it is done in structure formation studies. 
%In our setting the LSS are already present, formed via usual channels, and they belong to the background field configuration. 
%We also consider the sources of the GWs to be far away and study the propagation of the waves outside them. 
%In the case of an inflationary background, the sources would be given by the amplitude of the tensor mode at the horizon reentry. 
%
This assumption, which we dub {\it the classical matter approximation} (CM), explicitly consists of choosing,
\begin{equation}\label{eq:ClassicalMatter}
     \delta_\alpha \Theta_{\mu\nu} = 0\,,
\end{equation}
so that Einstein's equations~\eqref{eq:PerturbedEFEtracereversetransverse} become sourceless.
%Basically, we are considering the free propagation of GWs through a Universe described by the metric $\bar g_{\mu\nu}$ which we can choose as preferred. 
%
%
If one thinks of $\bar g_{\mu\nu}$ as a perturbed FLRW spacetime, Eq.~\eqref{eq:ClassicalMatter} can be understood also as assuming that the fluctuations in the matter sector induced by the GWs are negligible compared to the already present large scale structures.  Under this light, it becomes clear that the classical matter approximation is well suited to describe the late time Universe where relativistic species are subdominant.
Compatibly with Eq.~\eqref{eq:ClassicalMatter}, we also expand the observer's 4-velocity to liner order in $\alpha$ as 
\begin{equation}
    u^\mu = \bar u^\mu + \delta_\alpha u^\mu
\end{equation}
and require
\begin{equation}\label{eq:ClassicalMatterVelocity}
    \delta_\alpha u^\mu = 0\,.
\end{equation}
%namely, that the four-velocity of the fluid element is unperturbed by the GWs. 
%Again, one can also think that we are neglecting the contribution to the velocity of the matter fields induced by the traveling GWs compared to the Hubble flow and their peculiar velocity induced by large-scale structures.
This assumption has as a direct consequence that the contraction of $h_{\mu\nu}$ with the unperturbed velocity vanishes. This can be checked using
\begin{align}\label{eq:CompCondition}
u^\mu u_\mu = -1 &\qquad \to \qquad  h_{\mu \nu} \bar u^\mu \bar u^\nu = 0\,,
%u^\mu \nabla_\mu u^\nu = 0 &\to \qquad \bar u^\rho \bar \nabla_\rho \left( \bar u^\mu h^\nu_\mu\right) - \frac12 \bar u^\rho \bar u^\mu \bar \nabla^\nu h_{\mu\rho} = 0\,.
\end{align}
which we interpret as a compatibility condition of our approximation scheme.
Note that Eq.~\eqref{eq:ClassicalMatter} is not equivalent to neglecting the back-reaction of the waves on the spacetime, which is a second order contributions ($ \sim h^2$ ) to the equations of motion of the background $\bar g_{\mu\nu}$. Rather, it means neglecting the interaction between the GWs and the matter species, as the term $\delta_\alpha \Theta_{\mu\nu}$ in Eq.~\eqref{eq:PerturbedEFEtracereversetransverse} descends from quadratic terms of the form $\propto \, h^{\mu\nu} \delta_\alpha \Theta_{\mu\nu}$ in the second order action.

\subsection{Physical meaning of the CM approximation} \label{Sec:ExplanationCM}

In this Section, we illustrate the meaning of the assumption in Eq.~\eqref{eq:ClassicalMatter} by considering the simplified case where $\bar g_{\mu\nu}$ is a homogeneous and isotropic spacetime. 
Since we haven't characterized the metric perturbation $h_{\mu\nu}$ as a gravitational wave yet, at this stage $h_{\mu\nu}$ includes also scalar and vector modes. 
%We made this assumption to avoid accounting for the perturbations in the matter sector induced by the GWs however To understand better this statement, we illustrate what our formalism would predict in the simplified case where $\bar g_{\mu\nu}$ is an homogeneous and isotropic spacetime. 
%
%
In the previous Section we opted for the de-Donder gauge choice, but in principle we could change gauge, and use a more common one in cosmology. For instance, the Poisson or synchronous comoving gauges ~\cite{Ma:1995ey,Malik:2008im}, both tailored to exploit the symmetries of FLRW~\cite{Malik:2008im,Durrer:2008eom}. 
By choosing the Poisson gauge for $h_{\mu\nu}$, the perturbed line element takes the form
%\begin{align}\label{eq:MetricPoissonExample}
%    \text d s^2 &= (\bar g_{\mu\nu} + \alpha h_{\mu\nu}) \text  d x^\mu \text  d x^\nu = \nonumber \\
%    & = a^2 (\tau) \left\{ -(1 + 2 \alpha \psi) \text  d \tau^2 + 2 \alpha  w_i \, \text  d \tau \text  d x^i + [(1 - 2 \alpha \phi) \delta_{ij} + \alpha \gamma_{ij}] \text  d x^i \text  d x^j\right\}\,.
%\end{align}
%
\begin{align}\label{eq:MetricPoissonExample}
     \text d s^2 &= (\bar g_{\mu\nu} + \alpha h_{\mu\nu}) \text  d x^\mu \text  d x^\nu = \nonumber \\
     &= a^2 (\eta) \left\{ -(1 + 2 \alpha H_{00}) d \eta^2 + 2 \alpha  H_{0i} \, d \eta \, d x^i + [(1 - 2 \alpha H) \delta_{ij} + \alpha \gamma_{ij}] d x^i d x^j\right\} \,,
\end{align}
%
%where  $\gamma_{ij}$ is spatial and such that $\gamma^i_i = \partial_i \gamma^i_j = 0$.
%we follow the notations of \cite{Bertschinger:1993xt} but $\gamma_{ij}$ the tensor mode to avoid confusion with the total metric perturbation $h_{\mu\nu}$. 
where we have factored out an $a^2(\eta)$, so that $h_{00} = - 2 a^2 H_{00}$, $h_{0i} = - 2 a^2 H_{0i}$ and similarly for the spatial trace, $ \bar g^{ij} h_{ij } = -2 a^2H$. 
%In Eq.~\eqref{eq:MetricPoissonExample} there are: two scalar potentials, $\psi, \phi$, one transverse vector potential, $w_i$ such that $\partial^i w_i = 0$, and one transverse-traceless tensor potential,  $\gamma_{ij}$ such that $\partial^i \gamma_{ij} = 0$ and $\gamma^i_i = 0$, satisfying
It is known that in Eq.~\eqref{eq:MetricPoissonExample} there are: two scalar potentials, $ H_{00} $ and $ H $, one transverse vector potential, $H_{0i}$ such that $\partial^i H_{0i} = 0$, and one transverse-traceless tensor potential,  $\gamma_{ij}$ such that $\partial^i \gamma_{ij} = 0$ and $\gamma^i_i = 0$, satisfying~\cite{Bertschinger:1993xt}\footnote{For the metric perturbation, one can use the following dictionary between our notation and those of~\cite{Bertschinger:1993xt}: $ H_{00} \, \rightarrow \, \psi$,  $ H \, \rightarrow \, \phi$ and $H_{0i} \, \rightarrow \, w_i$ to write the system of equations above.}
%Indeed, in this section we are studying the dynamics of $h_{\mu\nu}$ over the cosmological background, which is nothing but standard cosmological perturbation theory. }
%\begin{align}
%     \Delta \phi &= 4 \pi G a^2 (\delta \rho + 3 \H \Phi_f) \,, \quad \mbox{with} \quad- \bm{ \nabla} \Phi_f = [(\bar \rho + \bar P) ({\bm v} + {\bm w})]_{\parallel}\,, \label{eq:PoissonEquationsScalar} \\ 
%     \Delta w_i &= 16 \pi G a^2 [(\bar \rho + \bar P) ({\bm v} + {\bm w})]_{\perp}\,,\label{eq:PoissonEquationsVector} \\
%     (\partial^2_\tau + 2 \H \partial_\tau - \Delta) \gamma_{ij} &=  8 \pi G a^2  \,  (\Sigma_{ij})_T \label{eq:WavesOnFLRW}
%\end{align}
\begin{align}
     \nabla^2 H &= 4 \pi G a^2 (\delta_\alpha T^0_0 + 3 \H H_f) \,, \qquad -  \partial_i H_f = [\delta_\alpha {T^0}_i]_{\parallel}\,, \label{eq:PoissonEquationsScalar} \\ 
      \nabla^2 H_{0i} &= 16 \pi G a^2 [\delta_\alpha {T^0}_i]_{\perp}\,,\label{eq:PoissonEquationsVector} \\
     (\partial^2_\tau + 2 \H \partial_\tau - \Delta) \gamma_{ij} &=  8 \pi G a^2  \, \delta_{ik} (\delta_\alpha {T^k}_{j} )_T \,,\label{eq:WavesOnFLRW}
\end{align}
where $\delta_\alpha {T^0}_i = (\bar \rho + \bar P ) (\delta_\alpha u_i + H_{0i})$ and with $(\delta_\alpha {T^k}_{j} )_T$ we mean the tensor part of $\delta_\alpha {T^i}_{j}$.  
%Note that we are using the same notation as~\cite{Bertschinger:1993xt} for the matter quantities, so that 
%The matter quantities on the right-hand-side of the equations above are defined through suitable projections of the perturbed stress-energy tensor, $\delta_\alpha T^\mu_{\nu}$. 
%following the notations of \cite{Bertschinger:1993xt} and where $(\Sigma_{ij})_T$ is the transverse-traceless part of the shear viscosity tensor. 
By imposing Eqs.~\eqref{eq:ClassicalMatter} and~\eqref{eq:ClassicalMatterVelocity}, the system of equations above admits as consistent solution $H = H_{0i} = 0$, and also $H_{00} = 0$, using the remaining perturbed Einstein's equations, if one also assumes that these components vanish on the boundary. Indeed, Eqs.~\eqref{eq:PoissonEquationsScalar} and~\eqref{eq:PoissonEquationsVector} become two sourceless Poisson's equations.
Contrary, being Eq.~\eqref{eq:WavesOnFLRW} a wave equation, it admits non-trivial solutions even if sourceless. 
Furthermore, one can compute the conservation equations for the stress-energy tensor, $\delta_\alpha \left[ \nabla_\mu T^\mu_\nu \right] = 0 $, using the line element in Eq.~\eqref{eq:MetricPoissonExample},
%\begin{align}
%    &\alpha \left \{ \partial_\tau (\delta \rho) + 3 (\H - \dot \phi) (\bar \rho + \bar P) + {\bm \nabla} \cdot \left[ (\bar \rho + \bar P) {\bm v}\right] \right\}= 0\,, \\
%    & \alpha \left \{  \partial_\tau \left[ (\bar \rho + \bar P) ({\bm v} + {\bm w})  \right] + 4 \H (\bar \rho + \bar P) ({\bm v} + {\bm w}) + {\bm \nabla} \delta P + {\bm \nabla} \cdot \Sigma + (\bar \rho + \bar P) {\bm \nabla} \psi \right\}= 0 \label{eq:RelativisticEuler}\,,
%\end{align}
%\begin{align}
%    &\alpha \left \{ \partial_\tau (\delta \rho) + 3 (\H - \dot H) (\bar \rho + \bar P) + {\bm \nabla} \cdot \left[ (\bar \rho + \bar P) {\bm v}\right] \right\}= 0\,, \\
%    & \alpha \left \{  \partial_\tau \left[ (\bar \rho + \bar P) (v_i + H_{0i})  \right] + 4 \H (\bar \rho + \bar P) (v_i + H_{0i}) + \partial_i \delta P + \partial^j  \Sigma_{ij} + (\bar \rho + \bar P) \, \partial_i H_{00} \right\}= 0 \label{eq:RelativisticEuler}\,,
%\end{align}
\begin{align}
    & \partial_\tau \left[ \delta_\alpha T^0_0\right] - 3  \dot H (\bar \rho + \bar P) + \partial_i \left[ \delta_\alpha {T^i}_0\right] = 0\,, \\
    &   \partial_\tau \left[ \delta_\alpha {T^0}_i \right] + 4 \H \left[ \delta_\alpha {T^0}_i \right] + \partial_j \left[ \delta_\alpha {T^j}_i\right]  + (\bar \rho + \bar P) \, \partial_i H_{00} = 0 \label{eq:RelativisticEuler}\,,
\end{align}
where ${\cal H} \equiv a' / a$ is the comoving Hubble parameter.
From the equations above, it is clear that the tensor modes $\gamma_{ij}$ don't contribute to  $\delta_\alpha \left[ \nabla_\mu T^\mu_\nu \right] = 0 $.
%, by the {\it scalar and vector potentials of $h_{\mu\nu}$ }~\cite{Bertschinger:1993xt}.
%either through the two equations above or directly via the Einstein's equations~\eqref{eq:PoissonEquationsScalar} and~\eqref{eq:PoissonEquationsVector}, as a result of Bianchi identities. 
Indeed, the linearized Einstein's equations, because of the Bianchi identities, develop non-trivial scalar and vector potentials  designed exactly to achieve the conservation of the stress-energy tensor, covariantly and at first order in $\alpha$. 
The tensor modes have a different role and are not uniquely fixed by Einstein's equation (one always has the freedom to add a freely propagating wave, satisfying the same equation of motion but sourceless). 
This is a usual result, considering that Einstein equations do not fix the Riemann tensor, $R_{\mu\nu\rho\sigma}$, but only its trace and GWs are contained in the Weyl part of it~\cite{wald1984general}. 
%
%This remark leads us to the main statement of this section, namely {\it the scalar and vector potentials switch on in order for the covariant conservation of energy-momentum to be satisfied at first order in perturbation theory}~\cite{Bertschinger:1993xt} and one can find such conservation equation by combining the energy and momentum constraints as
%\begin{equation}
%    \partial_0 \, (\delta_\alpha G^0_0) + \partial_i\, (\delta_\alpha G^i_{0}) = 0 \qquad \rightarrow \qquad\delta_\alpha  [ \nabla_\mu \, T^\mu_\nu ] = 0 \,.
%\end{equation}
%In general, gauge theories are equipped with Bianchi identities, directly associated to the conservation of the gauge symmetry current on shell.For instance in General Relativity the Bianchi identity $\nabla_\mu G^\mu_\nu = 0$ implies $\nabla_\mu T^\mu_\nu = 0$ or, in electromagnetism, $ \partial_\nu \partial_\mu F^{\mu\nu} = 0 $ implies the conservation of the electric current $ J^\mu$ on shell.  
We point out that the existence/absence of the scalar and vector potentials is not a gauge artifact. To illustrate this point we make the case of electromagnetism where Maxwell's equations are, 
\begin{equation}\label{eq:Maxwell}
   {\bm \nabla} \cdot {\bm E} = {\bm \nabla} \cdot {\bm E}_{\parallel} = 4 \pi \rho \,, \qquad - \partial_\tau {\bm E}_{\parallel} = 4 \pi {\bm J}_{\parallel} \,, \qquad  {\bm \nabla} \times {\bm B} -  \partial_\tau {\bm E}_{\perp} = 4 \pi {\bm J}_{\perp}\,,
\end{equation}
from which it is clear that ${\bm E}$, the electric field, develops a longitudinal component to guarantee charge conservation
\begin{equation}
    \partial_\tau \rho + {\bm \nabla} \cdot {\bm J} = \partial_\tau \rho + {\bm \nabla} \cdot {\bm J}_{\parallel} = 4 \pi \left( \partial_\tau {\bm \nabla} \cdot {\bm E}_{\parallel} - \partial_\tau {\bm \nabla} \cdot {\bm E}_{\parallel} \right)= 0\,.
\end{equation}
%which is instantaneously given by the electric charge. 
%Given that $E_i = - \partial_i \phi - {A_i}'$~\cite{jackson2012classical}, one can understand that ${\bm E}_{\parallel} \neq 0$ implies that the photon field, $A_\mu$, acquires a longitudinal mode.
Since the longitudinal component is present at the level of the observable electric field, it is not a gauge artifact~\cite{Brill67}. Rather, Maxwell's equations contain a redundant scalar equation so that charge conservation is directly built in into them. Similar reasoning can be done about General Relativity, where it can be shown that the scalar and vector cosmological potentials contribute to the Riemann tensor, thus to the geodesic deviation equation, which is a physical observable~\cite{Flanagan:2005yc}. 
Therefore, the classical matter approximation is designed to remove the conserved symmetry charges at first order in $\alpha$, i.e. $\delta_\alpha T^\mu_\nu$, so that the equations of motion do not need to develop extra components in $\alpha h_{\mu\nu}$ to guarantee their conservation. 
Indeed, when assuming Eqs.~\eqref{eq:ClassicalMatter} and~\eqref{eq:ClassicalMatterVelocity},  Eqs.~\eqref{eq:PoissonEquationsScalar}-~\eqref{eq:RelativisticEuler} admit as consistent solution $H = H_{0i} = H_{00} = 0$ and the only non-null part of $\alpha h_{\mu\nu}$ are the tensor modes. 
This mechanism is true regardless of the choice of background, $\bar g_{\mu\nu}$. In the case just examined, the symmetries of FLRW guarantee the separation of scalar, vector and tensor potentials, making the CM approximation not needed.  However,  for less symmetric spacetimes, these modes couple at leading order so that it is not consistent to study tensor modes alone. 

\smallskip
Given all of these considerations, we conclude that the role of the {\it classical matter approximation} is to isolate the freely propagating modes, which we call GWs, when the background spacetime is highly non-symmetric and one cannot rely on the scalar-vector-tensor decomposition.  

\smallskip
Under the CM approximation in Eq.~\eqref{eq:ClassicalMatter}, then Einstein equations in the de-Donder gauge choice, namely Eq.~\eqref{eq:PerturbedEFEtracereversetransverse},  becomes
\begin{equation}\label{eq:EFEClassicalMatterTraceRev}
    \bar \Box{\tilde h_{\mu\nu}} + 2 \bar R_{\lambda\mu\alpha\nu} h^{\lambda\alpha}  - \bar g_{\mu\nu} h^{\alpha\beta} \bar R_{\alpha \beta } = 0\,,
\end{equation}
which will be the starting point of the next Sections. 
Note that in this gauge choice, the constraint equations (time-time and time-space components of Eq.~\eqref{eq:EFEClassicalMatterTraceRev}) are not implemented through first order differential equations of motion as in Eqs.~\eqref{eq:PoissonEquationsScalar} and~\eqref{eq:PoissonEquationsVector}. This is due to the gauge choice: the system of perturbed equations implements causality automatically because the gauge choice made, namely de-Donder, is a covariant fixing.  On the contrary, Poisson's gauge, relying on the slicing of the spacetime, leads indeed to a-causal Poisson's equations~\cite{Bertschinger:1993xt}. 
Therefore, looking for components of the metric perturbation satisfying first order differential equations is not a good strategy to isolate GWs from scalar and vector potentials because it is a gauge dependent statement.

%A wave equation is clearly not elliptic since its solution is given in terms of retarded potentials. Hence, even if we removed the matter sources using Eq.~\eqref{eq:ClassicalMatter} we would not necessarily set to zero all the components of $h_{\mu\nu} $ except for the true propagating degrees of freedom.  
%However, the CM approximation still makes sense also in the context of covariant gauge choices.
%For instance, a scalar part of $h_{\mu\nu}$ is its trace and we can find its equation of motion by tracing Eq.~\eqref{eq:PerturbedEFEtracereversetransverse} 
%\begin{equation}
%    - \bar \Box h +  h^{\mu\nu} \bar R_{\mu\nu} + \delta_\alpha \Theta = 0 \,.
%\end{equation}
%where we used Eq.~\eqref{eq:ThetaImperfectFluid} for $\delta_\alpha \Theta_{\mu\nu}$. 
%We see that by removing the matter sources, namely by setting $\delta_\alpha \Theta = 0$, the trace of the metric perturbation $h$ might still have a source or not according to the choice of $\bar R_{\mu\nu}$ or equivalently of $\bar T_{\mu\nu}$ which we choose here. 

%%%%%%%%%%%%%%%%%%%%%%%%%%%%%%%%%%%%%%%%%%%%%%%%%%%%%%%%%%%%%%%%%%%%%%%%%%%%%%%%%%%%%%%%%%%%%%%%%%%%%%%%%%%%%%%%%%%%%%%%%%%%%%%%%%%%%%%%%%%%%

%%%%%%%%%%%%%%%%%%%%%%%%%%%%%%%%%%%%%%%%%%%%%%%%%%%%%%%%%%%%%%%%%%%%%%%%%%%%%%%%%%%%%%%%%%%%%%%%%%%%%%%%%%%%%%

%%%%%%%%%%%%%%%%%%%%%%%%%%%%%%%%%%%%%%%%%%%%%%%%%%%%%%%%%%%%%%%%%%%%%%%%%%%%%%%%%%%%%%%%%%%%%%%%%%%%%%%%%%%%%%%%%%%%%%%%%%%%%%%%%%%%%%%%%%%%%

\section{Linearization of equations of motion and perturbative scheme} \label{Section:PerturbedUniverse}
%As illustrated in the Introduction, the literature of wave-optics for gravitational waves is very extended. All the papers build up on the pioneering work~\cite{Nakamura1999WaveOI} where the authors solve the gravitational wave equation propagating in a static background containing a single spherically symmetric lens.
%\begin{equation}
%    d \bar s^2 = - (1 + 2 U({\mathbf x})) d t^2 + (1 - 2U ({\mathbf x})) d{\mathbf x}^2 \,.
%\end{equation}
%Wave effects are described as the interference of all possible path that the GW might take, even the non-geodesic ones, using the time delay distance to weight the various trajectories. 
%To arrive to such conclusion, though, the authors make a number of simplifying assumptions: first they neglect a second order spatial derivative term in the wave equation, claiming that such terms are smaller the first order derivatives ones (a sort of high-frequency approximation) and, secondly, they consider the GW as a scalar wave. 
%Albeit its strength, by treating the GWs as scalar waves, their formalism is intrinsically unable to capture any polarization information which we keep in high consideration in this work.
%Indeed, the latter can give us insights about the specific of the GWs sources or it is induced by the so-called secondary anisotropies and in this case it would map the distribution of dark matter and eventual tracers of it, as for the CMB.
In this Section, we set up the perturbative expansion needed to solve  Eq.~\eqref{eq:EFEClassicalMatterTraceRev}, without resorting to the geometric optics approximation nor neglecting the tensor structure of the GW.
%, as previously done in the literature. 
The first step is to make explicit the form of the  background metric, $\bar g_{\mu\nu}$, which follows from a choice of the matter content curving it. 

\subsection{Matter stress energy tensor and background metric}
Since our goal is to describe wave-optics effects, we include: an almost perfect fluid, accounting for the cosmological background and large scale structures of the Universe, plus a pressureless perfect fluid of lenses tracing the dark matter distribution,
\begin{equation} \label{eq:SET}
    \bar T_{\mu\nu} = \bar  \rho \,\bar  u_{\mu}\bar u_{\nu} + (P^{I} + P^{B}) \bar \Lambda_{\mu\nu} + \bar \Sigma_{\mu\nu}\,, \qquad \bar \rho =  \rho^{FLRW}+ \epsilon \, \rho^{LSS} + \rho^L
\end{equation}
where $\bar \Lambda_{\mu\nu} \equiv \bar  g_{\mu\nu} + \bar  u_\mu \bar  u_\nu$ is the orthogonal projector to $\bar u^\mu$, $P^{I} $  and $P^{B}$ are respectively the isotropic and bulk pressure contributions. $\bar \Sigma_{\mu\nu}$ represents the shear stresses which we take traceless and flow-orthogonal, i.e. $\bar  u^\mu \bar  \Sigma_{\mu\nu} =  0$~\cite{Bertschinger:1993xt}.  Note that we are introducing again the expansion parameter $\epsilon$, keeping track of the cosmic structures, and that we are neglecting any velocity bias between the lenses and the large sale structure.
%We stress that $\bar T_{\mu\nu}$ defined in Eq.~\eqref{eq:SET} contains already the large scale structures contributions so that, for instance, $\bar \rho = \bar \rho (x)$. 
%
Compatibly with this definition of $\bar T_{\mu\nu}$, we have that 
\begin{equation}\label{eq:BGPoisson}
    \text d \bar s^2 = \bar g_{\mu\nu} \text d x^\mu \text d x^\nu = a^2 (\tau) \left[ - (1 + 2 \epsilon \psi (x)) \text  d \eta^2 + (1 - 2 \epsilon \phi (x)) \text  d{\mathbf x}^2 \right]
\end{equation} 
following the notations of \cite{Ma:1995ey}.
%where we introduced the parameter $\epsilon$ to account for the large scale structures order, in contrast to $\alpha$ which regarded the gravitational wave expansion.
%At this point we introduced the expansion parameter $\epsilon$, which regulates the order of the {\it large scale structure expansion}, in order to avoid confusion with the previous linearization, which was regulated by the parameter $\alpha$ and regards the gravitational wave. 
%
%
%
To simplify the computations, we consider the case of GWs produced and propagating in the late time Universe so that we can neglect the anisotropic shear stress and set the two first order gravitational potentials equal, $\psi = \phi$.
This assumption is well justified both in the case of astrophysical GW background, whose sources were not present during the radiation dominated epoch,  and for the cosmological one because tensor modes that enter the horizon at early time are greatly damped. 
Moreover, the last primordial modes to reenter the horizon are those characterized by longer wavelength and hence could be the more suited to study wave-optics effects. 
%The late time Universe assumption allows us to neglect the anisotropic shear stress, related to the quadrupoles of the relativistic species distribution functions, and to set the two first order gravitational potentials equal $\psi = \phi$.
We stress that $\psi = \phi$ holds only for the first orders gravitational potentials, as the second orders ones would be different from each other regardless of the shear viscosity. Nonetheless, since we will expand Eq.\eqref{eq:EFEClassicalMatterTraceRev} only up to first order in $\epsilon$, it is consistent for us to set  $\psi = \phi$.
We also point out that
%even when the energy density perturbation is large, as it could be for in the case of the lenses, the two gravitational potentials are still small. Consequently,
our formalism is able to accommodate various perturbative approaches for the background metric, such as the standard weak field limit, or the post-Newtonian~\cite{Malik:2008im,Ma:1995ey}, the parameterized Post-Friedmann~\cite{Hu:2007pj,Hu:2008zd} or hybrids ones~\cite{Carbone:2004iv}.
Neglecting the radiation contribution make the classical matter approximation even more solid, since tensor modes interact with matter via quadrupole anisotropies.

%After all of the considerations above, the GW wave equation we will proceed with is 
%\begin{equation}\label{eq:EQGWnoSlip}
%    \bar \Box{\tilde h_{\mu\nu}} + 2 \bar R_{\lambda\mu\alpha\nu} h^{\lambda\alpha}  - \bar g_{\mu\nu} h^{\alpha\beta} \bar R_{\alpha \beta } + \bar \Lambda_{\mu\nu} \, \frac{ \zeta\, \bar u^\lambda}{2}   \bar \nabla_\lambda h   = 0\,.
%\end{equation}

\subsection{Conformal transformation and linearization of GW equations of motion}

%In this Section, we plug the background metric \eqref{eq:BGPoisson} into Eq.\eqref{eq:EFEClassicalMatterTraceRev} and expand to first order in $\epsilon$. 
The equations obtained by plugging Eq.~\eqref{eq:BGPoisson} into Eq.~\eqref{eq:EFEClassicalMatterTraceRev}, are linear in $\alpha$, but not yet in $\epsilon$. Given how complicated these are, and in the same spirit of~\cite{Nakamura1999WaveOI,Takahashi:2005sxa,Takahashi:2005ug}, we expand Eq.~\eqref{eq:EFEClassicalMatterTraceRev} up to linear order in $\epsilon$, namely in the gravitational potentials  $\psi, \phi$.  
%Indeed, our approach follows closely the logic of~\cite{} with the due generalizations.
Before doing so, it is convenient to perform a conformal transformation, defining the metric  $\hat g_{\mu\nu}$ as
\begin{equation}
     \bar g_{\mu\nu} =  a^2 \hat g_{\mu\nu} \,, \qquad  \bar g^{\mu\nu} =   \hat g^{\mu\nu} / a^2 \,,
\end{equation}
so that the background spacetime at order $\epsilon^0$ is flat.
%It is more convenient to perform a conformal transformation first and then expand up to first order in $\psi, \phi$ the equations of motion, this way, the zeroth order equation will be on Minkowski. 
We also introduce $H_{\mu\nu}$
\begin{equation}
    H^\mu_\nu \equiv h^\mu_\nu \,, \qquad H^{\mu\nu} \equiv \hat g^{\rho\nu} H^\mu_\rho = a^2 \, h^{\mu \nu} \,, \qquad  H_{\mu\nu} \equiv \hat g_{\mu \rho}\, H^\rho_\nu = \frac{1}{a^2 } \, h_{\mu \nu}\,,
\end{equation}
so that its  indices  are raised and lowered with the conformal metric, and 
%. Similarly, we also define
\begin{equation}
    \hat u^\mu \equiv  a \, \bar u^\mu \,, \qquad \hat u_\mu \equiv \hat g_{\mu\nu} \, \hat u^\nu  = \frac{\bar u_\mu}{a}\,,
\end{equation}
such that $\hat u^\mu$ is normalized to $-1$ with $ \hat g_{\mu\nu}$. 
The system of equations describing the GWs is composed of: the gauge condition in Eq.~\eqref{eq:GaugeDeDonder}, the compatibility condition in Eq.~\eqref{eq:CompCondition} and the equation of motion in Eq.~\eqref{eq:EFEClassicalMatterTraceRev}. 
After the conformal transformation, the first two of these read
\begin{eqnarray}
   \hat \nabla_\mu \tilde H^\mu_\nu - 4 \H \, \hat n^\mu \, H_{\mu\nu} + \H H \, \hat n_\nu &=& 0 \,, \label{eq:TransfConf} \\
    \hat u^\mu   \hat u^\nu  H_{\mu\nu} &=& 0 \label{eq:OrthConf}\,,
\end{eqnarray}
where $\hat \nabla_\mu$ is the covariant derivative associated to $\hat g_{\mu\nu}$, $\hat n^\mu \equiv (1,0,0,0)$ and $\tilde H_{\mu\nu} =  H_{\mu\nu} - \frac12 \hat g_{\mu\nu} H$ with $H \equiv \hat g^{\mu\nu} H_{\mu\nu}$. 
Similarly, the GW equation of motion~\eqref{eq:EFEClassicalMatterTraceRev} is  mapped into 
\begin{multline}\label{eq:EomConfomalTransform}
    \hat \Box \tilde H_{\mu\nu} - 2 \H  \hat n^\lambda \, \hat \nabla_{\lambda} \, \tilde H_{\mu\nu} + 2 \hat R_{\mu\alpha\nu\beta} H^{\alpha\beta} -\hat g_{\mu\nu}\, H^{\alpha\beta} \,\hat R_{\alpha\beta}  + 2 \hat g_{\mu\nu} \H^2 (\hat n^\alpha \hat n^\beta \tilde H_{\alpha\beta}) + 4 \H^2 H \hat n_\mu \hat n_\nu  + 
   \\  + \frac{2}{a} \left[\tilde H_\mu^\alpha \hat \nabla_\alpha \hat \nabla_\nu a  + \tilde H_\nu^\alpha \hat \nabla_\alpha \hat \nabla_\mu a \right] - 2 \H \hat n^\alpha \left[ \hat \nabla_\mu \tilde H_{\nu \alpha} + \hat \nabla_\nu \tilde H_{\mu \alpha}\right] + \hat g_{\mu\nu}  H \frac{\hat \Box a}{a} = 0\,,
\end{multline}
with $\hat \Lambda_{\mu\nu} \equiv \eta_{\mu\nu} + \hat n_\mu \hat n_\nu$, and  $\eta_{\mu\nu}$ the Minkwoski metric.
 %
%\subsection{Linearization}
%We introduce the bookkeeping parameter $\epsilon$ to keep track of the order of the LSS and we linearize in $\epsilon$ Eq.~\eqref{eq:EomConfomalTransform}. 
%For simplicity we choose to neglect the shear viscosity of the LSS: in this work we are mainly interested in describing the structures in the late-time universe where shear is negligible. 
%Since we are neglecting shear viscosity we set the two gravitational potentials of the metric perturbation in Newtonian gauge equal.
%Additionally, we consider only the scalar modes in the background metric, assuming that the vectors mode have already decayed and neglecting the tensor ones. 
%In practice, we are considering the conformal metric
%\begin{equation}
%    d \hat{s}^2 = - (1 + 2 \epsilon \phi(x)) d \tau^2 + (1 - 2 \epsilon \phi(x)) d {\bf x}^2\,,
%\end{equation}
%where we explicitely wrote the expansion parameter $\epsilon$.
%At this stage the conformally transformed metric is $d \hat{s}^2 = - (1 + 2 \epsilon \phi(x)) d \tau^2 + (1 - 2 \epsilon \phi(x)) d {\bf x}^2$. 

As previously mentioned, the next step is to linearize to first order in $\epsilon$ Eqs.~\eqref{eq:EomConfomalTransform},~\eqref{eq:TransfConf} and ~\eqref{eq:OrthConf}, similarly to what is done in, e.g., ~\cite{Takahashi:2005sxa}.
To do so, we introduce also the perturbation of the fluid 4-velocity
\begin{equation}
    \hat u^\mu = \hat n^\mu + \epsilon \delta \hat u^\mu = (1 - \epsilon \phi\,, \,\epsilon v^i) = (1 - \epsilon \phi\,, \,\epsilon\, \partial^i v) \,,
\end{equation} 
where, compatibly with the background metric in Eq.~\eqref{eq:BGPoisson}, we kept only the irrotational part of the peculiar velocity.
%We retained only the irrotational part of the peculiar velocity to be consistent with our background metric choice \eqref{eq:BGPoisson} where no vector modes are present. 
The expansion of Eq.~\eqref{eq:TransfConf} up to order ${\cal O} (\epsilon)$ leads to
 \begin{multline}\label{eq:TransfConfFirst}
    \left[ \partial^\mu \left( H_{\mu\nu} - \frac12 \eta_{\mu\nu} (\eta^{\alpha \beta } H_{\alpha\beta})\right) - 4 \H \, \hat n^\mu H_{\mu\nu} + H \, \H \,  \hat n_\nu \right]
     + \epsilon \bigg [ 2 \phi \, \partial^\mu \tilde H_{\mu\nu} + 8 \H \, \phi \, \hat n^\mu H_{\mu\nu}  + \\ + 2 \H \, \phi H \, \hat n_\nu  +    4 \H \, \phi \, \hat n_\nu 
     (\hat n^\alpha \hat n^\beta H_{\alpha \beta }) + 4 \phi' \, \hat n^\mu \, H_{\mu\nu }  + 4 \phi \, \hat n^\mu \, H'_{\mu\nu}   - 2 \phi \, \hat n^\alpha \hat n^\beta \,\partial_\nu H_{\alpha \beta }  \bigg] = 0 \,,
    \end{multline}
while the one of Eq.~\eqref{eq:OrthConf} gives
\begin{equation}\label{eq:CompConditionFirst}
    H_{\mu\nu} \hat n^{\mu} \hat n^{\mu} + 2 \epsilon  H_{\mu\nu} \hat n^{\mu} \delta {\hat u}^\nu= 0\,.
\end{equation}
The first order expression of Eq.~\eqref{eq:EomConfomalTransform} is very complicated, and we report explicitly in the Appendix in Eqs.~\eqref{app:EOMlinearized0} and ~\eqref{app:EOMlinearized1}.
Here, we write it schematically  as:
\begin{equation}\label{eq:EOMLinearized}
    \left[ {\cal O}_0 \, H \right]_{\mu\nu} + \epsilon \left[ {\cal O}_1 \, H \right]_{\mu\nu} = 0\,.
\end{equation}

\subsection{Perturbative scheme}\label{Section:PerturbedUniverseScheme}
Given the complexity of Eq.~\eqref{eq:EOMLinearized}, we solve it perturbatively by expanding also the gravitational wave in powers of $\epsilon$ 
\begin{equation}\label{eq:HepsilonExpansion}
    H_{\mu\nu} = H^{(0)}_{\mu\nu} + \epsilon H^{(1)}_{\mu\nu} + \epsilon^2 H^{(2)}_{\mu\nu} + \dots\,
\end{equation}
and solve the system of equations iteratively, as done also in~\cite{Takahashi:2005ug}  for instance.
The meaning of such decomposition is that the $i-th$ order GW is a solution of a particular equation at $i-th$ order in $\epsilon$. The expansion in Eq.~\eqref{eq:HepsilonExpansion} allows accounting for the effects of the matter inhomogeneities on the propagating GWs perturbatively in a weak-field approximation. We will show that $H^{(0)}_{\mu\nu}$ constitutes GW propagating on a FRLW Universe,  $H^{(1)}_{\mu\nu}$ accounts for the first order corrections induced by cosmic structures, and so on. 
Looking at the equations of motion in Eq.~\eqref{eq:EOMLinearized} we find that each $i-th$ modes satisfy
\begin{eqnarray}
    \left[ {\cal O}_0 \, H^{(i)} \right]_{\mu\nu} + \left[ {\cal O}_1 \, H^{(i-1)} \right]_{\mu\nu} = 0\,,
\end{eqnarray}
where $i =0, 1,2, \dots$ and with the convention $H^{(-1)}_{\mu\nu} = 0$ and we apply the same procedure to Eqs.~\eqref{eq:TransfConfFirst} and~\eqref{eq:CompConditionFirst}. 
Since we neglected terms of order $ \geq \epsilon^2$ to find Eqs.~\eqref{eq:CompConditionFirst}, ~\eqref{eq:TransfConfFirst} and~\eqref{eq:EOMLinearized}, we consider only $H_{\mu\nu} = H^{(0)}_{\mu\nu} + \epsilon H^{(1)}_{\mu\nu} $. As stated in~\cite{Takahashi:2005ug,Takahashi:2005sxa}, this is equivalent to perform a Born approximation.

\section{The zeroth order gravitational wave}\label{sec:0order}
In this Section, we derive the equation of motion for the zeroth order GW, namely $H^{(0)}_{\mu\nu}$, and solve it introducing its transfer function and the polarization basis. Although the results of this Section are not new, their derivation allows us to show explicitly the CM approximation at work and also introduce some notation, which we will use later on.

\subsection{Equation of motion}
The equations that $H^{(0)}_{\mu\nu}$ satisfies can be found taking the $\epsilon^0$ order of  Eqs.~\eqref{eq:TransfConfFirst},~\eqref{eq:CompConditionFirst} and~\eqref{eq:EOMLinearized}. 
The first two give
\begin{align}\label{eq:gaugeH0}
\hat n^\mu \hat n^\nu H^{(0)}_{\mu\nu} = 0 \,, \qquad 
\partial^\mu \tilde  H^{(0)}_{\mu\nu} - 4 \H \, \hat n^\mu H^{(0)}_{\mu\nu} + H^{(0)} \, \H \, \hat n_\nu= 0 \,,
\end{align}
where $H^{(0)} = \eta^{\alpha \beta } H^{(0)}_{\alpha\beta}$  and $\tilde  H^{(0)}_{\mu\nu} = H^{(0)}_{\mu\nu} - \frac12 \eta_{\mu\nu} H^{(0)} $.
%We can combine these two conditions to see that 
%\begin{equation}
%  \hat n^\nu (  \partial^\mu  \tilde H^{(0)}_{\mu\nu} + H^{(0)}\, \H \hat n_\nu ) = - \H  H = 0 \qquad \rightarrow \qquad H = 0\,,
%\end{equation}
%where we have used also $\hat n^\mu H^{(0)}_{\mu\nu} = 0$. 
We use these conditions in the equations of motion~\eqref{eq:EOMLinearized}, and find
\begin{align}\label{eq:UnperturbedGW}
      \bigg[ {\cal O}_0 \, H^{(0)} \bigg]_{\mu\nu}& = \,\Box_\eta  \tilde H^{(0)}_{\mu\nu} - 2 \H (\tilde H^{(0)}_{\mu\nu})' + \H \Big(  \hat n_\nu  \partial_\mu H^{(0)}+  \hat n_\mu \partial_\nu H^{(0)} \Big) - (\hat n_\mu \hat n_\nu  + \hat \Lambda_{\mu\nu}  )  \H' H^{(0)}  - \nonumber \\
 &- 2 \H \Big(  \hat n^\alpha \partial_\mu H^{(0)}_{\alpha \nu } + \hat n^\alpha \partial_\nu H^{(0)}_{\alpha \mu}\Big)  + 2 ( \H' + \H^2) \hat n^\alpha \Big( \hat n_\mu H^{(0)}_{\alpha\nu} + \hat n_\nu H^{(0)}_{\alpha\mu}\Big) \,.
\end{align}
The details of the computations can be found in Appendix~\ref{app:LinearizedEOM} and~\ref{app:GWeq} (see Eqs.~\eqref{app:EOMlinearized0},~\eqref{app:EOMlinearized1} and ~\eqref{app:O0H0}).
From Eq.~\eqref{eq:UnperturbedGW} we extract the time-time and time-space components
\begin{align}
    \frac12  \Box_\eta  H^{(0)} - 3 \H (H^{(0)})' -  \H' H^{(0)} &= 0   \qquad \qquad\mbox{Time-Time}\\
     \Box_\eta H^{(0)}_{0i} -4 \H ( H^{(0)}_{0i})' +  H^{(0)}_{0i} (6 \H^2 - 2 \H') - \H \, \partial_i H^{(0)} &= 0  \qquad \qquad \mbox{Space-Time} 
\end{align}
%
%\begin{align}
%    \Box H^{(0)} + \dots &= 0 \\
%     \Box H^{(0)}_{0i} +
%     \dots &= 0
%\end{align}
%
%
%\begin{align}
%    \Box H^{(1)} + \dots &= -4\, H^{(0)}_{ij} \partial^i \partial^j \phi \\
%     \Box H^{(1)}_{0i} + \dots &= -\frac{2}{a} \partial^k \phi \, (a H^{(0)}_{ki})'
%\end{align}
%
where $ \Box_\eta  = \eta^{\mu\nu} \partial_\mu \partial_\nu =  - \partial^2_\tau + \partial^i\partial_i$.
We choose 
\begin{equation}\label{eq:GW0TracelessSpatial}
    H^{(0)} = H^{(0)}_{0i} = 0 \,,
\end{equation}
as solution of the system above invoking also the fact that in General Relativity there are no isolated scalar or vector sources. 
In this case, Eq.~\eqref{eq:gaugeH0} tells us that the zeroth order GW is spatial and transverse. 
%From the second equation in the system above we see that $H^{(0)}$ cannot depend on time. Additionally, none of the wave equation above for  $H^{(0)}$ has source and there are no scalar sources in General Relativity. 
Using all of these results, the zeroth order GW equation is
\begin{equation}\label{eq:GW0Final}
 \Box_\eta H^{(0)}_{ij} - 2 \H (H^{(0)}_{ij})' = 0\,,
\end{equation}
as it is shown in Appendix~\ref{app:GWeq}.
%\begin{equation}
%         \bigg[ {\cal O}_0 \, H^{(0)} \bigg]_{\mu\nu} &= \,\Box_\eta  H^{(0)}_{\mu\nu} - 2 \H \, (H^{(0)}_{\mu\nu})'=0
%\end{equation}
As expected, we have found that on FLRW 
%out of the $8$ gauge conditions in Eq.\eqref{eq:gaugeH0} there are two that are degenerate so that also the trace is set to zero still preserving $2$ degrees of freedom. Therefore
$H^{(0)}_{\mu\nu}$ is purely spatial, traceless, transverse, and it satisfies the standard damped wave equation. 
This result is not surprising as, at the order $\epsilon^0$, we are studying metric fluctuations around the FLRW spacetime, as we did in Section~\ref{Sec:ExplanationCM}, but in de-Donder gauge rather than Poisson gauge and under the CM approximation. 
%Note, though, that Poisson's gauge is not an option for $H^{(0)}_{\mu\nu}$ because of the compatibility condition in Eq.~\eqref{eq:CompCondition} which sets $H^{(0)}_{00} = 0$ as it is shown in Eq.~\eqref{eq:gaugeH0}.

\subsection{Solution}\label{sec:FreeSolution}
%Given the results of the previous Section, we solve Eq.~\eqref{eq:GW0Final} with $H^{(0)}_{\mu\nu}$ transverse and traceless.
%\begin{equation}
%     \eta^{\mu\nu} H^{(0)}_{\mu\nu} =0 \,, \qquad \hat n^\mu H^{(0)}_{\mu\nu} =0 \,, \qquad \eta^{\mu\alpha} \partial_\alpha H^{(0)}_{\mu\nu} =0 \,, \qquad  \Box_\eta H^{(0)}_{\mu\nu} - 2 \H (H^{(0)}_{\mu\nu})' =0 \,. 
%\end{equation}
%Because of the symmetry under spatial translations, $H^{(0)}_{\mu\nu}$ has the form $\sim  e^{-i {\bf k} \cdot {\bf x}}\, H^{(0)}_{\mu\nu, {\bf k}}(\tau) $. 
 %
%\begin{equation}
%    H^{(0)}_{\mu\nu} (\tau, {\mathbf{x}})= \int d^3 k \, e^{i {\bf k} \cdot {\bf x}} \, H^{(0)}_{\mu\nu, {\bf k}}(\tau) 
%\end{equation}
We solve the $0$-th order system of equations in Fourier space, where each ${\bf k}$ mode satisfies
\begin{equation}\label{eq:H0FourierEqs}
     \eta^{\mu\nu} H^{(0)}_{\mu\nu, {\bf k}} =0 \,, \quad  H^{(0)}_{0\nu, {\bf k}} =0 \,, \quad k^i H^{(0)}_{i \nu, {\bf k}} =0 \,, \quad (H^{(0)}_{ij, {\bf k}} )'' + 2 \H (H^{(0)}_{ij, {\bf k}})' + k^2 H^{(0)}_{ij, {\bf k}} =0 \,, 
\end{equation}
where $k^2 \equiv \delta_{ij} k^i k^j$ and $H^{(0)}_{0\nu, {\bf k}} \equiv \hat n^\mu H^{(0)}_{\mu\nu, {\bf k}}$. At this order, if the  GW sources  are isotropically distributed, the GW depends only on the modulus of ${\bf k}$, hence we shall write $ H^{(0)}_{\mu\nu, {k}}$.
We define the 4-vector $k^\mu \equiv \{ k^0, k^i \}$ such that 
%$k^i$ is the Fourier transform mode and 
$k^0 \equiv - \hat n^\mu k_\mu$. Note that this vector is not necessarily null, as the zero component is  not fixed by a dispersion relation. 
%or  equivalently $k_0 \equiv  \hat n^\mu k_\mu$. 
%The dispersion relation will be fixed by the equations of motion as usual. 
However, since the $0 \nu$ components of the GW vanish, the GW is orthogonal to the entire 4-vector $k^\mu $.
%$k^\mu H^{(0)}_{\mu\nu, { k}} =0$.
%%%%%%%%%%%%%%%%%%%%%%%%%%%%%%%%%%%%%%%%%%%%%%%%%%%%%%%%%%%%%%%%%%%%%%%%%%%%%%%%%%%%%%%%
We introduce a vector basis $\{\hat e_\mu^{\hat a}\} = \{ \hat e_\mu^{\hat 0} ,\, m^{\pm}_\mu,\, \hat e_\mu^{\hat 3}\} $ such that 
\begin{equation}\label{eq:ComplexBasisVector}
    \hat e_\mu^{\hat 0} \equiv \hat n_\mu \,, 
    \qquad \hat e_\mu^{\hat 3} \equiv
    \,\frac{\left[ k_\mu - k^0 \hat n_\mu \right]}{ k } = [0, \hat k_i]\,,
    \qquad  m^{\pm}_\mu \equiv \frac{\hat e_\mu^{\hat 1} \pm i \hat e_\mu^{\hat 2}}{\sqrt{2}} \,,
\end{equation}
with $k_i = k \hat k_i$, where $\{ \hat e^\mu_{\hat 1}, \hat e^\mu_{\hat 2}\}$ are orthonormal between themselves and  $\{ \hat e^\mu_{\hat 0}, \hat e^\mu_{\hat 3}\}$~\cite{Newman566N,Eardley1973GravitationalwaveOA,Will:2018bme,chandrasekhar1998mathematical}. It is important to keep in mind that, since $ m^\pm_{\mu}$ are built orthogonal to $\hat e^\mu_{\hat 3}$, they depend on the direction of the propagating wave, $\hat k$. The scalar products between the elements of $\{\hat e^\mu_{\hat a}\} $ are summarized in
\begin{equation}\label{eq:ScalarProduct}
   \eta_{\hat a \hat b} = 
    \begin{bmatrix}
     -1 & 0 & 0 & 0 \\
     0 & 0 & 1 & 0 \\
     0 & 1 & 0 & 0 \\
     0 & 0 & 0 & 1 
    \end{bmatrix}\,.
\end{equation}
%
%The vectors $ m^{\mu} , \,\bar m^{\mu}$ can be built from an orthonormal tetrad $\{ e^\mu_{\hat 0} ,\,e^\mu_{\hat 1} , \,e^\mu_{\hat 2},\, e^\mu_{\hat 3}\} $ as
%\begin{equation}\label{eq:ComplexBasisVector}
%    m^\mu \equiv \frac{\hat e^\mu_{\hat 1} + i \hat e^\mu_{\hat 2}}{\sqrt{2}} \,, \qquad  \bar m^\mu \equiv \frac{\hat e^\mu_{\hat 1} - i \hat e^\mu_{\hat 2}}{\sqrt{2}} \,.
%\end{equation} 
With these vectors, we build a basis for rank-2 symmetric tensors, as 
\begin{equation}\label{eq:DefPolBasisTensors}
    Y^B_{\mu\nu} \equiv \:\frac{m^+_{( \mu} m^-_{ \nu)}}{\sqrt{2}} \,, \qquad  
    Y^0_{\mu\nu} \equiv \:\frac{{\hat e}_{(\mu}^{\hat 3} {\hat e}_{\nu)}^{\hat 3} -  m^+_{( \mu} m^-_{ \nu)}}{\sqrt{6}} \,, \qquad 
    Y^{\pm 1}_{\mu\nu} \equiv \:\mp \frac{m^\pm_{(\mu} {\hat e}_{\nu)}^{\hat 3} }{\sqrt{2}} \,, \qquad 
    Y^{\pm 2}_{\mu\nu} \equiv \: \frac{m^\pm_{(\mu} {m}^\pm_{\nu)} }{2} \,.
\end{equation}
Among the elements above, $Y^B_{\mu\nu}$ is the only non-traceless one, while the others constitute a basis for symmetric trace-free tensors.
In particular, these last five elements represent the $\ell=2$ and $m = 0, \pm 1, \pm 2$ irreducible representation of the rotation group, hence their names. Because of this, their transformation properties under a rotation are known~\cite{Pitrou:2019ifq,Durrer:2008eom}. 
We decompose each $k$ of the GW mode as 
\begin{equation}\label{eq:PolDecH0}
    H^{(0)}_{\mu\nu, { k}} (\tau) = H^{(0)}_{B , {\bf k}} (\tau)\: Y^{B }_{\mu\nu} (\hat k) + \, \sum_{\hat a} H^{(0)}_{\hat a , {\bf k}} (\tau)\: Y^{\hat a }_{\mu\nu} (\hat k) \,,
\end{equation}
where the sum runs over $\hat a = \{0, \pm 1, \pm 2\}$.
%The symmetry of the decomposition coefficients under the exchange of $\hat a$ and $\hat b$ justifies the factor $1/4$ in the definition of the polarization basis elements.
Because the basis elements in Eq.~\eqref{eq:DefPolBasisTensors} depends on $\hat k$, the decomposition coefficients $ H^{(0)}_{\hat a, {\bf k}}$ must depend on the direction of ${\hat k}$ in such a way that the left-hand side of Eq.~\eqref{eq:PolDecH0} doesn't depend on the direction of propagation.   
This is a key point in our discussion, and it will be at the core of the study of the polarization content of $H^{(1)}_{\mu\nu}$. 
Inserting the decomposition in Eq.~\eqref{eq:PolDecH0} into Eqs.~\eqref{eq:H0FourierEqs}, and using the scalar product rules~\eqref{eq:ScalarProduct}, it is straightforward to find
\begin{equation}
    H^{(0)}_{B, {\bf k}} = H^{(0)}_{0, {\bf k}} = H^{(0)}_{ \pm 1, {\bf k}} = 0\,.
\end{equation}
Therefore, the GW is decomposed on the polarization basis as
\begin{equation}\label{eq:GWpluscross}
    H^{(0)}_{\mu\nu, { k}} = H^{(0)}_{+2, \bf k} \: Y^{+2}_{\mu\nu}({\hat k}) + H^{(0)}_{-2, \bf k} \: Y^{-2}_{\mu\nu} ({\hat k}) = \sum_{\mathring{a}} H^{(0)}_{\mathring{a}, {\bf k}} (\tau)\: Y^{\mathring{a}}_{\mu\nu} (\hat k)  \,,
\end{equation}
where we have introduced the notation $\mathring{a} = \{ \pm 2\}$. The decomposition above is the standard expression of a gravitational wave in terms of left ($ H^{(0)}_{+2, \bf k}$) and right ($H^{(0)}_{-2, \bf k} $) helicity modes, whose equations of motion are
\begin{equation}\label{eq:0thWaveEq}
    (H^{(0)}_{\mathring{a}, {\bf k}} )'' + 2 \H (H^{(0)}_{\mathring{a}, {\bf k}})' + k^2 H^{(0)}_{\mathring{a}, {\bf k}} =0 \,.
\end{equation}
%where we introduce the index notation $\mathring{a} = \{ mm, \bar m \bar m\}$. 
%
%\begin{equation}
%    (H^{(0)}_{\lambda, {\bf k}} )'' + 2 \H (H^{(0)}_{\lambda, {\bf k}})' + k^2 H^{(0)}_{\lambda, {\bf k}} =0 \,
%\end{equation}
%\begin{equation}
%    H^{(0)}_{\lambda}(\tau,{\bf k})  \equiv H_{\lambda}({\tau_s}, {\bf k})  \, \T^H (\tau,{ k}) 
%\end{equation}
%
%\begin{equation}
%    (\T^H_{\lambda, {\bf k}} )'' + 2 \H (\T^H_{\lambda, {\bf k}})' + k^2 \T^H_{\lambda, {\bf k}} =0 \,
%\end{equation}
%\begin{equation}
%    \Big \langle H^*_\lambda ( {\bf k}, \tau^1_s) H_\lambda ( {\bf p}, \tau^2_s) \Big \rangle
%\end{equation}
We introduce the transfer function for gravitational waves as 
\begin{equation}
    H^{(0)}_{\mathring{a}}(\tau,{\bf k})  \equiv H_{\mathring{a}}({\tau_s}, {\bf k})  \, \T^H (\tau,{ k}) 
\end{equation}
where $H^{(0)}_{\mathring{a}}({\tau_s}, {\bf k})$ is the gravitational wave amplitude for the ${\bf k }$ mode at its source conformal time, such that 
\begin{equation}
    \T^H({\tau_s},{ k})= 1 \,.
\end{equation}
The transfer function satisfies the same differential equation as of $H^{(0)}_{\mathring{a}, {\bf k}}$, Eq.~\eqref{eq:0thWaveEq}, but with different boundary conditions. We absorbed the dependence on the direction $\hat k$, entirely due to the choice of elements in the tetrad, in the initial condition $H^{(0)}_{\mathring{a}, {\bf k}} ({\tau_s})$. In fact, once the choice is made, the way in which GWs propagate is independent of the spatial direction. Since the two polarization modes evolve identically, the transfer function  $\T^H (\tau,{k}) $ doesn't depend on the polarization state. 

\subsubsection{Gravitational wave transfer functions}
\begin{figure}[H]
   \centering
\includegraphics[width=\textwidth]{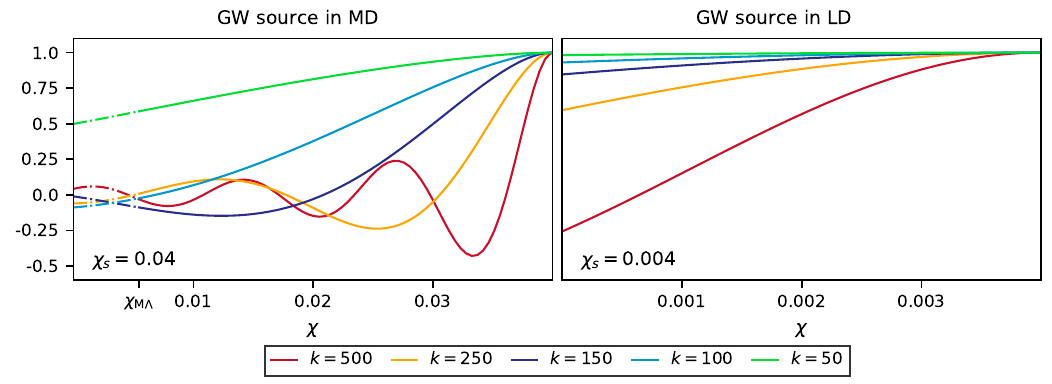}
    \caption{GW transfer functions for different values of $k$. On the left panel the source is located in matter domination, at a comoving distance of $\chi_s = 0.04$, while in the right panel it is in $\Lambda$ domination, at $\chi_s = 0.004$. Solid versus dashed-dot lines stands for the transfer function in the two epochs of the Universe, showing their matching at the moment of equivalence.  }
    \label{fig:Transfers}
\end{figure}

%In this Section we solve for the transfer function.
Since we are considering the Universe at late times, we solve Eq.~\eqref{eq:0thWaveEq} in two epochs:  during matter and $\Lambda$ domination. 
%Since we are considering the late time Universe, we will have two branches of solutions for the transfer function $\cal T$, according to whether the Universe is in matter or $\Lambda$ dominated epoch.
Using the expression of the scale factor of Appendix~\ref{app:BGeqFriedmann}, and changing the variable from conformal time to comoving distance, during matter domination we find
\begin{equation}\label{eq:TransferMDsolution}
        \T^H_{ {\rm MD}}( \chi, {k}) =  \sqrt{\frac{2 k}{\pi}} \, \frac{1}{\chi_i - \chi} \, \bigg[A \, j_1 [k ( \chi_i - \chi)] + B \, y_1 [k (\chi_i - \chi) ] \bigg]\,,
\end{equation}
while during $\Lambda$ domination
\begin{equation}\label{eq:TransferLDsolution}
    \T^H_{\Lambda {\rm D}}( \chi, {k}) =  \sqrt{\frac{2 k}{\pi}} \, (\chi + \chi_\Lambda)^2 \, \bigg[C \, j_1 [k ( \chi + \chi_\Lambda)] + D \, y_1 [k (\chi + \chi_\Lambda) ] \bigg]\,,
\end{equation}
$j_1(x)$ and $y_1(x)$ are the spherical Bessel functions of order one and of type one and two respectively, given by
\begin{eqnarray}
j_1 (x) &=& \frac{\sin x}{x^2} - \frac{\cos x}{x}\,, \qquad \qquad 
y_1 (x) = - \frac{\cos x}{x^2} - \frac{\sin x}{x}\,.
\end{eqnarray}
The two solutions presented here are well known in literature (see e.g.~\cite{Misner:1973prb,Kite:2021yoe} for matter dominated Universe and for solutions in de Sitter Universe~\cite{Guzzetti:2016mkm,Caprini:2018mtu}) and the values of $\chi_i$ and $\chi_\Lambda$ can be found in Table~\ref{tab:SpecialValues}.
%Fixing appropriate boundary conditions requires to distinguish two cases according to whether the GW source is located prior or after the moment of equivalence between matter and dark energy densities.
%We distinguish two cases to fix the boundary conditions: when the GW source is prior or after the matter-$\Lambda$ equality, set by $\chi_{{\rm M}\Lambda}$. Denoting with $\chi_s$ the GWs source comoving distance, when $\chi_s < \chi_{{\rm M}\Lambda}$ we impose
In the case of the GW source located after the moment of matter-$\Lambda$ equality, i.e. $\chi_s < \chi_{{\rm M}\Lambda}$ with $\chi_s$ is the comoving distance of the wave's source, we impose the boundary condition
\begin{equation}
    \T^H_{\Lambda {\rm D}}( \chi_s) = 1 \,, \qquad
    \frac{\partial \T_{\Lambda {\rm D}}^H}{\partial \chi}( \chi_s) = 0\,,
\end{equation}
assuming that the amplitude of the emitted GWs is maximum at the source position.  With these prescription we find 
\begin{equation}
    C_{\Lambda {\rm D}} = - \sqrt{\frac{\pi k}{2}} \frac{\cos [k(\chi_s + \chi_\Lambda)]}{(\chi_s + \chi_\Lambda)}\,, \qquad D_{\Lambda {\rm D}} = - \sqrt{\frac{\pi k}{2}} \frac{\sin [k(\chi_s + \chi_\Lambda)]}{(\chi_s + \chi_\Lambda)}\,.
\end{equation}
If the GW source is located prior to the equivalence, i.e.  $\chi_s > \chi_{{\rm M}\Lambda}$, we require
\begin{equation}
    \T^H_{\Lambda {\rm D}}  = \T^H_{ {\rm MD}} \, \Big|_{ \chi_{{\rm M} \Lambda}} \,,\qquad 
    \frac{\partial \T_{\Lambda {\rm D}}^H}{\partial \chi} = \frac{\partial \T_{ {\rm MD}}^H}{\partial \chi} \, \Big|_{ \chi_{{\rm M} \Lambda}}\,,\qquad
    \T^H_{ {\rm MD}}( \chi_s) = 1 \,, \qquad
    \frac{\partial \T_{ {\rm MD}}^H}{\partial \chi}( \chi_s) = 0\,,
\end{equation}
with the resulting boundary conditions
\begin{align}
    A_{\rm MD}&= - \sqrt{\frac{\pi k }{2}} (\chi_s - \chi_i)^2 \Big[ 3 y_1[k (\chi_s - \chi_i)]+\cos [k (\chi_s - \chi_i)]\Big] \,,\\
 %   B_{\rm MD}&= - \sqrt{\frac{\pi }{2}} \frac{1}{k^{3/2}} \left[\left(k^2 (\chi_s - \chi_i)^2-3\right) \sin (k (\chi_s - \chi_i))+3 k (\chi_s - \chi_i) \cos (k (\chi_s - \chi_i))\right] \\
     B_{\rm MD}&= - \sqrt{\frac{\pi k }{2}} 
    (\chi_s - \chi_i)^2 \Big[ \sin (k (\chi_s - \chi_i)) - 3  j_1 [k (\chi_s - \chi_i)] \Big] \,,\\
     C_{\rm MD}&= \: \frac{\Big[\Xi_1 \cos (k \chi_*) + \Xi_2 \sin (k \chi_*)  + \Xi_3 \cos (3 k \chi_*)  + \Xi_4 \sin (3 k \chi_*)  \Big]}{32 k^3 \chi^6_*} \,,\\ 
     D_{\rm MD}&= \: \frac{\Big[\Xi_2 \cos (k \chi_*) - \Xi_1 \sin (k \chi_*)  - \Xi_4 \cos (3 k \chi_*)  + \Xi_3 \sin (3 k \chi_*)  \Big]}{32 k^3 \chi^6_*} \,,
\end{align}
and where $\chi_*$ can be found in Table~\ref{tab:SpecialValues} and the coefficients $\Xi_i$ are 
\begin{align}
    \Xi_1 & = 3 (B_{\rm MD} - A_{\rm MD} \, k  \chi_*)\,, \qquad  \Xi_3 = 3 B_{\rm MD} - 9 A_{\rm MD} \, (k \chi_*) - 12 B_{\rm MD} (k \chi_*)^2 + 8 A_{\rm MD} (k \chi_*)^3 \\
    \Xi_2 & = 3 (A_{\rm MD} + B_{\rm MD} \, k  \chi_*) \,, \qquad   \Xi_4  = 3 A_{\rm MD} + 9 B_{\rm MD} \, (k \chi_*) - 12 A_{\rm MD} (k \chi_*)^2 - 8 B_{\rm MD} (k \chi_*)^3\,.
\end{align}
A plot of these transfer functions for different values of $k$ can be found in Fig.~\ref{fig:Transfers}.

%%%%%%%%%%%%%%%%%%%%%%%%%%%%%%%%%%%%%%%%%%%%%%%%%%%%%%%%%%%%%%%%%%%%%%%%%%%%%%%%%%%%%%%%%%%%%%%%%%%%%%%%%%%%%%%%%%%%%%%%%%%%%%

\section{The first order gravitational wave}\label{sec:1order}
In this Section, we derive the equation of motion for $H^{(1)}_{\mu\nu}$, following the procedure outlined in Section~\ref{Section:PerturbedUniverseScheme}. 
We will see that the GW at this order develops additional polarization components on top of the standard transverse and traceless modes. 
%As described in Section~\ref{Sec:ExplanationCM}, these modes arise to ensure the conservation of an effective source, built out of combinations of the free GW and the scalar potential describing the inhomogeneities in the Universe. 
%Then, we find a decoupled equation for the first order tensor modes, which we solve using a Green's function.

\subsection{Equations}
We start by working out the gauge and compatibility conditions. Inserting Eq.~\eqref{eq:HepsilonExpansion} into Eqs.~\eqref{eq:TransfConfFirst} and ~\eqref{eq:CompConditionFirst} and extracting the $\epsilon^1$ order, we find that  $H^{(1)}_{\mu\nu}$ satisfies
\begin{equation} \label{eq:gaugeH1}
\hat n^\mu \hat n^\nu H^{(1)}_{\mu\nu} = 0 \,, \qquad     \partial^\mu \tilde H^{(1)}_{\mu\nu} - 4 \H \, \hat n^\mu H^{(1)}_{\mu\nu} + H^{(1)} \, \H \,  \hat n_\nu  = 0 \,,
\end{equation}
where we have used also Eq.~\eqref{eq:GW0TracelessSpatial}.
Notice that the trace is defined as $H^{(1)} \equiv \eta^{\mu\nu} H^{(1)}_{\mu\nu}$ and also $ \tilde H^{(1)}_{\mu\nu} =  H^{(1)}_{\mu\nu} - \frac12  \eta_{\mu\nu} H^{(1)}$ .
From the first condition, we conclude that the time-time component of the first order GW is null, even at this order of perturbation.
%It is also interesting to notice that the transversality condition of $\tilde H^{(1)}_{\mu\nu}$, doesn't mix $H^{(1)}_{\mu\nu}$ and $H^{(0)}_{\mu\nu}$. This  is due to the fact that we are not including any shear viscosity in the large scale structures. In fact, if $\phi \neq \psi$ the same condition would mix them.
From Eq.~\eqref{eq:EOMLinearized}, on the other hand, we see that the structure of the equation of motion that $H^{(1)}_{\mu\nu}$ has the form
\begin{equation}\label{eq:EOMH1order}
    \Big[ {\cal O}_0 \,  H^{(1)} \Big]_{\mu\nu} = - \Big[ {\cal O}_1\, H^{(0)}\Big]_{\mu\nu}\,.
\end{equation}
The right-hand-side of the equation above does not depend on $H^{(1)}_{\mu\nu}$,  meaning that the adopted perturbation scheme produces an effective source, which accounts for the interaction between the freely propagating waves and the gravitational potentials.  In light of what discussed in Section~\ref{Sec:ExplanationCM}, we expect that $H^{(1)}_{\mu\nu}$ develops scalar and vector components whose role is to enforce the conservation of such source. Indeed, the left-hand-side of Eq.\eqref{eq:EOMH1order} is nothing but the first order Einstein tensor, evaluated in $H^{(1)}_{\mu\nu}$, which satisfies the linearized Bianchi identities. These additional components constitute extra polarizations but are not new and independent degrees of freedom, as their presence does not require new initial conditions. Indeed, their sources vanish in vacuum. 
By plugging $H^{(1)}_{\mu\nu}$ and $H^{(0)}_{\mu\nu}$ in the first order gravitational wave equation, namely Eq.~\eqref{eq:EOMH1order}, we find 
%\begin{align} 
 %   \bigg[ {\cal O}_0 \,  H^{(1)} + {\cal O}_1 \, H^{(0)} \bigg]_{\mu\nu} & =   \, \Box_\eta  \tilde H^{(1)}_{\mu\nu} - 2 \H  (\tilde  H^{(1)}_{\mu\nu})'   - 2 \H \hat n^\alpha \, \Big(\partial_\mu \tilde H^{(1)}_{\alpha\nu} + \partial_\nu \tilde H^{(1)}_{\alpha\mu} \Big)   \nonumber \\
 %   &\,\quad + 2 (\H' + \H^2) \, \hat n^\alpha \,  \Big( \hat n_\mu H^{(1)}_{\alpha \nu } + \hat n_\nu H^{(1)}_{\alpha \mu }\Big) + 2 H^{(0)}_{\mu\nu}\, \Big[ \Box_\eta \phi - 2 \H\, \phi'\Big]  \nonumber \\
 %   &\,\quad+ 4 \phi\, \Delta H^{(0)}_{\mu\nu} + 4 \,\partial^k \phi \, \partial_k\, H^{(0)}_{\mu\nu} + 2 \Big(\hat \Lambda_{\mu\nu}+ \hat n_\mu \hat n_\nu\Big) \, H^{(0)}_{\alpha \beta }\, D^\alpha D^\beta\, \phi \nonumber \\
 %   &\,\quad- 2 \hat n_\mu \, D^\alpha \phi \Big( {H^{(0)}}'_{\nu \alpha} + \H   H^{(0)}_{\nu \alpha} \Big)- 2 \hat n_\nu \, D^\alpha \phi \Big( {H^{(0)}}'_{\mu \alpha} + \H   H^{(0)}_{\mu \alpha} \Big) \nonumber \\
 %   &\,\quad  - 2 D_\mu \Big( H^{(0)}_{\alpha \nu} \, D^\alpha \phi \Big) - 2 D_\nu \Big( H^{(0)}_{\alpha \mu} \, D^\alpha \phi \Big) -  \H' H^{(1)} \Big( \hat \Lambda_{\mu\nu}+ \hat n_\mu \hat n_\nu\Big) \nonumber \\
 %   &=0 
%\end{align}
\begin{align} 
    \bigg[ {\cal O}_0 \, H^{(1)} \bigg]_{\mu\nu}& = \,\Box_\eta  \tilde H^{(1)}_{\mu\nu} - 2 \H (\tilde H^{(1)}_{\mu\nu})' + \H \Big(  \hat n_\nu  \partial_\mu H^{(1)}+  \hat n_\mu \partial_\nu H^{(1)} \Big) - (\hat n_\mu \hat n_\nu  + \hat \Lambda_{\mu\nu}  )  \H' H^{(1)}  - \nonumber \\
 &- 2 \H \Big(  \hat n^\alpha \partial_\mu H^{(1)}_{\alpha \nu } + \hat n^\alpha \partial_\nu H^{(1)}_{\alpha \mu}\Big)  + 2 ( \H' + \H^2) \hat n^\alpha \Big( \hat n_\mu H^{(1)}_{\alpha\nu} + \hat n_\nu H^{(1)}_{\alpha\mu}\Big) \label{eq:EOMH1orderO0H1} \\ 
 \left[ {\cal O}_1 \, H^{(0)} \right]_{\mu\nu} & = \, 4 \phi\, \Delta H^{(0)}_{\mu\nu} + 2 H^{(0)}_{\mu\nu}\, \Big[ \Box_\eta \phi - 2 \H\, \phi'\Big] + 4 \,\partial^k \phi \, \partial_k\, H^{(0)}_{\mu\nu} + 2 (\eta_{\mu\nu} + 2 \hat n_\mu \hat n_\nu) \, H^{(0)}_{\alpha \beta }\, D^\alpha D^\beta\, \phi \nonumber \\
    &- 2 \hat n_\mu \, D^\alpha \phi \Big( {H^{(0)}}'_{\nu \alpha} + \H   H^{(0)}_{\nu \alpha} \Big)- 2 \hat n_\nu \, D^\alpha \phi \Big( {H^{(0)}}'_{\mu \alpha} + \H   H^{(0)}_{\mu \alpha} \Big) \nonumber \\
    &  - 2 D_\mu \Big( H^{(0)}_{\alpha \nu} \, D^\alpha \phi \Big) - 2 D_\nu \Big( H^{(0)}_{\alpha \mu} \, D^\alpha \phi \Big)  \label{eq:EOMH1orderO1H0}
\end{align}
%
%\noindent
%By applying the  projector operator on Eq. (5.3) and (5.4) of arXiv:2210.0571  one can find the equation of the spatial, transverse and traceless mode
%\begin{equation}
 %   \Box_\eta \gamma^{(1)}_{ij} - 2 \H (\gamma^{(1)}_{ij})' = - 4 \phi\, \Delta H^{(0)}_{ij} - 2 H^{(0)}_{ij}\, \Big[ \Box_\eta \phi - 2 \H\, \phi'\Big] - 4 \,\partial^k \phi \, \partial_k\, H^{(0)}_{ij}\nonumber
%\end{equation}
%where all derivatives are partial derivatives.
%As in Peters (1974) we call 
%\begin{equation}
%    \tilde{H}_{ij} = H_{ij}(1 + 2 \phi) \approx H_{ij} + 2 H^{(0)}_{ij} \phi\nonumber
%\end{equation}
%and we extract the TT part of it, so that at first order we have
%\begin{equation}
%    \tilde{\gamma}^{(1)}_{ij} = \gamma^{(1)}_{ij} + 2 H^{(0)}_{ij} \phi\,.\nonumber
%\end{equation}
%Now we plug this into the first equation and you can find
%\begin{equation}
%    \Box_\eta \tilde \gamma^{(1)}_{ij} - 2 \H (\tilde \gamma^{(1)}_{ij})' = - 4 \phi\, \Delta H^{(0)}_{ij} \nonumber
%\end{equation}
%
where $D_\mu \equiv \hat \Lambda_{\mu}^\nu \partial_\nu$. The details of these computations can be found in Eqs.~\eqref{app:EOMlinearized0},~\eqref{app:EOMlinearized1} and ~\eqref{app:OoH1},~\eqref{app:O1H0} of Appendix~\ref{app:LinearizedEOM} and~\ref{app:GWeq}. 
We explicitly decompose $H^{(1)}_{\mu\nu}$ as
\[
\renewcommand\arraystretch{2}
H^{(1)}_{\mu\nu} \, = \,
\mleft[
\begin{array}{c|c}
  0 & H^{(1)}_{0i} \\
  \hline
  H^{(1)}_{i0} & \gamma^{(1)}_{ij} + E^{(1)}_{ij} + \frac13 \delta_{ij} H^{(1)}
\end{array}
\mright]
\]
where $\gamma^{(1)}_{ij} $ is such that $ \delta^{ij} \gamma^{(1)}_{ij} = \partial^i  \gamma^{(1)}_{ij} = 0$, and $E^{(1)}_{ij}$ is defined consequently. 
Using this decomposition of $H^{(1)}_{\mu\nu}$ into the equations of motion, we aim at finding a solution for all of its components.  
To single out the tensor mode $\gamma^{(1)}_{ij}$, we define the projector operator ${\hat{\mathcal P}^{ab}}_{ij}$ as in~\cite{Baumann:2007zm,Ananda:2006af}. Its action on a generic spacetime tensor is given by
\begin{equation}\label{eq:ActionProjector}
    \hat{\mathcal P}^{ab}_{ij} \,  A_{ab} (x) \equiv \int d^3 k \,  e^{i {\bf k} \cdot {\bf x}} \, \Big[ Y^{+2}_{ij} (\hat k) Y^{ab}_{-2} (\hat k)  +Y^{-2}_{ij} (\hat k) Y^{ab}_{+2} (\hat k) \Big] \, A_{ab} ({\bf k}, \tau)
\end{equation}
where $ A_{ij} ({\bf k}, \tau)$ is the Fourier transform of $A_{ij} (x)$. 
This way, the tensor mode is defined as $\gamma^{(1)}_{ij} \equiv \hat{\mathcal P}^{ab}_{ij} H^{(1)}_{ab}$.

\subsection{Scalar and vector modes}
To find the solutions of $H^{(1)}, H^{(1)}_{0i}$ and $E^{(1)}_{ij}$ we look at the time-time and the time-space components of Eq.~\eqref{eq:EOMH1order}. Using the explicit expressions in Eqs.~\eqref{eq:EOMH1orderO0H1} and~\eqref{eq:EOMH1orderO1H0} we find
\begin{align}
     \frac12 \Box_\eta H^{(1)} - 3 \H (H^{(1)})' -  \H' H^{(1)} &= -2 H^{(0)}_{ij} \partial^i \partial^j \, \phi\,,  \quad\quad\qquad\qquad \mbox{Time-Time} \label{eq:EOMH1orderTimeTime} \\
 %    \Box_\eta H^{(1)} - 2 \H (H^{(1)})' - (4 \H^2 + 2 \H') H^{(1)}  &= 0 \,,  \qquad \qquad \qquad \qquad  \quad\:\:\: \mbox{Trace}  \label{eq:EOMH1orderTrace} \\
    \Box_\eta  H^{(1)}_{0i} - 4 \H (H^{(1)}_{0i})'  - 2 (\H^2 + \H') H^{(1)}_{0i} &=  \H \partial_i H^{(1)} - \frac{2}{a} \partial^k \phi \, (a H^{(0)}_{ki})'\,,  \quad \mbox{Space-Time} \label{eq:EOMH1orderTimeSpace}
\end{align}
from which we see that the vector potential is sourced also by the spatial trace, $H^{(1)}$.
We note that these equations are second order in time derivatives, and not first, because of the covariant nature of the gauge choice made.
Nevertheless, we will call them {\it constraint equations} because they play the same role as them: to include the covariant conservation of the new source, $[{\cal O}_1 H^{(0)}]_{\mu\nu}$, directly into Einstein's equations. 
In addition, one should account for the gauge condition in Eq.~\eqref{eq:gaugeH1}, which can be rewritten as
\begin{equation}\label{eq:GaugeH1E}
    \partial^j E^{(1)}_{ij} = (H^{(1)}_{0i})' + 4 \H H^{(1)}_{0i} + \frac{\partial_i H^{(1)}}{6}\,.
\end{equation}
The constraint Eqs.~\eqref{eq:EOMH1orderTimeTime} and~\eqref{eq:EOMH1orderTimeSpace} and the gauge conditions~\eqref{eq:GaugeH1E}  relate the trace, the time-space component and the spatial divergence of $H^{(1)}_{\mu\nu}$, to (functions of) the projection of the zero-order gravitational wave along the gradient of the scalar gravitational potential. 
The only components of $H^{(1)}_{\mu\nu}$ that do not take part to the constraint equations, are the transverse-traceless and spatial ones, which we will address in the next section. Next, we compute the solutions of Eqs.~\eqref{eq:EOMH1orderTimeTime},~\eqref{eq:EOMH1orderTimeSpace} and~\eqref{eq:GaugeH1E}.

\subsubsection{First order spatial trace: solution}
We can solve Eq.~\eqref{eq:EOMH1orderTimeTime} by means of a Green function, $g^{s}(\tau, \tilde \tau)$. In general, Green's functions can be built from two solutions of the associated homogeneous problem~\cite{Baumann:2007zm}, $a(\tau)$ and $b(\tau)$,  as 
\begin{equation}\label{eq:DefGreenFunction}
    g_{k}(\tau, \tilde \tau) = \frac{a_{ k}(\tau) b_{ k}(\tilde \tau) -  b_{ k}(\tau) a_{ k}(\tilde \tau) }{a_{ k}'(\tilde \tau) b_{ k}(\tilde \tau) -  b_{ k}'(\tilde \tau) a_{ k}(\tilde \tau) }\,.
\end{equation}
Performing a Fourier transform, Eq.~\eqref{eq:EOMH1orderTimeTime} becomes 
\begin{equation}
    (H^{(1)}_{\bf k})'' + k^2 H^{(1)}_{\bf k} + 6 \H (H^{(1)}_{\bf k})' + 2\H'
(H^{(1)}_{\bf k}) = \S_{\bf k}^{H}(\tau)\,,
\end{equation}
with
\begin{equation}\label{eq:SourceH}
    \S_{\bf k}^{H}(\tau) \:\equiv\: - 4  \int d^3 p \, k^i k^j \, H^{(0)}_{ij} ({\bf p}, \tau) \, \phi_{{\bf k} - {\bf p}} (\tau)\,.
\end{equation}
The two homogeneous solutions, during matter and $\Lambda$ dominated eras (see Appendix~\ref{app:BGeqFriedmann} for the explicit expression of the Hubble parameter), are
\begin{eqnarray}
a_{\bf k}^{{\rm MD}}&=&  \,\frac{J_{\frac{\sqrt{137}}{2}} [k (\tau + \tau_i) ]}{(\tau + \tau_i)^{11/2}}\,, \qquad \, a_{\bf k}^{\Lambda {\rm D}}= ({\tau_\Lambda - \tau})^{7/2} \, J_{\frac{\sqrt{41}}{2}}[k(\tau_\Lambda - \tau) ]\\
b_{\bf k}^{{\rm MD}}&=& \,\frac{Y_{\frac{\sqrt{137}}{2}} [k (\tau + \tau_i) ]}{(\tau + \tau_i)^{11/2}} \,, \qquad b_{\bf k}^{\Lambda {\rm D}}=({\tau_\Lambda - \tau})^{7/2} \, Y_{\frac{\sqrt{41}}{2}}[k(\tau_\Lambda - \tau) ]\,,
\end{eqnarray}
where  $J_n, Y_n$ are the Bessel functions of order n and $\tau_i =  \chi_i - \tau_0 $ and $\tau_\Lambda = \tau_0 + \chi_\Lambda$, with $\tau_0$ today's conformal time. The special values of the comoving distance can be found in Table~\ref{tab:SpecialValues}.
Combining the solutions above, we find
\small{
\begin{equation}\label{eq:GreenFunctionS}
    g^H_k (\tau, \tilde \tau) =\frac{\pi}{2 } \left\{
\begin{aligned}
    &  \sqrt{\frac{(\tilde \tau+\tau_i)^{13}}{( \tau+\tau_i)^{11}} }\left[J_{\frac{\sqrt{137}}{2}} [k (\tilde \tau + \tau_i) ] Y_{\frac{\sqrt{137}}{2}} [k (\tau + \tau_i)] - J_{\frac{\sqrt{137}}{2}} [k (\tau + \tau_i) ] Y_{\frac{\sqrt{137}}{2}} [k (\tilde \tau + \tau_i)   ] \right] \: {\rm MD} \\
    &  \sqrt{\frac{(\tau_\Lambda - \tilde \tau)^{7}}{( \tau_\Lambda - \tau)^{5}} }\left[J_{\frac{\sqrt{41}}{2}} [k (\tau_\Lambda - \tilde \tau ) ] Y_{\frac{\sqrt{41}}{2}} [k (\tau_\Lambda - \tau ) ]- J_{\frac{\sqrt{41}}{2}} [k (\tau_\Lambda - \tau ) ] Y_{\frac{\sqrt{41}}{2}} [k (\tau_\Lambda - \tilde \tau )   ] \right] \:\: \Lambda{\rm D}\\
\end{aligned}\right.
\end{equation}
}
This way, $H^{(1)}_{\bf k}(\tau)$ is given by the convolution of the Green's function and the source, namely
\begin{equation}
    H^{(1)}_{\bf k}(\tau) = \int^\tau_{{\tau_s}} d \tilde \tau \, g^H_{k}(\tau, \tilde \tau')S_{\bf k}^{H}(\tilde \tau)\,,
\end{equation}
where the integral runs from the conformal time of the source of gravitational waves, ${\tau_s}$.

\subsubsection{First order time-space components: solution}
The solution of $H^{(1)}_{0i}$ can be found similarly from Eq.~\eqref{eq:EOMH1orderTimeSpace}. The latter,  in Fourier space, takes the form
\begin{equation}
    (H^{(1)}_{0i, {\bf k}})'' + k^2 H^{(1)}_{0i, {\bf k}}  + 4 \H (H^{(1)}_{0i, \bf k})' + 2(\H^2 + \H')(H^{(1)}_{0i, {\bf k}}) = \S_{i, {\bf k}}^{H_{0i}}(\tau)\,,
\end{equation}
with
\begin{equation}
    \S_{i, {\bf k}}^{H_{0i}}(\tau) \:\equiv\: - i\, k_i  \, \H \, H^{(1)}_{\bf k}(\tau) \,  + \frac{2i}{a}\, \int d^3 p \, k^j \, \left[ a H^{(0)}_{ij} ({\bf p}, \tau) \right]' \, \phi_{{\bf k} - {\bf p}} (\tau)\,.
\end{equation}
For each component $i$, the equations above admit 
\begin{eqnarray}
a_{\bf k}^{{\rm MD}}&=&  \,\frac{J_{\frac{\sqrt{33}}{2}} [k (\tau + \tau_i) ]}{(\tau + \tau_i)^{7/2}}\,, \qquad \, a_{\bf k}^{\Lambda {\rm D}}= ({\tau_\Lambda - \tau})^{3} \, j_1 [k(\tau_\Lambda - \tau) ]\\
b_{\bf k}^{{\rm MD}}&=& \,\frac{Y_{\frac{\sqrt{33}}{2}} [k (\tau + \tau_i) ]}{(\tau + \tau_i)^{7/2}} \,, \qquad b_{\bf k}^{\Lambda {\rm D}}= ({\tau_\Lambda - \tau})^{3} \, y_1 [k(\tau_\Lambda - \tau) ]\,,
\end{eqnarray}
as solutions of the associated homogeneous problem, so that  the Green's function, $g^{H_{0i}}_k (\tau, \tilde \tau)$, can be constructed as
\begin{equation}\label{eq:GreenFunctionV}
    g^{H_{0i}}_k (\tau, \tilde \tau) = \left\{
\begin{aligned}
    \:& \frac{\pi}{2} \sqrt{\frac{(\tilde \tau + \tau_i)^9}{( \tau + \tau_i)^7}} \left[J_{\frac{\sqrt{33}}{2}}[k (\tilde \tau + \tau_i) ] Y_{\frac{\sqrt{33}}{2}}[k ( \tau + \tau_i) ] - J_{\frac{\sqrt{33}}{2}}[k (\tau + \tau_i) ] Y_{\frac{\sqrt{33}}{2}}[k (\tilde \tau + \tau_i) ] \right] \quad {\rm MD} \\
    \:& \frac{k (\tau_\Lambda -  \tau)^3}{(\tau_\Lambda - \tilde \tau)} \Big[j_1 [k(\tau_\Lambda - \tilde \tau  )]y_1 [k( \tau_\Lambda - \tau )] - j_1 [k(\tau_\Lambda - \tau )]y_1 [k( \tau_\Lambda - \tilde\tau )] \Big]\quad\qquad\qquad\quad \Lambda{\rm D}\\
\end{aligned}\right.\,,
\end{equation}
using Eq.~\eqref{eq:DefGreenFunction}. Therefore, we have that
\begin{equation}\label{eq:SolTimeSpace}
    H^{(1)}_{0i} ({\bf k}, \tau) = \int^\tau_{{\tau_s}} \, d \tilde \tau \,  g^{H_{0i}}_k (\tau, \tilde \tau) \, \S_{i, {\bf k}}^{H_{0i}}(\tilde \tau)\,.
\end{equation}
In the chosen gauge, $H^{(1)}_{0i}$ has three components.  We can decompose it into a vector and a scalar as 
\begin{equation}
   H^{(1)}_{0i, {\bf k}} (\tau) = B^{(1)}_{i, {\bf k}} (\tau) + i \,  k_i \, B^{(1)}_{\bf k} (\tau) \,,
\end{equation}
with $k^i B^{(1)}_{i, {\bf k}} = 0$. Accordingly, the solutions for the vector and the scalar potentials, namely $B^{(1)}_{i}$ and $B^{(1)}$, can be found by convolving the same Green function, $g^{H_{0i}}_k (\tau, \tilde \tau)$, with, respectively, the following two  sources in Fourier space
\begin{eqnarray}
    \S^{B_{i}}_{i, {\bf k}} (\tau)&=& \frac{2 i}{a} \left[ \delta^j_i - \frac{k_ik^j}{k^2}\right]  k^l\, \int d^3 p \,  \left[ a H^{(0)}_{jl} ({\bf p}, \tau) \right]' \, \phi_{{\bf k} - {\bf p}} (\tau) \,,\label{eq:SourceBi}\\
    \S^{B}_{\bf k}(\tau) &=& - \H \,H^{(1)}_{\bf k}(\tau) +  \,\frac{2}{a}\frac{ k^i k^j}{  k^2} \, \int d^3 p \,  \left[ a H^{(0)}_{ij} ({\bf p}, \tau) \right]' \, \phi_{{\bf k} - {\bf p}} (\tau)\,. \label{eq:SourceB}
\end{eqnarray}

\subsubsection{First order longitudinal space-space components: solution}
Since $E^{(1)}_{ij}$ is a longitudinal and traceless tensor, we can decompose it as 
\begin{equation}
    E^{(1)}_{ij} = \frac{1}{2} \left( \partial_i E^{(1)}_j + \partial_j E^{(1)}_i \right) + (\partial_i \partial_j - \frac13 \delta_{ij}  \Delta ) E^{(1)}\,.
\end{equation}
It is straightforward to show that its components are entirely fixed by Eq.~\eqref{eq:GaugeH1E} to
\begin{eqnarray}
    E^{(1)}_{i, {\bf k}} &=& -\frac{2}{k^2} \left[ (B^{(1)}_{i, {\bf k}})' + 4 \H B^{(1)}_{i, {\bf k}}  \right]\label{eq:EVector} \,,\\
     E^{(1)}_{{\bf k}} &=& -\frac{3 }{2 k^2} \left[ (B^{(1)}_{ {\bf k}})' + 4 \H B^{(1)}_{{\bf k}}  
 + \frac{H^{(1)}_{ {\bf k}}}{6} \right]\,. \label{eq:EScalar}
\end{eqnarray}
These relations also clearly show that $H^{(1)}_{\mu\nu}$ is composed of a total of six independent components: one in the spatial trace $H^{(1)}$, three in the time-space components $H^{(1)}_{0i}$, and two tensor modes $\gamma^{(1)}_{ij}$. Indeed, $E^{(1)}_{ij}$ can be substituted everywhere in favor of $H^{(1)}$ and $H^{(1)}_{0i}$, using Eqs.~\eqref{eq:EVector} and~\eqref{eq:EScalar}.

\subsection{Tensor modes}
The equation of motion of $\gamma^{(1)}_{ij}$ can be obtained by projecting Eq.~\eqref{eq:EOMH1order} using the operator $\hat {\cal P}^{ab}_{ij}$, as in Eq.~\eqref{eq:ActionProjector}. After some manipulations, we obtain
%\begin{equation}\label{eq:GammaProjectors}
%     \hat{\mathcal P}^{ab}_{ij} \bigg[ {\cal O}_0 \,  H^{(1)} \bigg]_{ab} = -  \hat{\mathcal P}^{ab}_{ij} \bigg[ {\cal O}_1\, H^{(0)} \bigg]_{ab}\,,
%\end{equation}
%which, after some manipulations, gives
\begin{equation}\label{eq:GammaFirstRealSpace}
    \Box_\eta \gamma^{(1)}_{ij} - 2 \H (\gamma^{(1)}_{ij})' = - \hat{{\cal P}}^{ab}_{ij} \left[ 4 \phi\, \Delta H^{(0)}_{ab} + 2 H^{(0)}_{ab}\, \Big[ \Box_\eta \phi - 2 \H\, \phi'\Big] + 4 \,\partial^k \phi \, \partial_k\, H^{(0)}_{ab} \right]\,.
 \end{equation}
As in the previous cases, we solve it in Fourier space 
\begin{equation}\label{eq:Gamma1Coordinates}
   ( \gamma^{(1)}_{ij,{\bf k}} )'' + 2 \H ( \gamma^{(1)}_{ij,{\bf k}} )' + k^2  \gamma^{(1)}_{ij,{\bf k}}  \:=\: {\cal S}^{\gamma}_{ij, {\bf k}} (\tau)\,,
\end{equation}
identifying the tensor mode source
\begin{equation}\label{eq:SourceGamma}
{\cal S}^{\gamma}_{ij, {\bf k}} (\tau) \:\equiv\:   - 2 \hat{{\cal P}}^{ab}_{ij, {\bf k}}  \, \,    \int d^3 p   \,  H^{(0)}_{ij, {\bf  p}}  \, \Big[ \phi''_{|{\bf k} - {\bf p}|} + 2 \H \phi'_{|{\bf k} - {\bf p}|} +  \phi_{{\bf k} - {\bf p}}( p^2 + k^2)  \, \Big]  \, .
\end{equation}
According to Eq.~\eqref{eq:DefGreenFunction}, to build the Green function we must compute two solutions of the homogeneous associated problem, which in this case are 
\begin{eqnarray}
a_{\bf k}^{{\rm MD}}&=&  \,\frac{j_1 [k(\tau + \tau_{i} )] }{(\tau + \tau_{i})} \,, \qquad \, a_{\bf k}^{\Lambda {\rm D}}= ({\tau_\Lambda - \tau})^2 \, j_1 [k(\tau_\Lambda - \tau) ]\\
b_{\bf k}^{{\rm MD}}&=& \,\frac{y_1 [k(\tau + \tau_{i} )] }{(\tau + \tau_{i})}  \,, \qquad b_{\bf k}^{\Lambda {\rm D}}= ({\tau_\Lambda - \tau})^2 \, y_1 [k(\tau_\Lambda - \tau) ]\,,
\end{eqnarray}
%where $\tau_i =  \chi_i - \tau_0 $ and $\tau_\Lambda = \tau_0 + \chi_\Lambda$ with $\tau_0$ today's conformal time and the values of the comoving distance can be found in Table~\ref{tab:SpecialValues}.
so that the tensor modes Green's function take the form
\begin{equation}\label{eq:GreenFunctionT}
    g^\gamma_k (\tau, \tilde \tau) = \left\{
\begin{aligned}
    \:& \frac{k (\tilde  \tau + \tau_{i} )^3}{(\tau + \tau_{i} )}  \Big[j_1 [k(\tilde \tau + \tau_{i} )]y_1 [k( \tau + \tau_{i} )] - j_1 [k( \tau + \tau_{i} )]y_1 [k(\tilde \tau + \tau_{i} )] \Big] \quad\quad {\rm MD} \\
    \:&  k (\tau_\Lambda - \tau )^2  \Big[j_1 [k(\tau_\Lambda - \tau  )]y_1 [k( \tau_\Lambda - \tilde\tau )] - y_1 [k(\tau_\Lambda - \tau )]j_1 [k( \tau_\Lambda - \tilde\tau )] \Big]  \quad \Lambda{\rm D}\\
\end{aligned}\right.\,.
\end{equation}
With this, the solution of Eq.~\eqref{eq:Gamma1Coordinates} is given by
\begin{equation}\label{eq:SolutionGammaGreen}
    \gamma^{(1)}_{ij,{\bf k}} (\tau) =  \int^\tau_{{\tau_s}} d \tilde \tau \, g^\gamma_{k}(\tau,  \tilde \tau ) \, {\cal S}^{\gamma}_{ij, {\bf k}} (\tilde \tau) \,.
\end{equation}
where the integral runs from the conformal time of the source of gravitational waves, up to $\tau$, so that $ \gamma^{(1)}_{\mathring{a}} ({\tau_s}, {\bf k}) = 0$.
Indeed, we are choosing the first order GW to satisfy the boundary condition, $ H^{(1)}_{\mu\nu} ({\tau_s}, {\bf x}) = 0$ since we have normalized the zeroth order GW to the total emitted amplitude. This is also in line with our  previous statement about the degrees of freedom: the number of initial conditions needed to solve the problem doesn't increase when considering the first order GW as its source is the unperturbed wave. 
A summary of the solution of $H^{(1)}_{\mu\nu}$, with all of its components, is reported in Table~\ref{tab:SourceSummary}.

\subsection{Summary and considerations}
Table~\ref{tab:SourceSummary} summarizes the results found so far. Looking at the sources of the first order gravitational wave, namely Eqs.~\eqref{eq:SourceH},~\eqref{eq:SourceB}, ~\eqref{eq:SourceBi} and~\eqref{eq:SourceGamma}  it is clear that whenever $k^i$ and $p^i$ are aligned, then the sources of the scalar and  vector modes vanish, while the one of the tensor modes remains. This is simply due to the transversality of the free wave, since $k^i H^{(0)}_{ij, {\bf p}} \propto p^i H^{(0)}_{ij, {\bf p}} =0 $ when the two wave-vectors are parallel.
This means that, for gravitational waves traveling in the direction of the gradient of the gravitational potential well, thus pointing toward the center of the well, no scalar and vector polarization are produced. Only in the situation of a wave traversing the potential wells in a generic direction, these modes are produced. 

\small{
\begin{table}[h!]
    \centering
    \renewcommand{\arraystretch}{3}
    \begin{tabular}{|c|c|c|}
        \hline
        Mode & \shortstack{Green's \\ function}   & Source \\ 
        \hline
        $H^{(1)}_{{\bf k}} $ & $g^{H}_{ k}(\tau, \tilde \tau)$ & $\S_{\bf k}^{H}(\tau) \:\equiv\: - 4  \int d^3 p \, k^i k^j \, H^{(0)}_{ij, {\bf p}} ( \tau) \, \phi_{{\bf k} - {\bf p}} (\tau)$ \\
        \hline 
        $B^{(1)}_{{\bf k}}$& $g^{H_{0i}}_{ k}(\tau, \tilde \tau)$ & $\S_{\bf k}^{B}(\tau) \:\equiv\: - \H \,H^{(1)}_{\bf k}(\tau) + \,\frac{2}{a} \frac{ k^i k^j}{k^2} \, \int d^3 p \,  \left[ a H^{(0)}_{ij, {\bf p}} (\tau) \right]' \, \phi_{{\bf k} - {\bf p}} (\tau)$ \\
        \hline
        $B^{(1)}_{i, {\bf k}}$ & $g^{H_{0i}}_{ k}(\tau, \tilde \tau)$&  $\S_{i, \bf k}^{B_i}(\tau) \:\equiv\:\frac{2 i}{a} \left[ \delta^j_i - \frac{k_ik^j}{k^2}\right]  k^l\, \int d^3 p \,  \left[ a H^{(0)}_{jl, {\bf p}} (\tau) \right]' \, \phi_{{\bf k} - {\bf p}} (\tau) $ \\
        \hline
        $\gamma^{(1)}_{ij, {\bf k}}$ & $g^{\gamma}_{ k}(\tau, \tilde \tau)$ & ${\cal S}^{\gamma}_{ij, {\bf k}} (\tau) \:\equiv\:   -2 \hat{{\cal P}}^{ab}_{ij, {\bf k}}  \, \,    \int d^3 p   \,  H^{(0)}_{ab, {\bf p}}   \Big[ \phi''_{|{\bf k} - {\bf p}|} + 2 \H \phi'_{|{\bf k} - {\bf p}|} +  \phi_{{\bf k} - {\bf p}}( p^2 + k^2)  \, \Big]  \, $\\
        \hline
    \end{tabular}
    \caption{Summary of the sources and Green's function of the modes constituting $H^{(1)}_{\mu\nu}$. We recall that $H^{(1)}_{0i} = B^{(1)}_i + \partial_i B^{(1)}$ and that the remaining spatial components, namely $E_{ij}$, can be found using Eqs.~\eqref{eq:EVector} and~\eqref{eq:EScalar}. The Green functions are given by, respectively, Eqs.~\eqref{eq:GreenFunctionS},~\eqref{eq:GreenFunctionV} and~\eqref{eq:GreenFunctionT}.}
    \label{tab:SourceSummary}
\end{table}
}

\begin{comment}
\noindent
Starting from Eq.~\eqref{eq:Gamma1Coordinates}, it is possible to show that the quantity $ {\overset{\smile}{\gamma}^{(1)}_{ij}} = \gamma^{(1)}_{ij} + 2 H^{(0)}_{ij} \phi $ satisfies
\begin{equation}\label{eq:GammaSmile}
    \Box_\eta {\overset{\smile}{\gamma}^{(1)}_{ij}} - 2 \H ({\overset{\smile}{\gamma}^{(1)}_{ij}})' = - 4 \phi\, \Delta H^{(0)}_{ij} - 4 \phi' (H^{(0)}_{ij})'\,.
\end{equation}
This is the same result of~\cite{Peters1974} upon considering $\phi'=0$. Note that  in~\cite{Peters1974} a different gauge choice is made at the beginning for the GW, instead of de-Donder. As a consequence, while the equations for $\overset{\smile}{\gamma}^{(1)}_{ij}$ coincide, those for $H^{(1)}, H^{(1)}_{0i} $ and $E^{(1)}_{ij}$ are different. This is not a problem, since observables are gauge independent and will not be affected by the different choice. 
\end{comment}
We point out another work, namely~\cite{Chang:2022vlv}, where the authors study the evolution equation for second order tensor modes on a perturbed Universe which includes both scalar and tensor perturbations. Even though the settings of the two works are different, effectively the methodologies employed in this work in~\cite{Chang:2022vlv} and shares some similarities.  In our case, $H^{(1)}_{\mu\nu}$ is only sourced by second order combinations built from one scalar and one tensor mode, while~\cite{Chang:2022vlv} also considers the possibility of having scalar-scalar and tensor-tensor sources. Our formalism doesn't include these cases by construction, however tensor-tensor sources could be included by adding a background tensor mode to Eq.~\eqref{eq:BGPoisson}.  Additionally, the authors of~\cite{Chang:2022vlv} consider only second order tensor modes, hence they cannot account for the scalar and vector polarization which we find at the perturbed level, namely $H^{(1)}, H^{(1)}_{0i}, E^{(1)}_{ij}$. 
Regardless of these differences and similarities, the scalar-tensor source, computed in both works, show some different numerical factors.

\subsection{Tensor modes in polarization basis}
In order to better understand the dynamics of $\gamma^{(1)}_{ij}$, we decompose it on the polarization basis of Eq.~\eqref{eq:DefPolBasisTensors}. These computations will also set the stage for the construction of the power spectrum and the density matrix, topics of the next Section. 
Since $\gamma^{(1)}_{ij}$ is transverse and traceless, we decompose it as $\gamma^{(1)}_{ij} ({\bf k}, \tau) = \sum \gamma^{(1)}_{\mathring{a}, {\bf k}} (\tau) Y^{\mathring{a}}_{ij} (\hat k)$, where $\mathring{a} = \{ \pm 2\}$.
The equation of motions of each helicity component can be found from Eqs.~\eqref{eq:Gamma1Coordinates} and~\eqref{eq:SourceGamma}, by expanding the source on the same basis. 
In particular, from Eq.~\eqref{eq:SourceGamma}, we have that 
\begin{align}\label{eq:SourceGammaDecomposition}
     {\cal S}^{\gamma}_{ij, {\bf k}} (\tau) &= - 2 \,\hat{{\cal P}}^{ab}_{ij, {\bf k}}   \, \sum_{\mathring{c}}   \int d^3 p   \,  \: \Big[ \phi''_{|{\bf k} - {\bf p}|} + 2 \H \phi'_{|{\bf k} - {\bf p}|} +  \phi_{{\bf k} - {\bf p}}( p^2 + k^2)  \, \Big] \:  H^{(0)}_{\mathring{c}, {\bf  p}} \, Y^{\mathring{c}}_{ab}(\hat p)   \nonumber \\
    &= - 2 \,\hat{{\cal P}}^{ab}_{ij, {\bf k}} \,  \sum_{\hat{b}}  \left[ \sum_{\mathring{c}}  \int d^3 p   \,   \, \Big[ \phi''_{|{\bf k} - {\bf p}|} + 2 \H \phi'_{|{\bf k} - {\bf p}|} +  \phi_{{\bf k} - {\bf p}}( p^2 + k^2)  \, \Big] \, H^{(0)}_{\mathring{c}, {\bf  p}} \, {[{\cal R} (\hat p, \hat k)]^{\mathring{c}}}_{\hat b}     \right] \: Y^{\hat b}_{ab} (\hat k) \nonumber \\
    &= - \frac12 \,  \sum_{\mathring{b}}  \left[ \sum_{\mathring{c}}  \int d^3 p   \,   \, \Big[ \phi''_{|{\bf k} - {\bf p}|} + 2 \H \phi'_{|{\bf k} - {\bf p}|} +  \phi_{{\bf k} - {\bf p}}( p^2 + k^2)  \, \Big] \, H^{(0)}_{\mathring{c}, {\bf  p}} \, {[{\cal R} (\hat p, \hat k)]^{\mathring{c}}}_{\mathring b}     \right] \: Y^{\mathring b}_{ij} (\hat k) \,.
    %&= - 2 \,    \int d^3 p   \, \Big[ \phi''_{{\bf k} - {\bf p}} + 2 \H \phi'_{{\bf k} - {\bf p}} +  \phi_{{\bf k} - {\bf p}} ( p^2 + k^2)  \, \Big] \times  \nonumber \\ 
    %\qquad   \times \sum_{\mathring{c}} \,  H^{(0)}_{\mathring{c}, {\bf p}} \,   \Big[ \Theta^{m m}_{ij} (\hat k) \, {[{\cal R} (\hat p, \hat k)]^{\mathring{c}}}_{mm} +\Theta^{\bar m \bar  m}_{ij} (\hat k) \, {[{\cal R} (\hat p, \hat k)]^{\mathring{c}}}_{\bar m \bar m} \Big] \,,
\end{align}
In the computation above, we have used that, under the rotation which brings ${\hat p}$ aligned with ${\hat k}$, the polarization basis element rotate as~\cite{Durrer:2008eom,Pitrou:2019ifq}
\begin{equation} \label{eq:RotationMatrix}
   Y^{\mathring{a}}_{ ij} (\hat p) = \sum_{\hat b} {[{\cal R} (\hat p, \hat k)]^{\mathring{a}}}_{\hat b} \: Y^{\hat b}_{ ij} (\hat k) \,, \qquad \mbox{where} \qquad {[{\cal R} (\hat p, \hat k)]^{\mathring{a}}}_{\hat b} \equiv {\rm D}^{\ell = 2}_{  \hat b \, \mathring{a}} [\varphi,\theta,0] \,,
\end{equation}
%\begin{equation} \label{eq:RotationMatrix}
%   Y^{\mathring{a}}_{ ij} (\hat p) = \sum_{\hat b} \: {\rm D}^{\ell = 2}_{  \hat b \, \mathring{a}} [\varphi,\theta,0]  \: Y^{\hat b}_{ ij} (\hat k) \,,
%\end{equation}
with $(\theta, \varphi)$ are the coordinates of $\hat p$ in reference system where $\hat k$ is aligned with one of the axis, while the matrix ${\rm D}^{\ell}_{m' m}$ is the so-called  Wigner D-coefficient related to the spin-weighted spherical harmonics, $\prescript{}{s}{Y}_{m}^{\ell}$, (see Appendix~\ref{app:SphericalHarmonics} for their definition) as
\begin{equation}\label{eq:DWignerSpin}
    {\rm D}^{\ell}_{m s} [\varphi, \theta, 0] = (-1)^m  \sqrt{\frac{4 \pi}{2 \ell + 1}} \:\prescript{}{s}{Y}_{-m}^{\ell}  (\theta, \varphi)\,.
\end{equation}
In other words, Eq.~\eqref{eq:RotationMatrix} shows that a generic rotation mixes elements with different values of m ($\hat b = \{0, \pm 1, \pm 2\}$), in the same $\ell = 2$ representation. 

From Eq.~\eqref{eq:SourceGammaDecomposition}, we can read off the source of each helicity components of the first order gravitational wave's tensor modes, namely
\begin{equation}\label{eq:SourcePolarization}
    {\cal S}^{\gamma}_{\mathring{a}, {\bf k}} = - \frac{1}{2}\, \sum_{\mathring{c}} \,    \int d^3 p   \,   \Big[ \phi''_{|{\bf k} - {\bf p}|} + 2 \H \phi'_{|{\bf k} - {\bf p}|} +  \phi_{{\bf k} - {\bf p}}( p^2 + k^2)  \, \Big] \, H^{(0)}_{\mathring{c}, {\bf p}} \, {[{\cal R} (\hat p, \hat k)]^{\mathring{c}}}_{\mathring{a}} \,,
\end{equation}
so that each component satisfies
\begin{equation}\label{eq:Gamma1Polarization}
 (\gamma^{(1)}_{\mathring{a}, \bf k} )'' + 2 \H (\gamma^{(1)}_{\mathring{a}, \bf k})' + k^2 \gamma^{(1)}_{\mathring{a}, \bf k} = {\cal S}^{\gamma}_{\mathring{a}, {\bf k}}\,.
\end{equation}
Eqs.~\eqref{eq:Gamma1Polarization} and~\eqref{eq:SourcePolarization} are one of the main results of this paper, and they display interesting features: 
\begin{itemize}
   \item Using Eq.~\eqref{eq:DWignerSpin} and the spin weighted spherical harmonics in Appendix~\ref{app:SphericalHarmonics}, we rewrite
    \begin{equation}\label{eq:MatrixRotCircleCircle}
    {[{\cal R} (\hat p, \hat k)]^{\mathring{c}}}_{\mathring{a}} = \sqrt{\frac{4 \pi}{5}}
        \begin{bmatrix}
       \cos^4\left(\frac{\theta }{2}\right)e^{2 i \varphi} &  \sin^4\left(\frac{\theta }{2}\right)e^{-2 i \varphi} \\
         \sin^4\left(\frac{\theta }{2}\right)e^{2 i \varphi} &  \cos^4\left(\frac{\theta }{2}\right)  e^{-2 i \varphi}
    \end{bmatrix}
    \end{equation}
    where we can see the standard $ e^{\pm 2 i \varphi}$ exponential factors stemming for the spin-2 nature of these polarization modes and reflecting their dependence on the choice of $\{\hat e_{\hat 1}, \hat e_{\hat 2} \}$. 
    \item The integral over all possible 3-momenta in Eq.~\eqref{eq:SourcePolarization}  accounts for the interference between GW and gravitational potential modes.  
    For a monochromatic source, $H^{(0)}_{\mathring{c}, {\bf p}}$ is peaked on a certain wave vector and direction, making the integral over $d^3 p$ trivial. 
     %This problem doesn't occur in a stochastic background, being a superposition of many different waves by definition.
    %
     \item The rotation matrix ${[{\cal R} (\hat p, \hat k)]^{\mathring{c}}}_{\mathring{a}}$ is diagonal exclusively when $\cos \theta = 1$, so that the left- and right- polarization modes of $\gamma^{(1)}_{\mathring{a}}$, can be sourced by combinations of the two helicity eigenstates of the zero-order GW. 
    %For a single propagating wave with definite wave-vector ${\bf p}$,  as just claimed, there is a non-trivial mixing of modes when  ${\bf k}$ and  ${\bf p}$ are not aligned. 
    This means that, for a single propagating wave, there is a non-trivial mixing of the two polarization modes of $H^{(0)}_{\mathring a}$, whenever ${\bf k}$ and  ${\bf p}$ are not aligned. This non-trivial mixing will also be, in part, responsible for the production of Q- and U- polarization modes in the stochastic gravitational wave background. 
   We point out that, among the classical relativistic effects, the one that is responsible for orthogonal changes in direction of the GW's trajectory is gravitational lensing~\cite{Bertacca:2017vod,Schmidt:2012ne,Bonvin:2011bg,Jeong:2014ufa}. 
    Even though in the context of wave-optics the concept of geodesics is not well-defined, our result suggests that a similar effect to lensing on the CMB polarization~\cite{Lewis:2006fu,Dai:2013nda,Hu:2000ee} is taking place. 
\end{itemize}

\subsection{Geodesic deviation equation and interpretation of scalar and vector modes}\label{sec:geodesicRiemann}
\begin{figure}[h]
    \centering
    \includegraphics[width=0.7\textwidth]{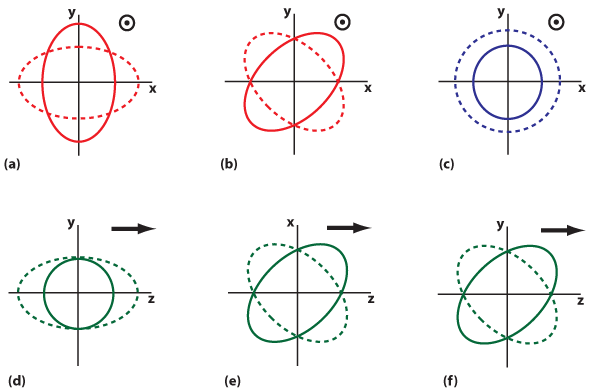}
    \caption{Possible polarization pattern which a GW can induce over a test of ring particles. The first row displays the transverse modes, as the  gravitational wave direction of propagation is out of the screen. In particular, the two red patters  corresponds to tensor polarization, sourced by $\delta_\alpha R^{\pm 2}_{\bf k}(\tau)$, while the blue one is the breathing mode, $\delta_\alpha R^{B}_{\bf k}(\tau)$. The bottom row collects the longitudinal modes (the gravitational waves propagates in the direction of the arrow). In this case, the first pattern is the longitudinal scalar polarization, sourced by $\delta_\alpha R^{L}_{\bf k}(\tau)$, while pattern (e) and (f) are the longitudinal vector modes, $\delta_\alpha R^{\pm 1}_{\bf k}(\tau)$.}
    \label{fig:GWpol}
\end{figure}

By solving Eqs.~\eqref{eq:EOMH1orderTimeTime},~\eqref{eq:EOMH1orderTimeSpace} and~\eqref{eq:GaugeH1E}, we linked $H^{(1)}, H^{(1)}_{0i}, E^{(1)}_{ij}$ to projections of the freely propagating wave along gradients of the gravitational scalar potentials, as shown in the explicit form of the sources in Table~\ref{tab:SourceSummary}.  
This means that the presence or absence of these modes is entirely determined by $ H^{(0)}_{ij}$. Indeed, these modes do not constitute new degrees of freedom, as no new initial condition is needed to solve the system of equations.
Even though their explicit forms are gauge dependent, these scalar and vector modes are physical, and as such they can contribute to specific observables. For instance, they source particular components of the electric part of the Riemann tensors, related to the relative motion between test particles~\cite{Flanagan:2005yc,wald1984general}.
One can check, that the first order (in $\alpha$), the Riemann tensor is given by
\begin{align}\label{eq:GeodRiemann}
    \delta_\alpha R_{i0j0} =& - \frac{1}{2} \left\{ {H^{(0)}_{ij}}'' + \H {H^{(0)}_{ij}}' \right\} - \frac{\epsilon}{2} \bigg\{ {\gamma^{(1)}_{ij}}'' + \H {\gamma^{(1)}_{ij}}' + {E^{(1)}_{ij}}'' + \H {E^{(1)}_{ij}}' - \nonumber \\
    & - 2 \, \partial_{(i} {H^{(1)}_{0j)}}'' -2 \H \, \partial_{(i}  {H^{(1)}_{0j)}}' + \frac{\delta_{ij}}{3} \left[ {H^{(1)}}'' + \H {H^{(1)}}' \, \right] + \nonumber \\
    & + 2 \left( \phi {H^{(0)}_{ij}} \right)'' + 2\, \H  \left( \phi {H^{(0)}_{ij}} \right)'  - \phi' {H^{(0)}_{ij}}' + 2 \partial_{(i} \left( H^{(0)}_{j)k}  \partial^k \phi \right)  \bigg\} \,,
\end{align}
where we kept contributions up to $\epsilon^1$.  Therefore, we see that all the components of $H^{(1)}_{\mu\nu}$ participate in the geodesic deviation equation.  The lowest order, namely ${\cal O}(\epsilon^0)$, corresponds to the standard result for gravitational waves, linking the Riemann tensor's components to the second order time derivative of the gravitational waves~\cite{Flanagan:2005yc,Eardley1973GravitationalwaveOA}.  However, at ${\cal O}(\epsilon^1)$, we note that two different types of terms contribute to Eq.~\eqref{eq:GeodRiemann}: either proportional to derivatives of  $H^{(1)}_{\mu\nu}$ or those of combinations of the free wave, $H^{(0)}_{\mu\nu}$, and the gravitational scalar potential, $\phi$ (last line of Eq.~\eqref{eq:GeodRiemann}).
Focusing first on the first kind of these, it is easy to realize that the contributions entering each of $\delta_\alpha R_{i0j0}$ are gauge invariant combinations of the metric perturbation (see e.g.~\cite{Flanagan:2005yc}). 
The second kind of terms, on the other hand, are a novelty compared to standard cosmological perturbation theory. Indeed, in the standard literature of cosmology, the sources of the metric perturbations are the matter perturbation, which do not enter directly the Riemann tensor. On the contrary, in the case examined in this work, the sources of $H^{(1)}_{\mu\nu}$ arise from the zeroth order metric perturbation, therefore they do enter the Riemann tensor directly, when expanding it to second order. 
These extra terms are gauge invariant on their own, as $H^{(0)}_{ij}$ is so and $\phi$ is part of a background which is considered fixed.
To understand better the role of each term in Eq.~\eqref{eq:GeodRiemann} in the motion of test particles, we decompose $\delta_\alpha R_{i0j0}({\bf k},\tau)$ on a polarization basis consisting of $\{ Y^{B}_{\mu\nu}, Y^{\pm 1}_{\mu\nu}, Y^{\pm 2}_{\mu\nu} \}$ and the new element
\begin{equation}
    Y^L_{\mu\nu} (\hat k)\equiv \frac{e^{\hat 3}_{( \mu} e^{\hat 3}_{ \nu)}}{2} \,,
\end{equation}
instead of $ Y^{0}_{\mu\nu}$ considered in Eq.~\eqref{eq:DefPolBasisTensors}. We will have that 
\begin{equation}\label{eq:Riemann}
    \delta_\alpha R_{i0j0} ({\bf k}, \tau) \: \equiv \: \delta_\alpha R^{B}_{\bf k}(\tau) \:Y^B_{ij} \:+\: \delta_\alpha R^{L}_{\bf k}(\tau) \: Y^L_{ij} \:+\: \delta_\alpha R^{\pm1}_{\bf k}(\tau)\: Y^{\pm1}_{ij}\: +\: \delta_\alpha R^{\pm2}_{\bf k}(\tau)\: Y^{\pm2}_{ij}
\end{equation}
where each component can be found by contracting the Riemann tensor with a polarization basis element. In particular, these are given by
\begin{eqnarray}
    \delta_\alpha R^{B}_{\bf k}(\tau) &\equiv&  4 Y_B^{ij} \: \delta_\alpha R_{i0j0} ({\bf k}, \tau) =  \nonumber \\
    &=& \epsilon \sum_{\mathring a}  \int d^3 p \: \phi^{\tau_{in}}_{{\bf k} - {\bf p}} \: H^{(0)}_{\mathring a, {\bf p}}({\tau_s}) \: {[ R(\hat p, \hat k)]^{\mathring a}}_0  \: {\cal G}^B({\bf k}, {\bf p}, \tau) \,,\label{eq:RiemannB}\\\nonumber \\ \nonumber \\
    \delta_\alpha R^{L}_{\bf k}(\tau) &\equiv&  4 Y_L^{ij} \: \delta_\alpha R_{i0j0} ({\bf k}, \tau) =  \nonumber \\
    &=& \epsilon \sum_{\mathring a}  \int d^3 p \: \phi^{\tau_{in}}_{{\bf k} - {\bf p}} \: H^{(0)}_{\mathring a, {\bf p}}({\tau_s}) \: {[ R(\hat p, \hat k)]^{\mathring a}}_0  \: {\cal G}^L({\bf k}, {\bf p}, \tau) \,,\label{eq:RiemannL}\\\nonumber \\  \nonumber \\
    \delta_\alpha R^{\pm 1}_{\bf k}(\tau) &\equiv&  - 4 Y_{\mp 1}^{ij} \: \delta_\alpha R_{i0j0} ({\bf k}, \tau) =  \nonumber \\
    &=& \epsilon \sum_{\mathring a}  \int d^3 p \: \phi^{\tau_{in}}_{{\bf k} - {\bf p}} \:H^{(0)}_{\mathring a, {\bf p}}({\tau_s}) \: {[ R(\hat p, \hat k)]^{\mathring a}}_{\pm 1}  \: {\cal G}^{V} ({\bf k}, {\bf p}, \tau) \,,\label{eq:RiemannV}\\\nonumber \\ \nonumber \\
    \delta_\alpha R^{\pm 2}_{\bf k}(\tau) &\equiv& 4 Y_{\mp 1}^{ij}\: \delta_\alpha R_{i0j0} ({\bf k}, \tau) =  \nonumber \\
    &=& -\frac12 \left\{  {H^{(0)}_{\pm 2}}'' + \H {H^{(0)}_{\pm2}}'\right\}+ \epsilon \sum_{\mathring a}  \int d^3 p \: \phi^{\tau_{in}}_{{\bf k} - {\bf p}} \:H^{(0)}_{\mathring a, {\bf p}}({\tau_s}) \: {[ R(\hat p, \hat k)]^{\mathring a}}_{\pm2}  \: {\cal G}^{T} ({\bf k}, {\bf p}, \tau) \,,\nonumber\\\label{eq:RiemannT}
\end{eqnarray}
up to first order in $\epsilon$. 
The functions ${\cal G}^{B}, {\cal G}^{L},{\cal G}^{\pm 1},{\cal G}^{\pm 2}$ are  given explicitly in Appendix~\ref{app:RiemannPol}. Additionally, $\phi^{\tau_{in}}$ is the value of the gravitational potential at the initial time, and $H^{(0)}_{\mathring a, {\bf p}}({\tau_s})$ the amplitude of the GW at the source position.
Each contribution above sources a different polarization pattern, i.e. different motion in a ring of test particles. Looking at Figure~\ref{fig:GWpol},  we have that 
\begin{itemize}
    \item $\delta_\alpha R^{\pm 2}_{\bf k}(\tau)$ source the two transverse tensor modes (patterns (a) and (b)).
    \item $\delta_\alpha R^{\pm 1}_{\bf k}(\tau)$ source the two longitudinal vector modes (patterns (e) and (f)). 
    \item $\delta_\alpha R^{B}_{\bf k}(\tau) $ sources the  breathing mode, sourcing transverse divergent motion (pattern (c)). 
    \item $\delta_\alpha R^{L}_{\bf k}(\tau)$ sources a longitudinal scalar motion in the direction of the wave's propagation (pattern (d)).
\end{itemize}
All of these are, in principle, observable with suitably shaped detectors, they are not gauge artifacts,  and their order of magnitude can be computed using the solutions provided in the previous Sections.  

\section{Power spectra}\label{sec:densityMatrix}
The formalism developed so far describes the propagation of GWs through the large scale structures of the Universe, without relying on the high-frequency approximation. In this Section, we consider the stochastic gravitational wave background (SGWB), namely an incoherent superposition of signals. To do so, we promote the amplitude of the GW to a random variable characterized by its n-point functions. In order to have more handy results, we slowly introduce different assumptions, e.g. statistical gaussianity or an unpolarized zeroth order background. Such hypotheses are more or less accurate according to the type of SGWB taken into consideration. Primordial gravitational waves generated in standard inflationary scenarios are expected to be well described by both of these assumptions. For gravitational waves of astrophysical origin, on the other hand, this hypothesis might break down depending on the type of sources. Nonetheless, the central limit theorem ensures that the statistics will approach that of a Gaussian random field if the background is sourced by a sufficiently large number of independent events and any high-signal-to-noise outliers have been subtracted from the detector time streams \cite{Renzini:2022alw,Renzini:2018vkx}. A detail characterization of each background type is beyond the scope of the present work and will be the subject of further investigations, here we are mainly interested in laying out the general formalism and we do so in the simplest possible scenario. 
%For simplicity, we will also carry on the analysis in the sub-horizon limit ($k \gg \H$).

\smallskip
\noindent
When performing GW observations, one is sensitive to the electric part of the Riemann tensor (Eq.~\eqref{eq:Riemann}), which contains various combinations of $H_{\mu\nu}$, rather than single components of the GW itself. To be closer to observations, instead of computing the power spectrum of the tensor, scalar and vector modes building the GW, we compute the one of $\delta_\alpha R^B, \delta_\alpha R^L, \delta_\alpha R^{\pm 1}$ and $\delta_\alpha R^{\pm 2}$. 

%In Eq.~\eqref{eq:Riemann} we have computed the components of the Riemann tensor which participate at the geodesic deviation equation, claiming that $\delta_\alpha R_{i0j0}$ also has contributions induced from the scalar and vector modes, on top of the tensor one, making them full-fledged observables.
%In this section, we focus on two of these contributions, namely $\delta_\alpha R^T_{i0j0}$ and $\delta_\alpha R^S_{i0j0}$, and we compute their power spectra. 

\subsection{Tensor modes power spectrum}
We start our analysis from the tensor modes and Eq.~\eqref{eq:RiemannT}.  These are two components, and thus they allow defining a $2 \times 2$ density matrix as in~\cite{Durrer:2008eom} 
%\footnote{Note that $\rho_{(ij)(ab)} ({\bf k}_1, \tau_1 ; {\bf k}_2, \tau_2) = \sum_{\mathring a \mathring b } \, \rho_{\mathring a \mathring b} ({\bf k}_1, \tau_1 ; {\bf k}_2, \tau_2)\, {Y}^{\mathring a}_{ij} (\hat k_1) Y^{\mathring{b}}_{ab} (\hat k_2)$.} 
\begin{align}\label{eq:DensityMatrix}
    \rho^T_{\scriptscriptstyle{\pm 2, \pm 2}} ({\bf k}_1, \tau_1 ; {\bf k}_2, \tau_2) 
    = \left[ \delta_\alpha R^{ \pm 2} ({\bf k}_1, \tau_1)\right]^* \, \delta_\alpha R^{ \pm 2 }  ({\bf k}_2, \tau_2)\,,
\end{align}
\begin{comment}
\begin{equation}
    \left \langle {\delta_\alpha R^{ \pm 1}}^* \, \delta_\alpha R^{ \pm 1 } \right\rangle = \frac{k^2_1 k^2_2}{8}\begin{bmatrix}
        I_V + V_V & Q_V - i U_V\\
        Q_V + i U_V & I_V - V_V
    \end{bmatrix}
\end{equation}
\end{comment}
and define the polarization tensor as
\begin{equation}\label{eq:PolarizationTensorDotBasis}
    {\cal P}^T_{\scriptscriptstyle{\pm 2, \pm 2}} ({\bf k}_1, \tau_1 ; {\bf k}_2, \tau_2) = \left \langle \rho^T_{\scriptscriptstyle{\pm 2, \pm 2}} ({\bf k}_1, \tau_1 ; {\bf k}_2, \tau_2)  \right\rangle\,,
\end{equation}
where the average is an ensemble average or a volume one upon invoking statistical homogeneity and isotropy~\cite{Romano:2016dpx,Renzini:2018vkx,Caprini:2018mtu}.
If one wants to compute the properties of the SGWB generated by astrophysical sources, then a summation over all possible unresolved sources is also understood~\cite{LISACosmologyWorkingGroup:2022kbp}.

\subsubsection*{zeroth order}
From its definition and Eq.~\eqref{eq:RiemannT}, it is clear that the polarization tensor has contributions at $\epsilon^0$, $\epsilon^1$ and $\epsilon^2$ orders, namely
\begin{equation}
    {\cal P}^T_{\scriptscriptstyle{\pm 2, \pm 2}} = {\cal P}^{T, (0)}_{\scriptscriptstyle{\pm 2, \pm 2}} \:+ \:{\cal P}^{T, (1)}_{\scriptscriptstyle{\pm 2, \pm 2}} \: + \:{\cal P}^{T, (2)}_{\scriptscriptstyle{\pm 2, \pm 2}}\,.
\end{equation}
From Eq.~\eqref{eq:RiemannT},  in the sub-horizon limit $k_1, k_2 \gg \H$, one can check that
\begin{align}
    {\cal P}^{T, (0)}_{\scriptscriptstyle{\pm 2, \pm 2}} ({\bf  k}_1, \tau_1 ; {\bf  k}_2, \tau_2) &= \frac{k^2_1 k^2_2}{4} \left \langle {H^{(0)}_{\pm 2}}^* ({\bf k}_1, \tau_1) \:\: H^{(0)}_{\pm 2}  ({\bf k}_2, \tau_2)   \right\rangle\,,
\end{align}
where we have also used the background equations of motions. 
This result suggests the definition of the Stokes parameters as the decomposition of the polarization tensor over the Pauli matrices basis
\begin{align}\label{eq:StokesParameters}
     {\cal P}^T_{\scriptscriptstyle{\pm 2, \pm 2}}  \equiv& \: \frac{k^2_1 k^2_2}{4} \: \frac12 \left( I_T \: \pmb{\sigma}^0 \, + Q_T\: \pmb{\sigma}^1 \, + U_T\: \pmb{\sigma}^2+ V_T\: \pmb{\sigma}^3 \right)=  \frac{k^2_1 k^2_2}{4}\:\frac12
   \begin{bmatrix}
        I_T + V_T & Q_T - i U_T\\
        Q_T + i U_T & I_T - V_T
    \end{bmatrix}\,,
\end{align}
where $\sigma_i$ are the Pauli matrices 
\begin{equation}
    \pmb{\sigma}^0 = 
    \begin{bmatrix}
        1&0\\
        0&1
    \end{bmatrix}\,, \quad
    \pmb{\sigma}^1 = 
    \begin{bmatrix}
        0&1\\
        1 &0
    \end{bmatrix}\,, \quad
    \pmb{\sigma}^2 = 
    \begin{bmatrix}
        0& -i\\
        i& 0
    \end{bmatrix}\,, \quad
     \pmb{\sigma}^3 = 
    \begin{bmatrix}
        1&0\\
        0&-1
    \end{bmatrix}\,.
\end{equation}
This definitions are such that, at order $\epsilon^0$, we find the usual background intensity and Stokes parameters~\cite{Anile1974GravitationalSP,Kosowsky:1994cy}), namely
\begin{align}
     \mbox I^{(0)}_T &= |H_{+2}|^2 + |H_{-2}|^2 \,, \,\quad \mbox V^{(0)}_T =|H_{+2}|^2 - |H_{-2}|^2\,, \\
     \mbox Q^{(0)}_T &= 2 {\mbox {Re}} (H^*_{+2} \, H_{-2})\,, \qquad \mbox U^{(0)}_T = 2 {\mbox {Im}} (H^*_{+2} \, H_{-2}) \,.
\end{align}
In particular, $ \mbox I^{(0)}_T$ is the background intensity, $\mbox V^{(0)}$ the net difference between the right- and left- helicity modes, while $ \mbox Q^{(0)}$ and $\mbox U^{(0)}$ represents a linear polarization of the SGWB due to a difference of the phases of the two circular ones.  Unless the GWs production mechanisms prefer one polarization mode over the other, as it could be the case in parity violating theories~\cite{Alexander:2009tp}, one should expect to observe an unpolarized SGWB~\cite{Renzini:2022alw} at  ${\cal O}(\epsilon^0)$. 
The intensity $\mbox I^{(0)}({\bf  k}_1, \tau_1 ; {\bf  k}_2, \tau_2)$ can be related to the tensor mode power spectrum as,
\begin{equation}
     \mbox I^{(0)}_T ({\bf  k}_1, \tau ; {\bf  k}_2, \tau) = 2 \pi^2 \, \delta^3({\bf k}_1 - {\bf k}_2) \, \frac{\Delta_h^2 ({\bf k}_1)}{|{\bf k}_1|^3}\,,
\end{equation}
in the case of a statistically homogeneous background.

\subsubsection*{First order}
We take Eq.~\eqref{eq:StokesParameters} as definition of the perturbed background's intensity and Stokes parameters. Looking at the first order, from Eq.~\eqref{eq:RiemannT} it is clear that 
\begin{align}
    {\cal P}^{T, (1)}_{\scriptscriptstyle{\pm 2, \pm 2}} ({\bf  k}_1, \tau_1 ; {\bf  k}_2, \tau_2)  \propto  \int d^3 p   \left \langle  \phi^{\tau_{in}} \, {H^{(0)}_{\mathring{a}} }^*(\tau_s^1) \,  \, H^{(0)}_{ \mathring c } (\tau_s^2) \right\rangle  \,.
\end{align}
According to the type of situation taken in consideration, the three point function can vanish or not. 
For simplicity, here we consider the case of uncorrelated Gaussian random fields, where the three point function above vanishes. For the astrophysical background, gaussianity follows from the emission from many uncorrelated regions, a condition which might not represent the case of sources tracing the underlying dark matter distribution~\cite{Caprini:2018mtu}. Modeling properly the statistics of astrophysical GW sources can be a highly non-trivial task, and it requires introducing the concept of GW bias~\cite{Libanore:2020fim}, a topic which is beyond the scope of this project. 
However, we take this opportunity to clarify that possible dark matter tracers, both playing the role of source or of lens, can be accommodated in our formalism at the level of transfer functions relating the specific tracer and the gravitational potential. 

\subsubsection*{Second order}
Looking at Eq.~\eqref{eq:RiemannT}, it is clear that we have to compute
\begin{comment}
\begin{equation}
   \left \langle \delta_\alpha R^{X}_{{\bf k}_1} \: \delta_\alpha R^{X}_{{\bf k}_2}\right\rangle \propto  \int d^3 p_1 d^3 p_2   \left \langle  \phi^{\tau_{in} *}_{{\bf k}_1 - {\bf p}_1} \,  {\phi_{{\bf k}_2 - {\bf p}_2}^{\tau_{in}}} \, H^{(0)*}_{{\mathring c},{\bf p}_1} (\tau^1_s ) \, H^{(0)}_{{\mathring d},{\bf p}_2} (\tau^2_s ) \right\rangle\,.
\end{equation}
\end{comment}
\begin{equation}
    {\cal P}^{T,(2)}_{\scriptscriptstyle{\pm 2, \pm 2}} ({\bf  k}_1, \tau_1 ; {\bf  k}_2, \tau_2)  \propto  \int d^3 p  \left \langle  \phi^{\tau_{in} *}_{{\bf k}_1 - {\bf p}_1} \,  {\phi_{{\bf k}_2 - {\bf p}_2}^{\tau_{in}}} \, H^{(0)*}_{{\mathring c},{\bf p}_1} (\tau^1_s ) \, H^{(0)}_{{\mathring d},{\bf p}_2} (\tau^2_s )  \right\rangle\,.
\end{equation}
In order to tackle the  four-points function above, we assume that  $\phi^{\tau_{in}}_{{\bf k}}$ and $H^{(0)}_{\pm 2} ({\bf p}, {\tau_s})$ are both Gaussian random fields, so that we can evaluate the expectation value using Wick's theorem.  Then, we further assume that the initial gravitational scalar potential and the gravitational wave sources are uncorrelated and that the GW background is unpolarized at $0^{th}$ order. This situation can represent a SGWB of cosmological origins, and taking this as an example, we fix the source of the GWs well outside the horizon ($k / \H({\tau_s}) \ll 1$). Introducing also the scalar potential power spectrum, we consider the following expectation values
\begin{eqnarray}
    \epsilon\: \Big \langle \phi^{\tau_{in} *}_{{\bf k}} \,  {\phi_{{\bf p}} ^{\tau_{in}}} \Big  \rangle &=& \, \delta^3({\bf k} - {\bf p}) \, \frac{2 \pi^2 }{k^3} \Delta_\phi^2 ({\bf k})\,, \label{eq:ScalPotPowerSpectrum}\\
    \epsilon\: \Big \langle \,H^{(0)*}_{{\mathring c},{\bf k}} (\tau_s^1 )\, {H^{(0)}_{\mathring d,{\bf p}}} (\tau_s^2 )  \Big \rangle  &=& \frac{\delta_{\mathring{c} \mathring{d} }}{2} \, \delta^3({\bf k} - {\bf p}) \, \frac{2 \pi^2 }{k^3} \Delta_h^2 ({\bf k}) \,,\label{eq:GWPowerSpectrum}\\
   \epsilon \Big \langle \,{\phi_{{\bf k}} ^{\tau_{in}}}\, {H^{(0)}_{\mathring d,{\bf p}}} (\tau_s^2 )  \Big \rangle &=& 0 \label{eq:PotGWUncorr}\,.
\end{eqnarray}
With these assumptions we have that, the equal time tensor polarization tensor is given by
\begin{multline}\label{eq:PolSecondIntermediateT}
    {\cal P}^{(2)}_{\scriptscriptstyle{\pm 2 \pm 2}} ({\bf  k}_1, \, \tau ; {\bf  k}_2, \, \tau) \: = \: \delta^3({\bf k}_1 - {\bf k}_2) \,  \frac{(2 \pi^2)^2}{2} \times    \\
     \qquad\times \int d^3 p \left| {\cal G}^T({\bf k}_1, {\bf p}, \tau)\right|^2 \: \frac{\Delta^2_\phi({\bf k}_1 - {\bf p}) \Delta^2_h({\bf p})}{|{\bf k}_1 - {\bf p}|^3 p^3}   \: \sum_{\mathring c} {\Big[{\cal R}^* (\hat p, \hat k_1) \Big]^{\mathring c}}_{\pm 2} {\Big[{\cal R} (\hat p, \hat k_1) \Big]^{\mathring c}}_{\pm 2} \,.
\end{multline}
To compute the product of the rotation matrices, one can use their definition in Eqs.~\eqref{eq:RotationMatrix} and~\eqref{eq:DWignerSpin} together with the summing rules of spin weighted spherical harmonics (see Appendix~\ref{app:SphericalHarmonics}).
In particular, we obtain
%\begin{align}\label{eq:RotationMatrixProduct}
%\sum_{\mathring c} {[{\cal R}^*  (\theta, \varphi) ]^{\mathring c}}_{\pm 2}  {[{\cal R} (\theta, \varphi)]^{\mathring c}}_{\pm 2} = 
% \renewcommand{\arraystretch}{1.6}
 %   \begin{bmatrix}
%\frac{4 \sqrt{\pi}}{5}\left(\, {Y}^{0}_{0}  + \frac{2 \sqrt{5}}{7} {Y}^{2}_{0} + \frac{{Y}^{4}_{0}}{42} \, \right) &\frac23 \sqrt{\frac{2 \pi}{35}}  {Y}^{4}_{4} \\
%\frac23 \sqrt{\frac{2 \pi}{35}}  {Y}^{4}_{-4}   & \frac{4 \sqrt{\pi}}{5}\left(\, {Y}^{0}_{0}  + \frac{2 \sqrt{5}}{7} {Y}^{2}_{0} + \frac{{Y}^{4}_{0}}{42} \, \right) \\
%    \end{bmatrix}
%\end{align}
\begin{align}\label{eq:RotationMatrixProduct}
\sum_{\mathring c} {[{\cal R}^*  (\theta, \varphi) ]^{\mathring c}}_{\pm 2}  {[{\cal R} (\theta, \varphi)]^{\mathring c}}_{\pm 2} = \frac18
 \renewcommand{\arraystretch}{1.6}
    \begin{bmatrix}
(\mu^4 + 6 \mu^2 + 1) \qquad &  (1-\mu^2)^2 e^{4 i \varphi}\\
(1-\mu^2)^2 e^{-4 i \varphi}  \qquad  &   (\mu^4 + 6 \mu^2 + 1) \\
    \end{bmatrix}
\end{align}
where $\mu = \cos \theta$. 
%all the functions in the matrix above are spin-$0$ spherical harmonics evaluated in $(\theta, \varphi)$.
Combining Eqs.~\eqref{eq:PolSecondIntermediateT} and~\eqref{eq:RotationMatrixProduct} allow us to write the intensity and the Stokes parameters at ${\cal O}(\epsilon^2)$.
From their definition in Eq.~\eqref{eq:StokesParameters} we write
\begin{align}
    I_T^{(2)}({\bf k}_1, \tau; {\bf k}_2, \tau) =& \quad \: \delta^3({\bf k}_1 - {\bf k}_2)\: \frac{(2 \pi^2)^2}{2} \: \frac{8}{k_1^2 k_2^2}  \times \nonumber\\
    &\qquad\times \int d^3 p \left| {\cal G}^T({\bf k}_1, {\bf p}, \tau)\right|^2 \, \frac{\Delta^2_\phi({\bf k}_1 - {\bf p}) \Delta^2_h({\bf p})}{|{\bf k}_1 - {\bf p}|^3 p^3} \:\times   \frac{(\mu^4 + 6 \mu^2 + 1) }{8}\label{eq:Stoke1I}\\ \nonumber\\
    V_T^{(2)}({\bf k}_1, \tau; {\bf k}_2, \tau) =& \quad \: 0 \label{eq:Stoke1V}\\  \nonumber\\
    (Q_T^{(2)} \pm i U_T^{(2)})({\bf k}_1, \tau; {\bf k}_2, \tau)  =& \quad \: \delta^3({\bf k}_1 - {\bf k}_2) \: \frac{(2 \pi^2)^2}{2} \: \frac{8}{k_1^2 k_2^2} \times \nonumber\\
    &\qquad\times \int d^3 p \left| {\cal G}^T({\bf k}_1, {\bf p}, \tau)\right|^2 \, \frac{\Delta^2_\phi({\bf k}_1 - {\bf p}) \Delta^2_h({\bf p})}{|{\bf k}_1 - {\bf p}|^3 p^3} \: \times \frac{(1-\mu^2)^2 }{8} e^{ \pm 4 i \varphi} \label{eq:Stoke1QpmiU}
\end{align}
Eqs.~\eqref{eq:Stoke1I},~\eqref{eq:Stoke1V} and~\eqref{eq:Stoke1QpmiU} constitute another of the key results of this paper. The information that they encode is
\begin{itemize}
    \item  Eq.~\eqref{eq:Stoke1V} clearly states that the Stokes parameter $V^{(2)}_T$ is null also at order ${\cal O}(\epsilon^2)$: the interaction with structures doesn't violate parity by creating a difference in the amounts of left- and right- states. This result was expected for two reasons: we assumed a GW generation mechanism that doesn't prefer one state over the other, and in our analysis the two tensor modes propagate in the same fashion. 

    \item Because of the orthogonality properties of the spherical harmonics, the angular integrals in Eqs.~\eqref{eq:Stoke1I} and~\eqref{eq:Stoke1QpmiU} select specific multipoles of the integrands. In the case of the latter equations, the dependence on the angle $\theta$ is contained in the functions ${\cal G}^T({\bf k}_1, {\bf p}, \tau)$ (see its definition in Eq.~\eqref{eq:DefGT}) and in  $|{\bf k}_1 - {\bf p}| = \sqrt{k^2_1 + p^2 - 2 \,p \,k \cos \theta}$.
    However, only the two spectra ($\Delta^2_\phi({\bf k} - {\bf p})$ and $\Delta^2_h({\bf p})$) can depend on $\phi$. Assuming statistical isotropy for the latter  (i.e. $\Delta^2_\phi({\bf k}) = \Delta^2_\phi(k)$), then the polarization $Q_T$ and $U_T$ patterns are produced at first order only if the tensor modes show an anisotropy of order $\ell = 4$. 
    Indeed, 
    \begin{equation}
     (1- \mu^2 )^2 e^{\pm 4 i \varphi} =  \frac{16}{3} \sqrt{\frac{2 \pi}{35}} \:  \prescript{}{0}{Y}_{\pm 4}^{4}(\theta, \varphi) \,.
    \end{equation}
    If both backgrounds are statistically isotropic (i.e. also $\Delta^2_h({\bf k}) = \Delta^2_h(k)$), then $V_T = Q_T = U_T = 0$. 

    \item The $\varphi$-dependence in Eq.~\eqref{eq:Stoke1QpmiU} reflects the spin-4 nature of these Stokes parameters, and the spin-2 one of the GWs vicariously. 
    The angle $\varphi$ depends on the choice of the tetrad's elements $\{ \hat e_{\hat 1} (\hat k), \hat e_{\hat 2} (\hat k)  \}$ made to construct the polarization basis for $H_{\mathring{a}, \bf k} (\tau)$. 
    To get rid of this arbitrariness, in parallel with CMB computations~\cite{Durrer:2008eom,Hu:1997hp}, one could define $E$ and $B$ modes for the SGWB. 
    The decomposition of $Q_T^{(2)}  \pm i U_T^{(2)}$ on spin weighted spherical harmonics can be found in~\cite{Chu:2020qiw}.
    
    \item Similarly to~\cite{Pizzuti:2022nnj}, we find the same spherical harmonics in the integrals of $I^{(2)} $ and $Q_T^{(2)}  \pm i U_T^{(2)}$, however the integrands are different. In our case, it is the interplay of the gravitational waves and the large scale structures which must be characterized by a hexadecapole anisotropy (i.e. $\ell = 4$) to produce a polarization pattern, while in their case this result applies exclusively to $I^{(0)}$. We comment that this is due to the great difference between our two formalism as the two are based on two different sets of assumptions.
    As in~\cite{Pizzuti:2022nnj}, we find that all values of $\ell$ from 0 to 4 contribute to  $I^{(2)}$.
  
    \item Starting from the set of equations above, one can compute the angular power spectrum by expanding the fields, evaluated on the $2-$ sphere, on the spherical harmonics basis. 
    Despite the exciting prospects~\cite{Baker:2019ync}, the poor angular resolution typical of GW detectors still poses a serious challenge in accessing the high multipoles of the SGWB anisotropies (ET is expected to probe multipoles up to $\ell \lesssim 50$, while LISA $\ell \lesssim 15$~\cite{Ungarelli:2001xu,Seto:2004np,LISACosmologyWorkingGroup:2022kbp}).
\end{itemize}

\subsection{Vector modes power spectrum}
Similarly to the tensor modes, also in this case we can define a polarization tensor 
\begin{align}
    \rho^V_{\scriptscriptstyle{\pm 1, \pm 1}} ({\bf k}_1, \tau_1 ; {\bf k}_2, \tau_2) 
    = \left[ \delta_\alpha R^{ \pm 1} ({\bf k}_1, \tau_1)\right]^* \, \delta_\alpha R^{ \pm 1 }  ({\bf k}_2, \tau_2)\,,
\end{align}
and define the polarization tensor as its ensemble average ${\cal P}^V_{\scriptscriptstyle{\pm 1, \pm 1}} \equiv \braket{\rho^V_{\scriptscriptstyle{\pm 1, \pm 1}}}$. Since $\delta_\alpha R^{\pm 1}$ is null at order $\epsilon^0$,  we have that ${\cal P}^V_{\scriptscriptstyle{\pm 1, \pm 1}} = {\cal P}^{V, (2)}_{\scriptscriptstyle{\pm 1, \pm 1}} $.
Using Eq.~\eqref{eq:RiemannV}, and under the same hypothesis just employed for the tensor modes,  we find
\begin{multline}\label{eq:PolSecondIntermediateV}
    {\cal P}^{V, (2)}_{\scriptscriptstyle{\pm 1, \pm 1}} ({\bf  k}_1, \, \tau ; {\bf  k}_2, \, \tau) \: = \: \delta^3({\bf k}_1 - {\bf k}_2) \,  \frac{(2 \pi^2)^2}{2} \times    \\
     \qquad\times \int d^3 p \left| {\cal G}^V({\bf k}_1, {\bf p}, \tau)\right|^2 \: \frac{\Delta^2_\phi({\bf k}_1 - {\bf p}) \Delta^2_h({\bf p})}{|{\bf k}_1 - {\bf p}|^3 p^3}   \: \: \frac12
 \renewcommand{\arraystretch}{1.6} 
    \begin{bmatrix}
1 - \mu^4  & \quad(1 - \mu^2)^2 e^{2 i \varphi} \\
(1 - \mu^2)^2 e^{-2 i \varphi}  & \quad1 - \mu^4 \\
    \end{bmatrix} \,,
\end{multline}
where the matrix is the result of $\sum_{\mathring c} {\Big[{\cal R}^* (\hat p, \hat k_1) \Big]^{\mathring c}}_{\pm 1} {\Big[{\cal R} (\hat p, \hat k_1) \Big]^{\mathring c}}_{\pm 1}$. 
In analogy to the tensors, we define also the vector modes intensity and Stokes parameters as 
\begin{equation}
    {\cal P}^V_{\scriptscriptstyle{\pm 1, \pm 1}}  \equiv \frac{k^2_1 k^2_2}{4}\:\frac12
  \begin{bmatrix}
      I_V + V_V & Q_V - i U_V\\
       Q_V + i U_V & I_V - V_V
   \end{bmatrix}\,,
\end{equation}
so that we can read off the components directly from Eq.~\eqref{eq:PolSecondIntermediateV}. In particular, we find 
\begin{align}
    &I_V^{(2)}({\bf k}_1, \tau; {\bf k}_2, \tau) =  \: \delta^3({\bf k}_1 - {\bf k}_2) \: \frac{8(2 \pi^2)^2}{ 2 k_1^2 k_2^2}  \int d^3 p \left| {\cal G}^V({\bf k}_1, {\bf p}, \tau)\right|^2 \, \frac{\Delta^2_\phi({\bf k}_1 - {\bf p}) \Delta^2_h({\bf p})}{|{\bf k}_1 - {\bf p}|^3 p^3} \:   \frac{(1 -\mu^4 ) }{2} \,,\nonumber\\\label{eq:Stoke1IV}\\ 
    &V_V^{(2)}({\bf k}_1, \tau; {\bf k}_2, \tau) = 0\,, \label{eq:Stoke1VV}\\ \nonumber\\
    &(Q_V^{(2)} \pm i U_V^{(2)})({\bf k}_1, \tau; {\bf k}_2, \tau)  =  \: \delta^3({\bf k}_1 - {\bf k}_2) \:\frac{(2 \pi^2)^2}{2} \: \frac{8}{ k_1^2 k_2^2} \times \nonumber\\
    &\qquad\qquad\qquad\times \int d^3 p \left| {\cal G}^V({\bf k}_1, {\bf p}, \tau)\right|^2 \, \frac{\Delta^2_\phi({\bf k}_1 - {\bf p}) \Delta^2_h({\bf p})}{|{\bf k}_1 - {\bf p}|^3 p^3} \: \times \frac{(1-\mu^2)^2 }{2} e^{ \pm 2 i \varphi} \label{eq:Stoke1QpmiUV}\,.
\end{align}
From this results we see that propagation effects can produce a vector component in the SGWB. Eq.~\eqref{eq:Stoke1IV} regulates the intensity of such background, while Eqs.~\eqref{eq:Stoke1QpmiUV} its linear polarization content. As in the case of tensor modes, the interaction with matter structures does not generate a net difference between the two vector polarization, namely $\pm 1$. 
Eqs.~\eqref{eq:Stoke1QpmiUV} clearly displays the angular dependence $e^{ \pm 2 i \varphi} $, typical of a polarization tensor of a spin 1 field. Because of the orthogonality properties of the spherical harmonics, we have that, in order to produce $Q_V$ and $U_V$ polarization pattern, a quadrupole or a hexadecapole anisotropy are required. Indeed
\begin{equation}
    (1-\mu)^2 e^{\pm 2 i \varphi} = \frac{24}{7} \sqrt{\frac{2 \pi}{15}} \left[ \prescript{}{0}{Y}_{\pm 2}^{2}(\theta, \varphi) - \frac{1}{3 \sqrt{3}} \prescript{}{0}{Y}_{\pm 2}^{4}(\theta, \varphi) \right]  \,.
\end{equation}
In the case of a statistically isotropic background, then $(Q_V^{(2)} = U_V^{(2)}) = 0$, as the integral in $d \varphi$ in Eqs.~\eqref{eq:Stoke1QpmiUV} vanishes.

%%%%%%%%%%%%%%% 
\subsection{Scalar modes power spectra}
Given the similar structure of $\delta_\alpha R^B$ and $\delta_\alpha R^L$ in Eqs.~\eqref{eq:RiemannB} and~\eqref{eq:RiemannL}, we treat them jointly. Since these are scalar modes, we define only their intensity as
\begin{eqnarray}
   I_{B/L} ({\bf k}_1 \tau ; {\bf k}_2, \tau) &\equiv& \frac{8}{k^2_1 k^2_2} \Big\langle [\delta_\alpha R^{B/L}({\bf k_1},\tau)]^* \delta_\alpha R^{B/L}({\bf k_2},\tau) \Big \rangle\,,
\end{eqnarray}
which explicitly are
\begin{align}\label{eq:IntensityBL}
 I_{B/L} ({\bf k}_1 \tau ; {\bf k}_2, \tau) =&  \: \delta^3({\bf k}_1 - {\bf k}_2) \: \frac{(2 \pi^2)^2}{2}\frac{8}{ k_1^2 k_2^2} \times \nonumber\\
 &\times \int d^3 p \left| {\cal G}^{B/L}({\bf k}_1, {\bf p}, \tau)\right|^2 \, \frac{\Delta^2_\phi({\bf k}_1 - {\bf p}) \Delta^2_h({\bf p})}{|{\bf k}_1 - {\bf p}|^3 p^3} \:   \frac{3(1 -\mu^2 )^2 }{4} \,,
\end{align}
under the hypothesis of gaussianity and Eqs.~\eqref{eq:ScalPotPowerSpectrum},~\eqref{eq:GWPowerSpectrum} and~\eqref{eq:PotGWUncorr} as before.

\subsection{Numerical results}
\begin{figure}[H]
   \centering
\includegraphics[width=\textwidth]{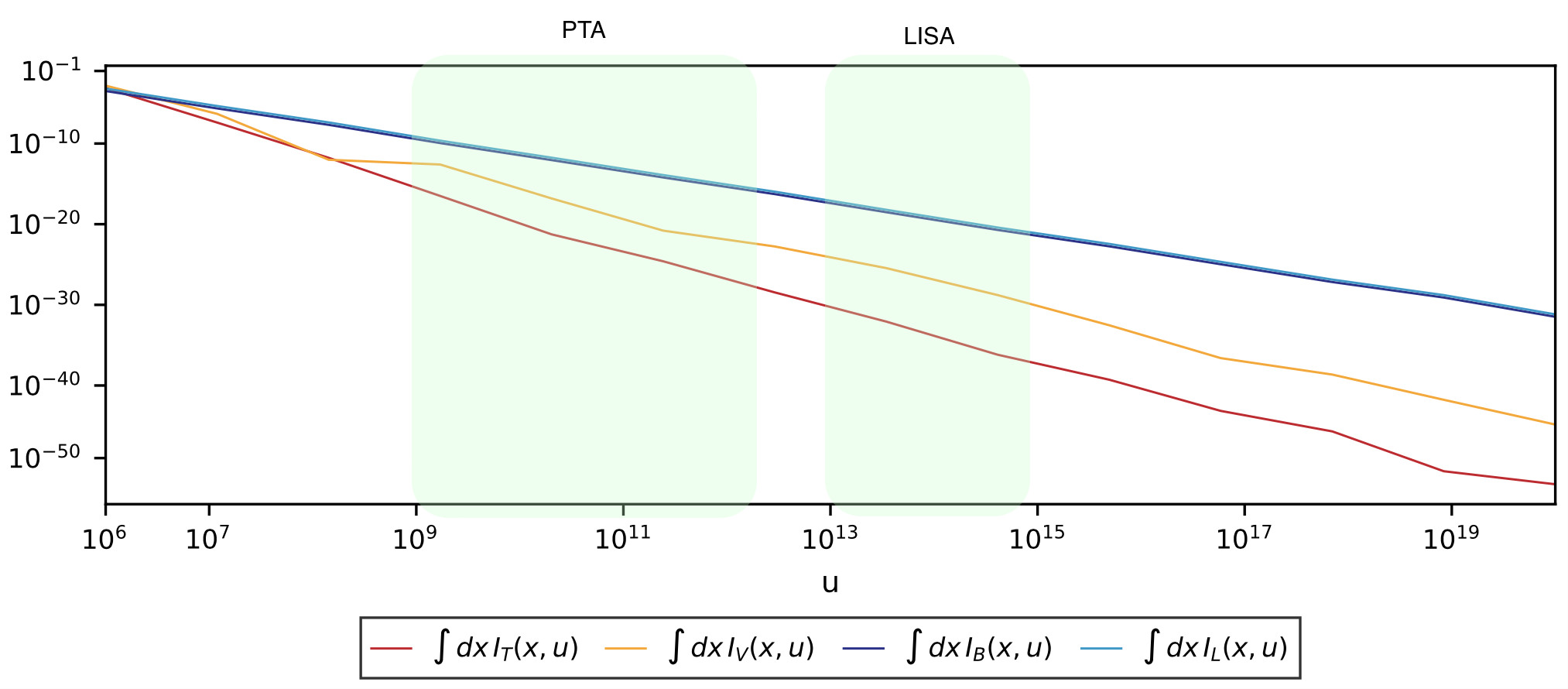}
    \caption{Integrals $\int  I_X(x,u)$  as a function of $u = 2 k/\H$. Using $k \approx u \cdot 10^{-18}$ Hz, we can convert from $u$ to the gravitational wave frequency today. We have that GWs in the frequency band probed by Pulsar Timing Arrays, i.e. $k \approx (10^{-9} - 10^{-6})$, correspond to $u \approx (10^{9}-10^{12})$, while for LISA, at $k \approx (10^{-5} - 10^{-3})$ Hz, we have $u \approx (10^{13}-10^{15})$. }
    \label{fig:Integrals}
\end{figure}

\begin{figure}[H]
   \centering
\includegraphics[width=1\textwidth]{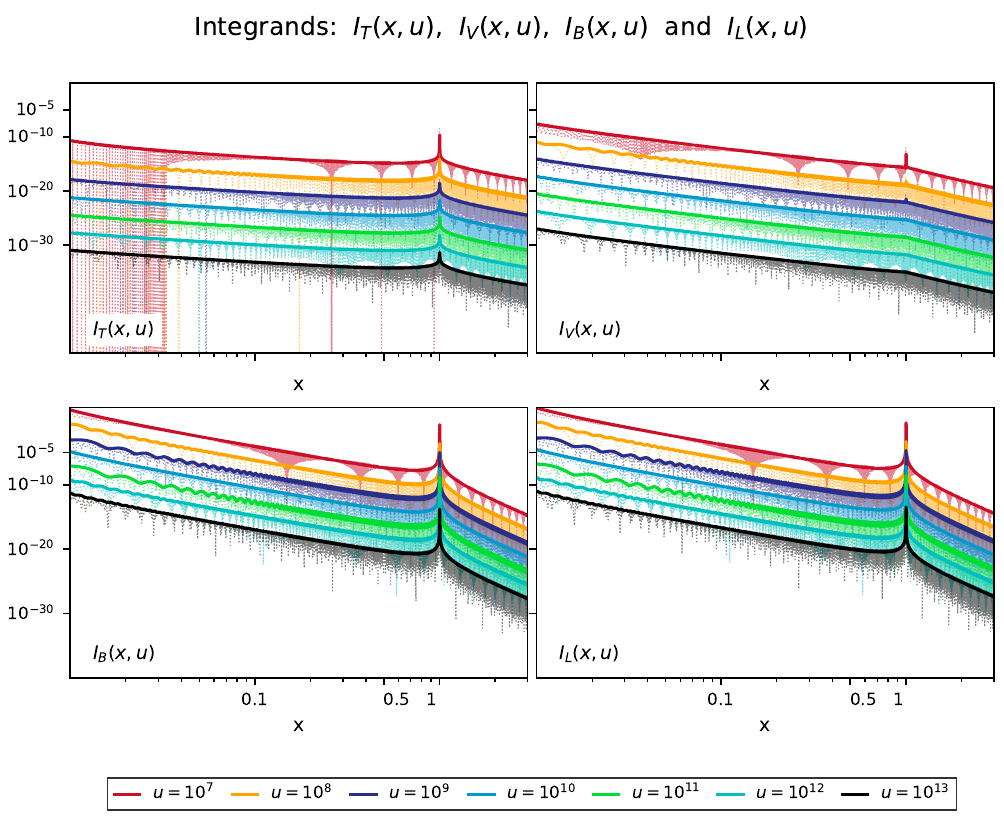}
    \caption{Integrands  $I_X(x,u)$ for  various values of $u$, as a function of $x$. The solid lines represent a smoothed average  of the functions. The peak at $x=1$ corresponds to the resonance at $p = k$.}
    \label{fig:Integrand}
\end{figure}

We numerically integrate the intensities $I^{(2)}_T$, $I^{(2)}_V$, $I^{(2)}_B$  and $I^{(2)}_L$ under the assumption of scale-invariant and isotropic spectra, i.e. $\Delta^2_\phi({\bf k}) = \Delta^2_\phi$ and $\Delta^2_h({\bf k}) = \Delta^2_h$ constants.  In this case,  also the second order SGWB is statistically isotropic, and the polarization Stokes parameter vanish, i.e. $Q_T \pm i U_T =Q_V \pm i U_V =0$.
For simplicity, we neglect the cosmological constant and consider a purely matter dominated Universe. Since we are also neglecting radiation, we have that $\T^{\phi}_{|{\bf k} - {\bf p}|}  \approx const$. 
We then rewrite the relative variation of the intensities, due to the intervening structures, as
\begin{eqnarray}
    \frac{I_X^{(2)}(k, \tau)}{I^{(0)}(k, \tau)} &=&  \: \left[  \T^{\phi}_{|{\bf k} - {\bf p}|} \right]^2 \times \Delta^2_\phi \times \int^{\infty}_0 d x \, {\mathcal I}_X(x,u)\,,
\end{eqnarray}
where $X = T, V, B, L $ and $u \equiv k (\tau + \tau_i)$ and $x \equiv p/k$. The functions ${\mathcal I}_T,  {\mathcal I}_V,  {\mathcal I}_B$ and ${\mathcal I}_L$ can be easily found from the previous results
\begin{eqnarray}
    {\mathcal I}_T(x,u) &\equiv& 2 \pi^3 \int^1_{-1}  d \mu \:\frac{|{\cal G}^T({\bf k}, {\bf p}, \tau)|^2}{\left[ k^2 \T^{\phi}_{|{\bf k} - {\bf p}|} \right]^2}\: \frac{\mu^4 + 6 \mu^2 + 1}{(1 + x^2 - 2 x \mu )^{3/2}}\,, \\
    {\mathcal I}_V(x,u) &\equiv&  8 \pi^3 \int^1_{-1}  d \mu \:\frac{|{\cal G}^V({\bf k}, {\bf p}, \tau)|^2}{\left[ k^2 \T^{\phi}_{|{\bf k} - {\bf p}|} \right]^2}\: \frac{(1-\mu^4)}{(1 + x^2 - 2 x \mu )^{3/2}} \,,\\
    {\mathcal I}_B(x,u) &\equiv&  12 \pi^3 \int^1_{-1}  d \mu \:\frac{|{\cal G}^B({\bf k}, {\bf p}, \tau)|^2}{\left[ k^2 \T^{\phi}_{|{\bf k} - {\bf p}|} \right]^2}\: \frac{(1 - \mu^2)^2}{(1 + x^2 - 2 x \mu )^{3/2}}\,, \\
    {\mathcal I}_L(x,u) &\equiv&  12 \pi^3 \int^1_{-1}  d \mu \:\frac{|{\cal G}^L({\bf k}, {\bf p}, \tau)|^2}{\left[ k^2 \T^{\phi}_{|{\bf k} - {\bf p}|} \right]^2} \:\frac{(1 - \mu^2)^2}{(1 + x^2 - 2 x \mu )^{3/2}} \,.
\end{eqnarray}
Figure~\ref{fig:Integrand} shows a plot of the integrands ${\cal I}_X$ for various values of $u$. The plots show a resonance peak at $x=1$, corresponding to the value of $p = k$. Indeed, in this case the frequencies of the two stochastic processes match, causing a constructive interference. The integrands approach $\to 0$ in the limit $x \gg 1$, indicating that the propagation effects become negligible when the two scales present in the system ($p$ and $k$) are different. The behavior close to $x\to 0$ is more subtle, as the integrands diverge in this limit. Since we are interested in the contribution of the resonance peak, and a treatment of the infrared divergence is beyond the scope of this work, we place an IR cutoff when integrating the functions ${\cal I}_X$. Additionally, some of the approximation made in order to numerically compute the integrals, might not hold in this limit.  Because of the level of the divergence, the result of the integral highly depends on this cutoff. The criteria that we follow to fix such cutoffs is to have $\int d x \, {\cal I}_X \sim 1$ at $u \approx 10^{4}$.
The reason behind this choices is that, in this way, at $u \approx 10^{4}$, roughly corresponding to the CMB scales today, the level of anisotropy is of order $\Delta^2_\phi$.  
Indeed, to translate from $u$ to the frequency of the gravitational wave, $k$,  one can consider the value of $u$ today. Given that $u = k (\tau+\tau_i)$, we have that $k \approx u \cdot 10^{-18}$ Hz or  $k \approx u \cdot 10^{-4}$ (Mpc$^{-1}$). For instance, GWs in the frequency band probed by CMB experiments corresponds to $k \approx (10^{-16} - 10^{-13})$ Hz thus $u \approx (10^2 - 10^5)$. 
With this guidance, we fix $x = 0.00005$, $x = 0.0001$ and $x= 0.003$ as IR cutoffs when integrating ${\cal I}_T$, ${\cal I}_V$ and ${\cal I}_{B/L}$ respectively. 
Also, as $u$ increases, the integrands ${\cal I}_X$ decrease. This behavior reflects the decay in time of the source of the first order modes (see Table~\ref{tab:SourceSummary}) and of the other second order combination entering the Riemann tensor in Eq.~\eqref{eq:Riemann}. Indeed, each polarization component of the Riemann tensor, namely Eqs.~\eqref{eq:RiemannB}-~\eqref{eq:RiemannT}, is sourced by one free gravitational wave $H^{(0)}_{\mu\nu}$ and one scalar potential. Even if the latter is constant during matter domination, the decay of $H^{(0)}_{\mu\nu}$ in time suppresses the various components of the Riemann tensor. Since $u = k (\tau + \tau_i) = 2 k/\H$, modes with higher $u$ have entered the horizon sooner and have had more time to decay (for fixed $k$, we have a smaller $\H$ and hence a larger $u$). 
Moreover, the behavior at smaller values of $u$ must be taken into consideration with some care, as the numerical results shown are valid in the sub-horizon limit, i.e. when $u\gg1$. 
Figure~\ref{fig:Integrals} shows the results of the numerical integration, namely  $\int d x \, {\cal I}_X$ for $X= T, V, B, L$ as a function of $u$. 
In order of magnitude, we have first the two scalar integrals, then the vector one and finally the tensor. This difference in amplitude is due to the fact that scalars are sourced by monopole, vectors by dipole and tensors by quadrupole anisotropies of second order combinations of $H^{(0)}_{\mu\nu}$ and $\phi$. 
Additionally, in Figure~\ref{fig:Integrals} the two scalar polarization modes, namely the breathing and the longitudinal one, have very similar amplitude. This is due to the fact that, in the sub-horizon limit, $|{\cal G}^L({\bf k}, {\bf p}, \tau)|^2 \to |{\cal G}^B({\bf k}, {\bf p}, \tau)|^2$, as it can be checked from their expression in Appendix~\eqref{app:RiemannPol}.
Even though the order of magnitude of the effect is very small at the energy scales probed by interferometers, we stress that the results of this numerical integration depend greatly on the number of assumptions made. Above all, the choice of the two power spectra: $\Delta^2_\phi$ and $\Delta^2_h$. We expect that, in models where these two present some peaked scale dependent features, the intensities of the modes predicted in this work, will be greatly amplified. The choices made here served as a proof of principle to show that, if propagations effects alter the tensor modes, then they also produce physical scalar and vector modes whose observational prospects may be more feasible than the anisotropies of the tensorial part of the SGWB.

%%%%%%%%%%%%%%%%%%%%%%%%%%%%%%%%%%%%%%%%%%%%%%%%%%%%%%%%%%%%%%%%%%%%%%%%%%%%%%%%%%%%%%%%%%%%%%%%%%%%%%%%%%%%%%%%%%%%%%%%%%%%%%%%

\section{Discussion and Conclusions}\label{sec:Conclusion}

Detecting the stochastic gravitational wave background is a primary objective for future gravitational wave observatories~\cite{Romano:2016dpx,Regimbau:2011rp,Renzini:2021iim,Hotinli:2019tpc}. Multi-band gravitational wave observations offer unique opportunities in cosmology by providing insights into the diverse behaviors of waves across different energy regimes, possibly enabling the disentanglement of intrinsic and induced properties of the SGWB by breaking degeneracies between the various effects at play.
In this work, we developed a formalism able to describe the intensity and the polarization of the SGWB, across the entire frequency spectrum, accounting for its anisotropies induced by the interaction with the matter structures present in the Universe. 
Working under the {\it classical matter approximation}, justified in Section~\ref{Sec:ExplanationCM}, we found and solved the equations of motion of the metric perturbation, $H_{\mu\nu}$, using the iterative scheme described in Section~\ref{Section:PerturbedUniverseScheme}. 
This perturbative approach yields a source term for the first-order gravitational wave, denoted as $H^{(1)}_{\mu\nu}$, specifically $\left[ {\cal O}_1  H^{(0)} \right]_{\mu\nu}$, whose explicit expression provided in Eq.~\eqref{eq:EOMH1orderO1H0}.
The source term consists of a freely propagating wave, $H^{(0)}_{\mu\nu}$, and the gravitational scalar potentials, $\phi$, which describe the matter structures in the Universe. Notably, the matter overdensities responsible for generating these potential wells are not bound to be small, allowing $\phi$ to also represent the gravitational potential of a compact lens.
Due to the presence of this source term, $H^{(1)}_{\mu\nu}$ develops scalar and vector components, in addition to tensor components, as explained in Section~\ref{Sec:ExplanationCM}. The evolution equations governing these components are given by Eqs.~\eqref{eq:EOMH1orderTimeTime},~\eqref{eq:EOMH1orderTimeSpace}  and~\eqref{eq:GaugeH1E} in the chosen gauge.
In Section~\ref{sec:1order}, we provide the complete solution for $H^{(1)}{\mu\nu}$, and a summary of it can be found in Table~\ref{tab:SourceSummary}.  In the latter equations, we can see the interference effect at play:  the sources of the scalar, vector and tensor modes of $H^{(1)}{\mu\nu}$ are given in terms of  an integral over all possible momenta, ${\bf p}$, of the free GW, convoluted with the matter gravitational potential. 
The appearance of the scalar and vector modes,  included in $H^{(1)}, H^{(1)}_{ij}$ and $E^{(1)}_{ij}$, is related to a change in the direction of propagation between the perturbed and unperturbed GW. Indeed, when their wave vectors, $k^i$ and $p^i$, are parallel, the source of such modes vanishes. We claimed that, even though the concept of a trajectory becomes unclear in the wave-optics regime, this result can be interpreted under the light of lensing: an effect that changes the GWs direction of propagation (toward the center of the gravitational potential well) and that it is known to be responsible for the production of B-modes in the CMB~\cite{Lewis:2006fu,Dai:2013nda,Hu:2000ee,Durrer:2008eom}. 
Another way of stating this result is that $H^{(1)}_{\mu\nu}$ doesn't develop extra components only when the free GW, $H^{(0)}_{\mu\nu}$, falls straight inside the gravitational potential well, toward its center. In less symmetric situation, non-trivial mixing of the polarization modes of $H^{(0)}_{\mu\nu}$ arise, and $H^{(1)}_{\mu\nu}$ gains extra components in order for the gauge conditions to be satisfied and its source (i.e. $\left[ {\cal O}_1 \, H^{(0)} \right]_{\mu\nu}$) to be conserved.
To demonstrate that all the components of the first order GW are related to physical observables, we compute the electric components of the Riemann tensor induced by the waves in Section~\ref{sec:geodesicRiemann}. As expected, these components depend on gauge-invariant combinations of $H^{(1)}{\mu\nu}$ (see to~\cite{Flanagan:2005yc}) as well as different combinations involving products of $H^{(0)}_{\mu\nu}$ and $\phi$.
We emphasize that these latter types of sources are novel, arising from the fact that the sources of the first-order wave are the propagating metric perturbation (the free gravitational wave) rather than matter sources.
Next, we perform an expansion of the Riemann tensor on a polarization basis. This decomposition enables us to categorize the tensor according to the type of motion induced on a ring of test particles. The results are presented in Eqs.~\eqref{eq:RiemannB}-~\eqref{eq:RiemannT} and encompass one transverse breathing mode, one longitudinal scalar mode, two longitudinal vector modes, and two transverse tensor modes. Since these modes represent the observable quantities, instead of computing the power spectrum of the individual constituents of $H^{(1)}{\mu\nu}$, we opted for computing the power spectra of $\delta_\alpha R^B$, $\delta_\alpha R^L$, $\delta_\alpha R^{\pm 1}$, and $\delta_\alpha R^{\pm 2}$ in Section~\ref{sec:densityMatrix}. Indeed, during observations, we measure the complete polarization pattern rather than specific contributions to it.
To be able to draw conclusions on the SGWB, we increasingly introduced assumptions; we required that the unperturbed SGWB is unpolarized, statistically homogeneous and that all the present fields are Gaussian random fields.  we also assumed that $\phi$ and $H^{(0)}_{ij}$ are uncorrelated at their respective initial times, namely Eq.~\eqref{eq:PotGWUncorr}. This assumption, which is well justified in the case of a cosmological SGWB,  might have to be revisited in the case of the astrophysical one. 
Under these hypothesis, we were able to compute the four point functions present in ${\cal P}^T_{\scriptscriptstyle{\pm 2, \pm 2}}$, ${\cal P}^V_{\scriptscriptstyle{\pm 1, \pm 1}}$, ${\cal P}^B$ and ${\cal P}^L$. Note that, because of the presence of two vector and two tensor modes, in each case we defined a $2 \times 2$ polarization tensor, so that the tensor and vector components of the SGWB are described both by their intensity and polarization content.  
The theoretical results are presented in Eqs.~\eqref{eq:Stoke1I},~\eqref{eq:Stoke1V} and~\eqref{eq:Stoke1QpmiU} for the tensor modes, Eqs.~\eqref{eq:Stoke1IV},~\eqref{eq:Stoke1VV} and~\eqref{eq:Stoke1QpmiUV} for the vector modes and ~\eqref{eq:IntensityBL} for the two scalars modes. 
In particular, Eqs.~\eqref{eq:Stoke1V} and~\eqref{eq:Stoke1IV} show that the interaction with structures does not generate a net differences between the tensor $\pm 2$ modes nor the vector $\pm 1$ ones. This result was expected, as none of the processes included in this work is parity violating.
Then Eqs.~\eqref{eq:Stoke1QpmiU} and~\eqref{eq:Stoke1QpmiUV} show that the SGWB can be linearly polarized in its tensor and vector components if the product $\Delta^2_\phi({\bf k}-{\bf p}) \times \Delta^2_h({\bf p})$ presents some specific angular dependence. 
In particular, $Q_T$ and $U_T$ are different from zero if the latter shows a hexadecapole anisotropy (i.e. $\ell = 4$), while $Q_V, U_V \neq 0$ also in the case of a quadrupole anisotropy (i.e. $\ell = 2$). 
Finally, we assume also statistically isotropic backgrounds and investigate numerically our theoretical results. 

\smallskip
The aim of this work is to provide a new tool to describe the SGWB, opening up the possibility of using new detection channels, such as its polarization content or the frequency dependent wave-optics effects.
The amount of possible ramifications and applications that can be addressed starting from our results is remarkable, and they all revolve around the two key ingredients of our analysis: the statistics of the GW's sources and the one of the gravitational potential wells. 
For instance, by changing the scalar potential power spectrum, one could use the result of this work to probe scale dependent features in the primordial power spectrum, usually linked to the formation of primordial black holes~\cite{Saito:2009jt,Bartolo:2018evs,Bavera:2021wmw,Wang:2021djr}. 
Another avenue is to use our results to reconstruct the redshift distribution of compact objects, which is the main bottleneck in the application of the dark sirens method to constrain cosmological parameters~\cite{Mastrogiovanni:2022hil}. 
To break the degeneracy among the effects of the two stochastic processes occurring, namely the generation of the GW and the propagation effects induced by matter overdensities, one could perform cross-correlations with other dark matter tracers, such as galaxy surveys or CMB experiments. 
Characterizing the order of magnitude of the effects predicted in this work, in different scenarios, will be the topic of future investigations.

%%%%%%%%%%%%%%%%%%%%%%%%%%%%%%%%%%%%%%%%%%%%%%%%%%%%%%%%%%%%%%%%%%%%%%%%%%%%%%%%%%%%%%%%%%%%%%%%%%%%%%%%%%%%%%%%%%%%%%%%%%%%%%%%
\appendix

\section{Expansion of equations of motion}\label{app:LinearizedEOM}
Here we write explicitly the equations used in the main text. 
In particular, we need the linearized form of Eq.~\eqref{eq:EomConfomalTransform}, i.e. of 
\begin{multline}
    \hat \Box \tilde H_{\mu\nu} - 2 \H  \hat n^\lambda \, \hat \nabla_{\lambda} \, \tilde H_{\mu\nu} +2 \hat R_{\mu\alpha\nu\beta} H^{\alpha\beta} -\hat g_{\mu\nu}\, H^{\alpha\beta} \,\hat R_{\alpha\beta}+ 2 \hat g_{\mu\nu} \H^2 (\hat n^\alpha \hat n^\beta \tilde H_{\alpha\beta}) + \hat g_{\mu\nu}  H \frac{\hat \Box a}{a}  + 
    \\ + 4 \H^2 H \hat n_\mu \hat n_\nu + \frac{2}{a} \left[\tilde H_\mu^\alpha \hat \nabla_\alpha \hat \nabla_\nu a  + \tilde H_\nu^\alpha \hat \nabla_\alpha \hat \nabla_\mu a \right] - 2 \H \hat n^\alpha \left[ \hat \nabla_\mu \tilde H_{\nu \alpha} + \hat \nabla_\nu \tilde H_{\mu \alpha}\right] = 0\,,
\end{multline}
 with $\hat \Lambda_{\mu\nu} = \eta_{\mu\nu} + \hat n_\mu \hat n_\nu$  the orthogonal projector to $\hat n^\mu$. To arrive to the latter expression one needs to confromally transform the equation of motion, and we used the gauge condition in Eq.~\eqref{eq:TransfConf}.
The equations above the background metric is still generic. 
We choose Eq.~\eqref{eq:BGPoisson} as background metric, setting $\phi = \psi$, and organize the equation as $\left[ {\cal O}_0 \, H \right]_{\mu\nu} + \epsilon \left[ {\cal O}_1 \, H \right]_{\mu\nu} = 0$.
The explicit expressions of the two terms are
\begin{align}
    \left[ {\cal O}_0 \, H \right]_{\mu\nu} &=  \, \Box_\eta  \Big(H_{\mu\nu} - \frac12 \eta_{\mu\nu} H \Big) - 2 \H \Big(H_{\mu\nu} - \frac12 \eta_{\mu\nu} H \Big)' +  \hat n_\mu \hat n_\nu ( 4 \H^2 - \H') H - \hat \Lambda_{\mu\nu} \H' H \nonumber \\
    & + 2 \H \Big( \hat n_\nu \partial^\alpha H_{\alpha_\mu } + \hat n_\mu \partial^\alpha H_{\alpha_\nu } - \hat n^\alpha \partial_\mu H_{\alpha \nu } - \hat n^\alpha \partial_\nu H_{\alpha \mu}\Big) + 2 \eta_{\mu\nu} \H^2 \Big( \hat n^\alpha \hat n^\beta H_{\alpha \beta}\Big)  \nonumber \\
    &  + (2 \H' - 6 \H^2) \hat n^\alpha \Big( \hat n_\mu H_{\alpha\nu} + \hat n_\nu H_{\alpha\mu}\Big) \,,\label{app:EOMlinearized0} 
\end{align}
and
\begin{align}
    \Big[ {\cal O}_1 \, H \Big]_{\mu\nu} &= 2 H_{\mu\nu} \Big[\Box_\eta \phi - 2 \H \phi' \Big]     + 2 \phi \, \Box_\eta {\tilde H}_{\mu\nu} + 4 \phi \, {\tilde H}''_{\mu\nu} +  2 \hat n_\mu \hat n_\nu  \phi \, \Box_\eta H  - 2 \eta_{\mu\nu} \phi \, \Box_\eta  (\hat n^\alpha \hat n^\beta H_{\alpha\beta} ) \nonumber \\ 
    & +4 \eta_{\mu\nu} \Big[ \H \, \hat n^\gamma  \phi -  {D^\gamma \phi }\Big] \partial_\gamma (\hat n^\alpha \hat n^\beta H_{\alpha\beta} )  + (\partial_\mu H_{\nu\alpha} + \partial_\nu H_{\mu\alpha}) \Big[ 4 \H \, \phi' \hat n^\alpha- 2 \partial^\alpha \phi\Big] + \nonumber \\
    & + 4 \partial_\alpha H_{\mu\beta} \Big[\hat n^\alpha \hat n^\beta (\partial_\nu \phi - 2 \H \hat n_\nu \phi)  +\hat n_\nu (\hat n^\beta D^\alpha \phi - \hat n^\alpha D^\beta \phi )\Big] +\nonumber \\
    & + 4 \partial_\alpha H_{\nu\beta} \Big[\hat n^\alpha \hat n^\beta (\partial_\mu \phi - 2 \H \hat n_\mu \phi)  +\hat n_\mu (\hat n^\beta D^\alpha \phi - \hat n^\alpha D^\beta \phi )\Big] +\nonumber \\
    & + 4 \partial_\alpha {\tilde H}_{\mu\nu} \Big[ \H \hat n^\alpha \phi + D^\alpha \phi \Big] - 4 \H \, \phi \, \hat n_\mu \hat n_\nu \hat n^\alpha \partial_\alpha H + \nonumber \\
    & +2\,  \partial^\alpha H_{\alpha \mu } \Big[\partial_\nu \phi + 2 \H\, \hat n_\nu \phi  \Big] + 2\,  \partial^\alpha H_{\alpha \nu } \Big[\partial_\mu \phi + 2 \H\, \hat n_\mu \phi  \Big] + H \Big[ 2 \hat \Lambda_{\mu\nu} \phi'' - 2 \eta_{\mu\nu} \Delta \phi + \nonumber \\
   & + \phi (2 \H' \eta_{\mu\nu} + 8 \H^2 \hat n_\mu \hat n_\nu) + \H \phi' (4 \eta_{\mu\nu} - 4  \hat n_\mu \hat n_\nu)  - 2 \hat n_\mu D_\nu \phi' - 2 \hat n_\nu D_\mu \phi'\Big] + \nonumber \\
    & +   (H^{\alpha\beta} \hat n_\alpha \hat n_\beta) \Big[\, \phi   \H^2(12 \eta_{\mu\nu} - 8 \hat n_\mu \hat n_\nu  ) + \phi \H' (4 \eta_{\mu\nu} + 8 \hat n_\mu \hat n_\nu  ) + \eta_{\mu\nu}  \phi''  \Big] \nonumber + \\
   &+H^{\alpha\beta}  \Big[ 4 \H \, \hat n_\alpha \, D_\beta \phi (\eta_{\mu\nu} + 2 \hat n_\mu \hat n_\nu  ) + 4  \hat n_\alpha \hat n_\beta  \Big[D_\nu D_\mu \phi - H \hat n_\mu D_\nu \phi - H \hat n_\nu D_\mu \phi  \Big] \Big] + \nonumber \\
   & + H^{\alpha\beta} \Big[ (2 \eta_{\mu\nu} + 4 \hat n_\mu \hat n_\nu  ) D_\alpha D_\beta \phi - 4 \hat n^\alpha \Big[\hat n_\mu D_\beta D_\nu \phi + \hat n_\nu D_\beta D_\mu \phi \Big] \Big] + \nonumber \\
   & + H_\mu^\alpha \Big[ \hat n_\alpha \hat n_\nu (12 \H^2 \phi  - 4 \H' \phi + 4  \H \phi' - 4 \phi'' + 2 \Delta \phi) - 2 D_\alpha D_\nu \phi \Big] + \nonumber \\
   &+ H_\mu^\alpha \Big[ 2 \H \hat n^\alpha D_\nu \phi + 4 \hat n_\alpha D_\nu \phi' - 2 \H \hat n_\nu D_\alpha \phi \Big] + \nonumber \\
   & + H_\nu^\alpha \Big[ \hat n_\alpha \hat n_\mu (12 \H^2 \phi  - 4 \H' \phi + 4  \H \phi' - 4 \phi'' + 2 \Delta \phi) - 2 D_\alpha D_\mu \phi \Big] + \nonumber \\
   &+ H_\nu^\alpha \Big[ 2 \H \hat n^\alpha D_\mu \phi + 4 \hat n_\alpha D_\nu \phi' - 2 \H \hat n_\mu D_\alpha \phi \Big]\,, \label{app:EOMlinearized1}
\end{align}
where $H = \eta^{\mu\nu} H_{\mu\nu}$, ${\tilde H}_{\mu\nu} \equiv H_{\mu\nu} - \eta_{\mu\nu} H/2$  and where we introduced the notation ${\cal  D}_\mu \equiv \hat \Lambda^\nu_\mu \,\partial_\nu$. The derivatives ${\cal  D}_\mu$ are simply a covariant way of writing spatial partial derivatives, i.e. ${\cal  D}_\mu = \delta^i_\mu \partial_i$. 
Such expressions have been computed using the Mathematica package {\it xPand}~\cite{Pitrou:2013hga}.

\section{zeroth and first order equations}\label{app:GWeq}
Now we can plug in  $ H_{\mu\nu} = H^{(0)}_{\mu\nu} + \epsilon  H^{(1)}_{\mu\nu}$ into Eq.s~\eqref{app:EOMlinearized0} and~\eqref{app:EOMlinearized1}. The two equations of motion are
\begin{equation}\label{app:EqApprox}
    \left[ {\cal O}_0 \, H^{(0)} \right]_{\mu\nu} = 0 \,, \qquad \mbox{and} \qquad \left[ {\cal O}_0 \, H^{(1)} \right]_{\mu\nu} = - \left[ {\cal O}_1 \, H^{(0)} \right]_{\mu\nu} \,,
\end{equation}
while the compatibility and gauge conditions are given by
\begin{align}
\hat n^\mu \hat n^\nu   H^{(0)}_{\mu\nu} &= 0 \,, \qquad  \partial^\mu \tilde H^{(0)}_{\mu\nu} - 4 \H \, \hat n^\mu H^{(0)}_{\mu\nu}  + H^{(0)} \,\H\, \hat n_\nu = 0\,,  \label{app:GaugeH0} \\ 
\hat n^\mu \hat n^\nu  H^{(1)}_{\mu\nu} + 2 \hat n^\nu \delta {\hat u}^\mu  H^{(0)}_{\mu\nu} &= 0 \,, \qquad       \partial^\mu  \tilde H^{(1)}_{\mu\nu} - 4 \H \, \hat n^\mu H^{(1)}_{\mu\nu} + H^{(1)} \, \H \,  \hat n_\nu  = 0\,, \label{app:GaugeH1}
\end{align}
where the trace is $H^{(i)} = \eta^{\mu\nu} H^{(i)}_{\mu\nu}$ and $\tilde H^{(i)}_{\mu\nu} = H^{(i)}_{\mu\nu} - \frac12 \eta_{\mu\nu} H^{(i)} $. Combining these equations, we obtain for the left-hand-sides of equations in Eq.~\eqref{app:EqApprox}
\begin{align}
 \bigg[ {\cal O}_0 \, H^{(0)} \bigg]_{\mu\nu}& = \,\Box_\eta  \tilde H^{(0)}_{\mu\nu} - 2 \H (\tilde H^{(0)}_{\mu\nu})' + \H \Big(  \hat n_\nu  \partial_\mu H^{(0)}+  \hat n_\mu \partial_\nu H^{(0)} \Big) - (\hat n_\mu \hat n_\nu  + \hat \Lambda_{\mu\nu}  )  \H' H^{(0)}  - \nonumber \\
 &- 2 \H \Big(  \hat n^\alpha \partial_\mu H^{(0)}_{\alpha \nu } + \hat n^\alpha \partial_\nu H^{(0)}_{\alpha \mu}\Big)  + 2 ( \H' + \H^2) \hat n^\alpha \Big( \hat n_\mu H^{(0)}_{\alpha\nu} + \hat n_\nu H^{(0)}_{\alpha\mu}\Big)\label{app:O0H0}
 \end{align}
 \begin{align}
  \bigg[ {\cal O}_0 \, H^{(1)} \bigg]_{\mu\nu}& = \,\Box_\eta  \tilde H^{(1)}_{\mu\nu} - 2 \H (\tilde H^{(1)}_{\mu\nu})' + \H \Big(  \hat n_\nu  \partial_\mu H^{(1)}+  \hat n_\mu \partial_\nu H^{(1)} \Big) - (\hat n_\mu \hat n_\nu  + \hat \Lambda_{\mu\nu}  )  \H' H^{(1)}  - \nonumber \\
 &- 2 \H \Big(  \hat n^\alpha \partial_\mu H^{(1)}_{\alpha \nu } + \hat n^\alpha \partial_\nu H^{(1)}_{\alpha \mu}\Big)  + 2 ( \H' + \H^2) \hat n^\alpha \Big( \hat n_\mu H^{(1)}_{\alpha\nu} + \hat n_\nu H^{(1)}_{\alpha\mu}\Big) + \nonumber \\
 &+ 2 \eta_{\mu\nu} \H^2 (\hat n^\alpha \hat n^\beta H^{(1)}_{\alpha\beta})\label{app:OoH1}.
\end{align}
The expression of $[ {\cal O}_1 \, H^{(0)}]_{\mu\nu}$ depends on the solution of $H^{(0)}_{\mu\nu}$, hence we first solve its equations. Taking  the time-time and  space-time component of Eq~\eqref{app:O0H0} we  obtain
\begin{align}
    \frac12  \Box_\eta  H^{(0)} - 3 \H (H^{(0)})' -  \H' H^{(0)} &= 0   \qquad \qquad\mbox{Time-Time}\\
     \Box_\eta H^{(0)}_{0i} -4 \H ( H^{(0)}_{0i})' +  H^{(0)}_{0i} (6 \H^2 - 2 \H') - \H \, \partial_i H^{(0)} &= 0  \qquad \qquad \mbox{Space-Time} 
\end{align}
of which we take the solution $H^{(0)} = H^{(0)}_{0i} =0 $ as explained in the main text. This way Eq.~\eqref{app:O0H0} gives
\begin{equation}
 \Box_\eta H^{(0)}_{ij} - 2 \H (H^{(0)}_{ij})' = 0\,.
\end{equation}
%We can look at the different components of such equation along and perpendicular to $\hat n^\mu$
%\begin{align}
%    \frac12  \Box_\eta  H^{(0)} + \H (H^{(0)})' -  \H' H^{(0)} &= 0   \qquad \qquad\mbox{Time-Time}\\
%     - \H \, \partial_i H^{(0)} &= 0  \qquad \qquad \mbox{Space-Time} \\
% -\frac12 \Box_\eta H^{(0)} + \H (H^{(0)})' - 3 \H'  H^{(0)} + \frac32 \hat Z (H^{(0)})' &= 0  \qquad \qquad \mbox{Space Space trace} \\
% \Box_\eta H^{(0)}_{ij} - 2 \H (H^{(0)}_{ij})' - \frac13 \delta_{ij} \left[ \Box_\eta H^{(0)} - 2 \H (H^{(0)})' \right] &= 0  \qquad \qquad\mbox{Space-Space traceless}\end{align}
%From the second equation in the system above we see that $H^{(0)}$ cannot depend on time. Additionally, none of the wave equation above for  $H^{(0)}$ has source and there are no scalar sources in General Relativity. 
Given this solution for the zeroth order gravitational  wave we compute the source of the first order one, namely
\begin{align}
    \left[ {\cal O}_1 \, H^{(0)} \right]_{\mu\nu} =& \, 4 \phi\, \Delta H^{(0)}_{\mu\nu} + 2 H^{(0)}_{\mu\nu}\, \Big[ \Box_\eta \phi - 2 \H\, \phi'\Big] + 4 \,\partial^k \phi \, \partial_k\, H^{(0)}_{\mu\nu} + 2 (\eta_{\mu\nu} + 2 \hat n_\mu \hat n_\nu) \, H^{(0)}_{\alpha \beta }\, D^\alpha D^\beta\, \phi \nonumber \\
    &- 2 \hat n_\mu \, D^\alpha \phi \Big( {H^{(0)}}'_{\nu \alpha} + \H   H^{(0)}_{\nu \alpha} \Big)- 2 \hat n_\nu \, D^\alpha \phi \Big( {H^{(0)}}'_{\mu \alpha} + \H   H^{(0)}_{\mu \alpha} \Big) \nonumber \\
    &  - 2 D_\mu \Big( H^{(0)}_{\alpha \nu} \, D^\alpha \phi \Big) - 2 D_\nu \Big( H^{(0)}_{\alpha \mu} \, D^\alpha \phi \Big) \label{app:O1H0}
\end{align}
%
%\begin{equation}
%    \text d s^2  = a^2 (\eta) \Big\{ \left[ - \text d \eta^2 + \text d {\bf x}^2 \right] - 2 {\color{red}{\epsilon}} \phi \left[  \text d \eta^2 + \text d {\bf x}^2\right] +  {\color{blue}{\alpha}} \, H_{\mu\nu} \text d x^\mu \text d x^\nu \Big \}
%\end{equation}
Note also that, when $H^{(0)}_{0i} = 0$ then, from Eq.~\eqref{app:GaugeH1}, we conclude that  $H^{(1)}_{00} = 0$.
Using now Eqs.~\eqref{app:OoH1} and ~\eqref{app:O1H0} we  can compute  the time-time and space-time component of the equation of motion of the first order gravitational wave
\begin{align}
    \frac12 \Box_\eta H^{(1)} - 3 \H (H^{(1)})' -  \H' H^{(1)} &= - 2 H^{(0)}_{ij} \partial^i \partial^j \, \phi\,,  \qquad \mbox{Time-Time} \\
 %    \Box_\eta H^{(1)} - 2 \H (H^{(1)})' - (4 \H^2 + 2 \H') H^{(1)}  &= 0 \,,  \qquad \qquad \qquad \qquad  \quad\:\:\: \mbox{Trace}  \label{eq:EOMH1orderTrace} \\
    \Box_\eta  H^{(1)}_{0i} - 4 \H (H^{(1)}_{0i})' - \H \partial_i H^{(1)} - 2 (\H^2 + \H') H^{(1)}_{0i} &= - \frac{2}{a} \partial^k \phi \, (a H^{(0)}_{ki})'\,,  \quad \mbox{Space-Time}  
\end{align}
which are the two constraint equations.
%While for $H^{(1)}_{\mu\nu}$
%\begin{align} \label{app:EOMlinearized1Approx}
%    \left[ {\cal O}_0 \, H^{(1)} \right]_{\mu\nu} =&  \, \Box_\eta  \tilde H^{(1)}_{\mu\nu} - 2 \H  (\tilde  H^{(1)}_{\mu\nu})'   - 2 \H \hat n^\alpha \, \Big(\partial_\mu \tilde H^{(1)}_{\alpha\nu} + \partial_\nu \tilde H^{(1)}_{\alpha\mu} \Big) -  \H' H^{(1)} \Big( \hat \Lambda_{\mu\nu}+ \hat n_\mu \hat n_\nu\Big)  \nonumber \\
%    &  + 2 (\H' + \H^2) \, \hat n^\alpha \,  \Big( \hat n_\mu H^{(1)}_{\alpha \nu } + \hat n_\nu H^{(1)}_{\alpha \mu }\Big)  + \frac12 \hat \Lambda_{\mu\nu} \hat Z {H^{(1)}}'  \nonumber \\
 %   \left[ {\cal O}_1 \, H^{(0)} \right]_{\mu\nu} =& \, 4 \phi\, \Delta H^{(0)}_{\mu\nu} + 2 H^{(0)}_{\mu\nu}\, \Big[ \Box_\eta \phi - 2 \H\, \phi'\Big] + 4 \,\partial^k \phi \, \partial_k\, H^{(0)}_{\mu\nu} + 2 (\eta_{\mu\nu} + 2 \hat n_\mu \hat n_\nu) \, H^{(0)}_{\alpha \beta }\, D^\alpha D^\beta\, \phi \nonumber \\
 %   &- 2 \hat n_\mu \, D^\alpha \phi \Big( {H^{(0)}}'_{\nu \alpha} + \H   H^{(0)}_{\nu \alpha} \Big)- 2 \hat n_\nu \, D^\alpha \phi \Big( {H^{(0)}}'_{\mu \alpha} + \H   H^{(0)}_{\mu \alpha} \Big) \nonumber \\
 %   &  - 2 D_\mu \Big( H^{(0)}_{\alpha \nu} \, D^\alpha \phi \Big) - 2 D_\nu \Big( H^{(0)}_{\alpha \mu} \, D^\alpha \phi \Big) 
%\end{align}
%where we have also used Eq.~\eqref{app:EOMlinearized0Approx} in order to further  simplify the expression and $\tilde H^{(1)}_{\mu\nu} \equiv H^{(1)}_{\mu\nu} - \frac12 \eta_{\mu\nu} H^{(1)}  $.

\section{Friedmann equations}\label{app:BGeqFriedmann}
We work in matter and $\Lambda$ domination, hence we will include in Einstein equations these two contributions.
Defining $\Omega_m  \equiv {\bar \rho}_m (\tau) / \rho^0_{crit} $ and $\Omega_\Lambda  \equiv {\bar \rho}_\Lambda (\tau) / \rho^0_{crit} $ where $\rho_{crit} = 3 H^2_0 / 8 \pi G = 6 H^2_0$ (in our units $G = 1/ 16 \pi$), Friedmann's equations, and the conservation laws, read
\begin{align}
    \H^2 &= a^2 H^2_0 \, ( \Omega_m  + \Omega_\Lambda  )\,,  \\
    (\H^2 + 2 \H') & =   3 a^2 H^2_0 \, \Omega_\Lambda \,, \\ %- \frac{a^2}{2} P_0  =
    \Omega'_m &= - 3 \H \, \Omega_m \,, \\
    \Omega_\Lambda &= \Omega^0_\Lambda \,.
\end{align}
In the equations above we have used the equations of states $w = 0$ and $w = -1$ for the matter (cold dark matter and baryons) and $\Lambda$ respectively. Additionally,  $\Omega^0_\Lambda$ is the value of the density parameter today and the solution of the third equation in the system above is the standard $\Omega_m = \Omega^0_m a^{-3}$. Since we are neglecting radiation, cold dark matter and baryons behave in the same way, hence $\Omega^0_m$ is the value of the total matter density parameter today. Because all the matter components are described by fluids, we find the evolution equations of the matter content via the conservation of the stress-energy tensor, $\nabla_\mu T^\mu_\nu = 0$, instead of the Boltzmann equations. Both of the procedures leads to the same results.
We rewrite the first Friedmann's equation in terms of the scale factor as
\begin{equation}\label{eq:F1noRad}
    a' - H_0 \, \sqrt{\Omega^0_{\rm m} a +  \Omega^0_\Lambda a^4} = 0\,.
\end{equation}
It is convenient to make a change of variable and use the comoving distance $\chi = \tau_0 - \tau = \int^1_a da'/(a' \H(a'))$, where $\tau_0$ is the value of conformal time today. From this definition, it is  clear that $d \chi = - d \tau$.  The scale factor and the comoving distance belong respectively to the ranges $[0,1]$ and $[0, \chi_{i}]$, where $\chi_{i}$ is the value of $\chi$ corresponding to $a = 0$. Such initial value  does not have any particular physical meaning, and it is merely a consequence of having neglected the radiation contribution. 
Although a solution of  Eq.~\eqref{eq:F1noRad} interpolating between matter and dark energy domination eras exist \cite{Sazhin:2011ze}, it is quite intricate and it won't allow us to solve the transfer function differential equation. Therefore, we solve Eq.~\eqref{eq:F1noRad} separately in the two epochs and match the solutions at their equivalence, $a_\Lambda (\chi_{{\rm M} \Lambda} ) = a_{\rm M} (\chi_{{\rm M} \Lambda} ) = a_{{\rm M} \Lambda} $ and $a_{{\rm M} \Lambda} = \sqrt[3]{\Omega^0_{\rm m} /\Omega^0_\Lambda}$. 
We find
\begin{equation}\label{eq:HubbleComoving}
a(\chi) = \left\{ 
\begin{aligned}
\: &\frac{\big[\chi - \chi_{i} \big]^2 }{4 \chi^2_{\rm M}} \quad& \chi \in [\chi_{{\rm M} \Lambda}, \chi_i ]  \\
\:& \frac{\chi_\Lambda}{\chi_\Lambda + \chi} \quad& \chi  \in [0,\chi_{{\rm M} \Lambda}] 
\end{aligned}\right.\,,
\qquad 
\H(\chi) = - \frac{1 }{a} \, \frac{\partial a}{\partial \chi} = \left\{ 
\begin{aligned}
\:& \frac{2}{\chi_i - \chi } \qquad \chi \in [\chi_{{\rm M} \Lambda}, \chi_i ] \\
\:&\frac{1}{ \chi + \chi_\Lambda}  \qquad \chi  \in [0,\chi_{{\rm M} \Lambda}] 
\end{aligned}\right.
\end{equation}
where all the quantities defined in the equation above can be found in the Table~\ref{tab:SpecialValues}. 
\begin{table}
    \centering
     \renewcommand{\arraystretch}{1.6}
    \begin{tabular}{|c|c|c|c|c|c|c|}
        \hline
        & $\chi_{\rm M}$ &  $\chi_\Lambda$ & $\chi_* $& $\chi_{{\rm M}\Lambda}$ & $\chi_i $  \\
        \hline
       \hline
       Def & $[H_0 \sqrt{\Omega^0_m}]^{-1}$ & $[H_0 \sqrt{\Omega^0_\Lambda}]^{-1}$ & $[H_0 (\Omega^0_m)^{1/3} (\Omega^0_\Lambda)^{1/6}]^{-1} $& $\chi_* - \chi_\Lambda $ & $3 \chi_* - \chi_\Lambda$  \\
       Value & $0.027$ & $0.017$ & $0.023$& $0.0054$ & $0.052$\\
        \hline
    \end{tabular}
    \caption{List of special comoving distance values enetering in Eq.~\eqref{eq:HubbleComoving}. The numerical values are computed considering $H_0= 67.3$, $\Omega^0_m =0.31 $, $\Omega^0_\Lambda = 0.69 $. }
    \label{tab:SpecialValues}
\end{table}

\section{Perturbed Einstein equations and transfer functions}\label{app:BGeqPerturbation}
Considering the perturbed line element 
\begin{equation}
    d s^2 = a^2(\tau) [- (1 + 2 \epsilon \phi) d \tau^2 + (1 - 2 \epsilon \phi) d {\bf x}^2]
\end{equation}
and the matter perturbation overdensity, $\delta = ( \rho_m - {\bar \rho}_m) / {\bar \rho}_m$, the linearized Einstein's equations in Fourier space take the form
\begin{eqnarray}
     &\phi_k'' + 3 \H \phi_k' +  (\H^2 + 2 \H') \phi_k  = 0\,, \qquad &\delta_k' - k^2 v_k =  3  \, \phi_k' \, \qquad
    v_k' +  \H  v_k = - \phi_k \\
     &\phi_k' + \H \phi_k = \displaystyle{- \frac{3 H^2_0 \, \Omega^0_m }{2 a}\, v_k }\,,\qquad  &\phi_k  = -\frac{3  H^2_0 \,  \Omega^0_m }{2 a k^2 }\left[ \delta_k - 3 \H  v_k \right]  \,.
\end{eqnarray}
Note that we are already considering the velocity divergence, $v^i_k =  i k^i \, v_k$. 
We introduce the following relations
\begin{align}
    v(a, k) &= - \frac{9}{10} T_m (k) \, {\cal G}_v (a, k) \phi^{in}_k \\
    \delta^C_m (a, k) &= - \frac{9}{10} T_m (k) {\cal G}_m (a, k) \phi^{in}_k \\ 
    \phi(a, k) & = \frac{9}{10} T_m (k) \frac{{\cal G}_\phi (a, k) }{a} \phi^{in}_k
\end{align}
where $\delta^C_m (a, k)   \equiv   \delta_k - 3 \H  v_k $ is the gauge-invariant comoving density contrast and $\phi^{in}_k$ is the primordial value of the gravitational potential.
In the equations above $T_m (k)$ is the matter transfer function which is computed numerically by Einstein-Boltzmann solvers codes such as \texttt{ CAMB}, \texttt {CLASS}. Alternatively, one can use suitable fitting formulas \cite{Eisenstein:1997ik,Eisenstein:1997jh,dodelson2020modern,Hamilton:2000tk}. 
For a $\Lambda$ CDM model we have 
\begin{equation}
    {\cal G}_\phi  = D_m\,, \qquad {\cal G}_m  = \frac{2 k^2}{ 3 H^2_0 \Omega^0_m} \, D_m\,, \qquad {\cal G}_v = f(a) \, \frac{\H}{k^2}\, {\cal G}_m
\end{equation}
%\begin{align}
%    {\cal G}_\phi  &= D_m \\
%    {\cal G}_m & = \frac{2 k^2}{ 3 H^2_0 \Omega^0_m} \, D_m \\
%    {\cal G}_v &= f(a) \, \frac{\H}{k}\, {\cal G}_m
%\end{align}
where $D_m (a)$ is the growth factor of the matter fluctuations and $f(a)$ the growth rate defined as $f = {\text d} \ln \delta^C_m / {\text d} \ln a = {\text d} \ln D_m / {\text d} \ln a$.
Therefore, the gravitational potential transfer function, defined as  $\phi(\tau, k)  = \T^\phi_k(\tau) \phi^{in}_k$ and used in the main text, is 
\begin{align}
    \T^\phi_k (a) &= \frac{9}{10} T_m (k) \frac{{\cal G}_\phi (a, k) }{a}\,. \label{eq:PhiTransf}
 %   \T^v_k (a) &= - \frac{9}{10} T_m (k) \, {\cal G}_v (a, k)\,. \label{eq:vTransf}
\end{align}
The growth factor for modes inside the horizon can be approximated in the late time universe as \cite{dodelson2020modern,Linder:2003dr,Carroll:1991mt} 
\begin{equation}
    D_m(a) = \frac{5 H^2_0 \Omega^0_m}{2} \frac{\H(a)}{a} \int^a_0 \, \frac{d a}{ [ \H(a)]^3}\,. 
\end{equation}
Note that in our notations we use $\H = (a'/a) (\tau)$ as the Hubble parameter in conformal time, while in \cite{dodelson2020modern} they use $H = (\dot a / a ) (\tau)$. A more handy fitting formula is \cite{Linder:2003dr,Carroll:1991mt}
\begin{equation}\label{eq:FittingMatterGrowth}
    D_m(a) =\frac{5 \Omega^0_m}{2 a^2 
 E^2(a)} \left[ \left( \frac{\Omega^0_m}{a^3 E^2(a)}\right)^{4/7} - \frac{\Omega^0_\Lambda}{E^2(a)} +  \left(1 + \frac{\Omega^0_m}{2 a^3 E^2(a)} \right)\left(1 + \frac{\Omega^0_\Lambda}{70 E^2(a)} \right)\right]\,,
\end{equation}
with $E^2(a) = \H^2 / (a^2 H^2_0) = (\Omega^0_m / a^3 + \Omega^0_\Lambda)$.
%\begin{figure}
%    \centering
%    \includegraphics[width=0.6\textwidth]{Plots/Growth-Factor.png}
%    \caption{ DM/a is gravitational potential growth factor. It is constant in matter domination and decays in Lambda domination more or less. This is obtained through the fitting formula Eq.~\eqref{eq:FittingMatterGrowth} and choosing $\Omega^0_m = 0.27$, $\Omega^0_\Lambda = 0.7$.}
%    \label{fig:GrowthFactor}
%\end{figure}

\section{Components of Riemann tensor on polarization basis} \label{app:RiemannPol}
In this section we give the explicit expressions of ${\cal G}^{B},{\cal G}^{L},{\cal G}^{V}$ and ${\cal G}^{T}$.
In order to do so it is convenient to define first the function
{\small 
\begin{eqnarray}
    {\rm A}_{1} ({\bf p}, {\bf k} - {\bf p}, \tau) &\equiv& \: 2 {\T^\phi }''_{|{\bf k} - {\bf p}|}  \T^H_{p}  + { \T^\phi}'_{|{\bf k} - {\bf p}|} \left[ {\T^H_{p}}' + 2 \H {\T^H_{p}} \right]  + \T^\phi_{|{\bf k} - {\bf p}|}  \left[ 2 {\T^H_{p}}'' + 2 \H {\T^H_{p}}' + {\T^H_{p}} \, {\bf p}\cdot ({\bf k} - {\bf p})\right] \,, \nonumber\\\\
    {\rm A}_{2} ({\bf p}, {\bf k} - {\bf p}, \tau) &\equiv& \T^\phi_{|{\bf k} - {\bf p}|} {\T^H_{p}}\,,
\end{eqnarray}
}
where $\T^\phi$ and $\T^h$ are the gravitational potential and GW transfer functions, which depend on time.
Indeed, the expressions of the interested functions are
\begin{align}
    {\cal G}^{B} ({\bf k}, {\bf p}, \tau) &=   \sqrt{\frac{1}{3}}\left\{ \int^\tau_
    {{\tau_s}} d \tilde \tau \: {\rm T}_B ({\bf k},{\bf p},\tau, \tilde \tau)  +{\rm A}_{1} ({\bf p}, {\bf k} - {\bf p}, \tau)\right\} \label{eq:DefGB}\\\nonumber\\
    {\cal G}^{L} ({\bf k}, {\bf p}, \tau) &=   \sqrt{\frac{2}{3}}\left\{\int^\tau_
    {{\tau_s}} d \tilde \tau \: {\rm T}_L ({\bf k},{\bf p},\tau, \tilde \tau)  - \frac12 {\rm A}_{1} ({\bf p}, {\bf k} - {\bf p}, \tau) - k^2{\rm A}_{2} ({\bf p}, {\bf k} - {\bf p}, \tau) \right\} \label{eq:DefGL}\\\nonumber\\
    {\cal G}^{V} ({\bf k}, {\bf p}, \tau) &=  \int^\tau_
    {{\tau_s}} d \tilde \tau \: {\rm T}_V ({\bf k},{\bf p},\tau, \tilde \tau)  - \frac12 {\rm A}_{1} ({\bf p}, {\bf k} - {\bf p}, \tau) - \frac{k^2}{2}{\rm A}_{2} ({\bf p}, {\bf k} - {\bf p}, \tau) \label{eq:DefGV}\\\nonumber\\
    {\cal G}^{T} ({\bf k}, {\bf p}, \tau) &=   \frac14 \int^\tau_
    {{\tau_s}} d \tilde \tau \: {\rm T}_\gamma({\bf k},{\bf p},\tau, \tilde \tau)  - \frac12 {\rm A}_{1} ({\bf p}, {\bf k} - {\bf p}, \tau) \label{eq:DefGT}
\end{align}
with
{\small 
\begin{align}   
    &{\rm T}_B({\bf k},{\bf p},\tau, \tilde \tau) = k^2  \left[ \T^H_{p} (\tilde \tau)  \T^\phi_{|{\bf k} - {\bf p}|}(\tilde \tau)\right] \times 
 \left[ \frac{d^2 g_k^H}{d \tau^2} + \H_\tau \frac{d g_k^H}{d \tau}\right] + \nonumber\\
    &\qquad\qquad\qquad+ \left[\T^\phi_{|{\bf k} - {\bf p}|}(\tilde \tau) \left[ \frac{ d \T^H_{p} (\tilde \tau) }{d \tilde \tau}  +  \H_{\tilde \tau} \T^H_{p} (\tilde \tau)\right] - 2k^2 \H_{\tilde \tau} \int^{\tilde \tau}_{{\tau_s}} d \hat \tau g^H_{k}(\tilde \tau, \hat \tau) \T^H_{p} (\hat \tau)  \T^\phi_{|{\bf k} - {\bf p}|}(\hat \tau) \right] \times\nonumber \\
    &\qquad\qquad\qquad\times \Bigg[\frac{d^3 g_k^{H_{0i}}}{d \tau^3} + 5 \H_{\tau}\frac{d^2 g_k^{H_{0i}}}{d \tau^2} +\Big[ 4 (\H_{\tau}^2 + 2 \H_{\tau}')  \Big]\frac{d g_k^{H_{0i}}}{d \tau} +\left[ 4 (\H_{\tau}  \H_{\tau}' +  \H_{\tau}'')  \right] g_k^{H_{0i}} \Bigg]  \nonumber \\ \\\nonumber \\
    &{\rm T}_L({\bf k},{\bf p},\tau, \tilde \tau) = k^2  \left[ \T^H_{p} (\tilde \tau)  \T^\phi_{|{\bf k} - {\bf p}|}(\tilde \tau)\right] \times\left[ \frac{d^2 g_k^H}{d \tau^2} + \H_\tau \frac{d g_k^H}{d \tau}\right] - \nonumber\\
    &\qquad\qquad\qquad - \left[\T^\phi_{|{\bf k} - {\bf p}|}(\tilde \tau) \left[ \frac{ d \T^H_{p} (\tilde \tau) }{d \tilde \tau}  +  \H_{\tilde \tau} \T^H_{p} (\tilde \tau)\right] - 2k^2 \H_{\tilde \tau} \int^{\tilde \tau}_{{\tau_s}} d \hat \tau g^H_{k}(\tilde \tau, \hat \tau) \T^H_{p} (\hat \tau)  \T^\phi_{|{\bf k} - {\bf p}|}(\hat \tau) \right]\times \nonumber\\
     &\qquad\qquad\qquad\times \Bigg[\frac{d^3 g_k^{H_{0i}}}{d \tau^3} + 5 \H_{\tau}\frac{d^2 g_k^{H_{0i}}}{d \tau^2} +\Big[ 4 (\H_{\tau}^2 + 2 \H_{\tau}')  + 2 k^2 \Big]\frac{d g_k^{H_{0i}}}{d \tau} +\left[ 4 (\H_{\tau}  \H_{\tau}' +  \H_{\tau}'') + 2 k^2   \H_\tau \right] g_k^{H_{0i}} \Bigg]   \nonumber \\ \\ \nonumber\\
     &{\rm T}_V({\bf k},{\bf p},\tau, \tilde \tau) =  \T^\phi_{|{\bf k} - {\bf p}|}(\tilde \tau) \left[ \frac{ d \T^H_{p} (\tilde \tau) }{d \tilde \tau}  +  \H_{\tilde \tau} \T^H_{p} (\tilde \tau)\right] \times \Bigg[\frac{d^3 g_k^{H_{0i}}}{d \tau^3} + 5 \H_{\tau}\frac{d^2 g_k^{H_{0i}}}{d \tau^2} + \nonumber\\
     &\qquad\qquad\qquad +\Big[ 4 (\H_{\tau}^2 + 2 \H_{\tau}') - k^2 \Big]\frac{d g_k^{H_{0i}}}{d \tau} +\left[ 4 (\H_{\tau}  \H_{\tau}' +  \H_{\tau}'') - \H_{\tau} k^2 \right] g_k^{H_{0i}}\Bigg]  \\ \nonumber\\\nonumber \\
     &{\rm T}_\gamma({\bf k},{\bf p},\tau, \tilde \tau) = \T^H_{p} (\tilde \tau)\left[ {\T^\phi }''_{|{\bf k} - {\bf p}|}(\tilde \tau) + 2 \H_{\tilde \tau} {\T^\phi }'_{|{\bf k} - {\bf p}|}(\tilde \tau) + \T^\phi_{|{\bf k} - {\bf p}|}(\tilde \tau) (p^2 + k^2)\right] \times  \left[ \frac{d^2 g_k^\gamma}{d \tau^2} + \H_\tau \frac{d g_k^\gamma}{d \tau}\right] \nonumber \\
\end{align}}
where $\H_{ \tau} =\H( \tau)$ and  $\H_{\tilde \tau} = \H(\tilde \tau)$ and all the Green's functions ($g^H_k, g^{H_{0i}}_k, g^\gamma_k$) are evaluated in $(\tau, \tilde \tau)$ unless explicitly stated.
These expressions simplify when considering a matter dominated Universe. Indeed, in such situation, $\H_{\tau}^2 + 2 \H_{\tau}' = \H_{\tau}  \H_{\tau}' +  \H_{\tau}'' = 0$ and also the scalar gravitational potential become constant (we are neglecting radiation). In this case, we write ${\T^\phi }_{|{\bf k} - {\bf p}|} \approx \T^\phi $ constant.

\section{Spin weighted spherical harmonics and Clebsh-Gordan coefficients}\label{app:SphericalHarmonics}
The content of this Section follows~\cite{Durrer:2008eom,Hu:1997hp}.
The spin-0 spherical harmonics are given by
\begin{align}
    Y^{\ell}_m (\theta, \varphi) = (-1)^m \sqrt{\frac{(2 \ell + 1)}{4 \pi} \frac{(\ell - m)!}{(\ell + m)!} } \, e^{i m \varphi} P_{\ell m } (\cos \theta)
\end{align}
where $P_{\ell m } (x)$ is the associated Legendre polynomial, while the spin weighted spherical harmonics are
\begin{align}
    \prescript{}{s}{Y}^{\ell}_m (\theta, \varphi)  &= (-1)^m \sqrt{\frac{(2 \ell + 1)}{4 \pi} \frac{(\ell + m)!(\ell - m)!}{(\ell + s)!(\ell - s)!}  } \left[ \sin  \left( \frac{\theta}{2} \right)\right]^{2 \ell}  e^{i m \varphi}  \times \nonumber \\
    & \times \sum_r \left(
    \begin{matrix}
        \ell - s \\
        r
    \end{matrix}\right)\, 
    \left(
    \begin{matrix}
        \ell + s \\
        r + s - m
    \end{matrix}\right)
    \, (-1)^{\ell - r - s } \,  \left[ \cot  \left( \frac{\theta}{2} \right)\right]^{2 r + s - m} \,. 
\end{align}
The sum over $r$ runs over those values for which the binomial coefficients are non-vanishing, namely $\mbox{max}\{ 0, m-s\} \leq \,r \,\leq \mbox{min} \{ \ell- s, \ell + m \}$. 
The spin weighted spherical harmonics are defined for $|s| \leq \ell$ and $|m| \leq \ell$ and are such that 
\begin{equation}
    \left(  \prescript{}{s}{Y}^{\ell}_m (\theta, \varphi) \right)^* = (-1)^{s+m}  \prescript{}{-s}{Y}^{\ell}_{- m} (\theta, \varphi)\,,
\end{equation}
and under parity
\begin{equation}
     \prescript{}{s}{Y}^{\ell}_m (\pi - \theta, \pi + \varphi) = (-1)^\ell \,  \prescript{}{-s}{Y}^{\ell}_m (\theta, \varphi) \,.
\end{equation}
Also the following relation holds
\begin{align}
    \prescript{}{s_1}{Y}^{\ell_1}_{m_1} (\theta, \varphi) \prescript{}{s_2}{Y}^{\ell_2}_{m_2} (\theta, \varphi) =& \frac{\sqrt{(2 \ell_1 + 1) (2 \ell_2+2)}}{4 \pi}  \sum^{j = \ell_1 + \ell_2}_{j = |\ell_1 - \ell_2| }  \sqrt{\frac{4 \pi}{(2 j + 1)}} \, \prescript{}{s}{Y}^{j}_{m} (\theta, \varphi)  \nonumber \\
    & \times \, \langle \ell_1, \ell_2 ; m_1, m_2 | \, j, m \rangle \, \langle \ell_1, \ell_2 ; -s_1, -s_2 | \, j, -s  \rangle 
\end{align}
where $s = s_1 + s_2$, $m = m_1 + m_2$ and the Clebsh-Gordan coefficients are given by
\begin{align}
     \langle \ell_1, \ell_2 ; m_1, m_2 &| \, j, m_1 + m_2  \rangle = \nonumber \\
      =& \sqrt{\frac{(\ell_1 + \ell_2 - j)! (j + \ell_1 - \ell_2)! (j + \ell_2 - \ell_1)! (2 j + 1)}{(j + \ell_1 + \ell_2+1)!}}\, \times  \nonumber \\
     & \times \sum_k \Bigg[ \frac{(-1)^k \sqrt{(\ell_1 + m_1)! (\ell_1 - m_1)! (\ell_2 + m_2)! (\ell_2 - m_2)! }}{k! (\ell_1 + \ell_2 - j - k)! (\ell_1 - m_1 -k)! } \nonumber\\
     & \times  \frac{\sqrt{(j + m_1 + m_2 )! (j - m_1 -m_2)!}}{(\ell_2 + m_2 -k)!(j - \ell_2 + m_1 + k )!(j - \ell_1 - m_2 + k )!}\Bigg]
\end{align}
where the sum over $k$ goes from $ \mbox{max} \{ 0, \ell_2 - j - m_1, \ell_1 - j + m_2 \} \, \leq k \leq \mbox{min} \{ \ell_1 + \ell_2 - j , \ell_1 - m_1, \ell_2 + m_2\}$. 
Some spin-2 spherical harmonics are
\begin{align}
    \prescript{}{\pm 2}{Y}_{0}^{2} (\theta, \varphi) &= \frac14 \sqrt{\frac{15}{2\pi}} \sin^2 \theta \,, \label{eq:SpinHarm0}\\
    \prescript{}{\pm 2}{Y}_{\pm 1}^{2} (\theta, \varphi) &= \mp \, \frac12 \sqrt{\frac{5}{\pi}} \sin \theta \, \sin^2 \left(\frac{\theta}{2} \right)\, e^{ \pm i \varphi} \label{eq:SpinHarm1Sin}\\
    \prescript{}{\mp 2}{Y}_{\pm 1}^{2} (\theta, \varphi) &= \pm \, \frac12 \sqrt{\frac{5}{\pi}} \sin \theta \, \cos^2 \left(\frac{\theta}{2} \right)\, e^{ \pm i \varphi} \label{eq:SpinHarm1Cos}\\
    \prescript{}{\pm 2}{Y}_{\pm 2}^{2} (\theta, \varphi) &=  \, \frac12 \sqrt{\frac{5}{\pi}} \, \sin^4 \left(\frac{\theta}{2} \right)\, e^{ \pm 2 i \varphi} \label{eq:SpinHarm2Sin}\\
    \prescript{}{\mp 2}{Y}_{\pm 2}^{2} (\theta, \varphi) &= \, \frac12 \sqrt{\frac{5}{\pi}} \, \cos^4 \left(\frac{\theta}{2} \right)\, e^{ \pm 2 i \varphi} \label{eq:SpinHarm2Cos}\,.
\end{align}

\acknowledgments
AG thanks Alessandra Silvestri, Subodh Patil, Gen Ye, Anna Negro, Fabrizio Renzi, Angelo Ricciardone and Ryuichi Takahashi for the useful discussions and support throughout the completion of the work. 
AG acknowledges support from the NWO and the Dutch Ministry of Education.

\bibliographystyle{JHEP}
\bibliography{bibliography.bib}

\end{document}